\documentstyle[11pt,english,epsf]{report}

\newlength{\lun}
\textheight=\lun
\large
\newcommand{\be}{\begin{equation}}
\newcommand{\ee}{\end{equation}}
\newcommand{\bea}{\begin{eqnarray}}
\newcommand{\eea}{\end{eqnarray}}
\newcommand{\ba}{\begin{array}}
\newcommand{\ea}{\end{array}}
\newcommand{\beas}{\begin{eqnarray*}}
\newcommand{\eeas}{\end{eqnarray*}}
\newcommand{\ov}{\overline}

\newcommand{\bvec}{\mathbf}
\newcommand{\rt}{\rightarrow}
\newcommand{\lrt}{\leftrightarrow}

\newcommand{\gapproxeq}
{\lower .7ex\hbox{$\;\stackrel{\textstyle >}{\sim}\;$}}
\newcommand{\lapproxeq}
{\lower .7ex\hbox{$\;\stackrel{\textstyle <}{\sim}\;$}}
\newcommand{\p}{{\mathrm p}}
\newcommand{\rr}{\rho}

\newcommand{\nni}{n_i}
\newcommand{\pri}{\p_i}
\newcommand{\ri}{\rr_i}
\newcommand{\sgi}{S_i}
\newcommand{\si}{s_i}
\newcommand{\ei}{E_i}
\newcommand{\mi}{m_i}
\newcommand{\gi}{g_i}

\newcommand{\gs}{g_\ast}
\newcommand{\gss}{g_{\ast s}}
\newcommand{\td}{T^i_D}

\newcommand{\reac}{(\ref{cb1}), (\ref{cb2}), (\ref{cb3}) }

\newcommand{\om}{\omega}
\newcommand{\omr}{\omega_R}
\newcommand{\ac}{\hat{a}}
\newcommand{\bc}{\hat{b}}
\newcommand{\cc}{\hat{c}}
\newcommand{\apc}{\hat{a}^\prime}
\newcommand{\bpc}{\hat{b}^\prime}
\newcommand{\cpc}{\hat{c}^\prime}

\newcommand{\tpunto}{\frac{dT}{dt}}
\newcommand{\aphie}{\frac{\partial \rr_e}{\partial T} \, + \,
 \frac{\partial \rr_e}{\partial \phi_e} \frac{\partial \phi_e}{\partial T}}
\newcommand{\bphie}{\frac{\partial \rr_e}{\partial \phi_e}
 \frac{\partial \phi_e}{\partial n_B} \frac{d n_B}{dt} \, + \,
 \frac{\partial \rr_e}{\partial \phi_e} \frac{\partial \phi_e}{\partial q_B}
 \frac{d q_B}{dt}}
\newcommand{\tet}{3 \, \frac{1}{R} \frac{dR}{dt}}

\def\ne{\hbox{$\nu_e \!$ }}
\def\neb{\hbox{$\ov{\nu}_e \!$ }}
\def\nm{\hbox{$\nu_\mu \!$ }}
\def\nt{\hbox{$\nu_\tau \!$ }}

\def\ca{{C_{\scriptscriptstyle A}}}
\def\cv{{C_{\scriptscriptstyle V}}}

\newcommand{\mod}{\sum_{\mathrm spins} |M|^2}
\newcommand{\four}{(2\pi)^4 \,\delta^4\left( \sum_\alpha
 {\mathrm p}_\alpha\right)}
\newcommand{\tre}{\left( \cv^2 + 3 \ca^2 \right)}

\newcommand{\PTP}{\it Prog. Theo. Phys.}
\newcommand{\Nature}{\it Nature}
\newcommand{\PRep}{\it Phys. Rep.}
\newcommand{\ApJ}{{\it Astrophys. J.}}
\newcommand{\ApJS}{\it Astrophys. J. Suppl.}
\newcommand{\ARAA}{\it Annu. Rev. Astron. Astrophys.}
\newcommand{\ARNS}{\it Annu. Rev. Nucl. Sci.}
\newcommand{\NP}{\it Nucl. Phys.}
\newcommand{\PR}{\it Phys. Rev.}

\newcommand{\PRS}{\it Proc. Roy. Soc.}
\newcommand{\PL}{\it Phys. Lett.}
\newcommand{\RMP}{\it Rev. Mod. Phys.}
\newcommand{\RPP}{\it Rep. Prog. Phys.}
\newcommand{\etal}{{\it et al.}}
\begin{document}

\baselineskip=0.7cm
%-------------------------------------------------------------------------
%-------------------------------------------------------------------------
\thispagestyle{empty}

%\vspace*{.1truecm}
\centerline{\bf \Huge University of Naples ``Federico II''}
\vspace*{0.1 cm}
\begin{figure}[h]
  \begin{center}
    \mbox{\epsfysize=3cm\epsffile{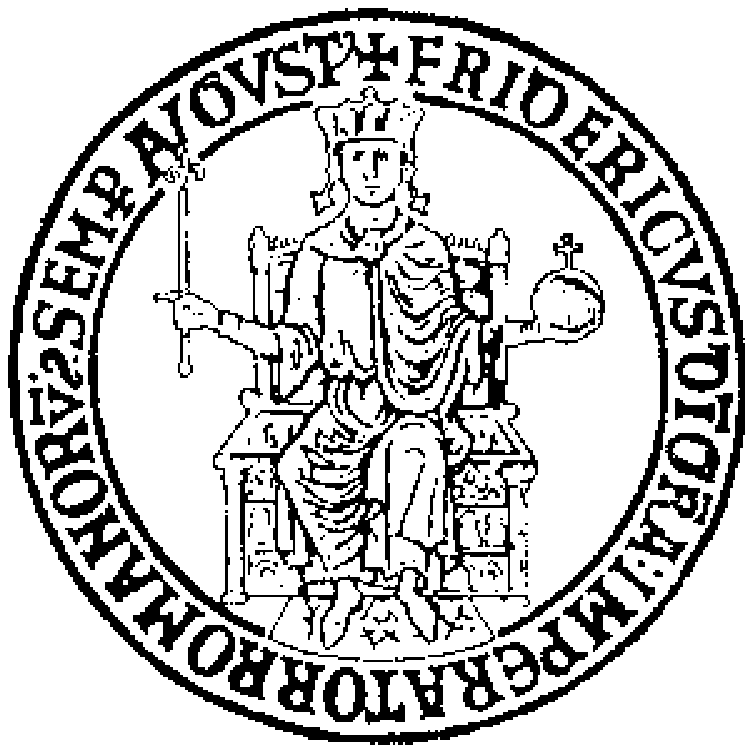}}
%   \leavevmode
%   \epsfysize=3cm
%   \epsffile{feder.eps}
  \end{center}
\end{figure}
\vspace*{0.2 cm}
\centerline{\bf \Large Faculty of Mathematical, Physical and Natural Sciences}
\vspace*{0.8 cm}
\centerline{\bf \large Ph.D. Thesis in Physics}
\vspace*{1 cm}
\begin{center}
{\LARGE {\bf Primordial Nucleosynthesis: accurate predictions
for light element abundances }}
\end{center}
\vspace*{.8 cm}
\centerline{\Large \bf Salvatore Esposito}
\vspace*{3.0cm}
\begin{flushleft}
  {} \hfill {Advisors:}\\ [.2truecm]
  {} \hfill {\large \bf Dr. G. Mangano} \\
  {} \hfill {\large \bf Dr. G. Miele} \\
  {} \hfill {\large \bf Prof. F. Buccella}\\
\end{flushleft}
%-------------------------------------------------------------------------
\newpage\thispagestyle{empty}\mbox{}\newpage\thispagestyle{empty}
\mbox{}
\vspace{2cm}

\begin{verse}

\hspace{5cm} The heavens declare the glory of God;
\\ \hspace{5cm} and the firmament sheweth his handywork. \\ \mbox{} \\

\hspace{5cm} Day unto day uttereth speech,
\\ \hspace{5cm} and night unto night sheweth knowledge. \\ \mbox{} \\

\hspace{5cm} No speech nor language,
\\ \hspace{5cm} their voice is not heard. \\ \mbox{} \\

\hspace{5cm} Their line is gone out through all the earth,
\\ \hspace{5cm} and their words to the end of the world. \\ \mbox{} \\

{} \hfill{(Psalms 19)}
\end{verse}

\newpage\thispagestyle{empty}\mbox{}\newpage\thispagestyle{empty}

\pagestyle{plain}
\pagenumbering{roman}
\setcounter{page}{1}
\tableofcontents
%-------------------------------------------------------------------------
%-------------------------------------------------------------------------
\newpage
\noindent {\huge {\bf Introduction}}
\addcontentsline{toc}{chapter}{{Introduction}}

\vspace{2cm}

\noindent
Primordial Big Bang Nucleosynthesis (BBN) \cite{pee71}-\cite{Sarkrev}
represents one of the greatest successes of the hot Big Bang model, along
with the Hubble expansion and the Cosmic Microwave Background Radiation. Of
the three, BBN probes the Universe to the earliest times, from a fraction
of a second to thousands of seconds from its born, and its formulation
predicted the existence of cosmic microwave background radiation
\cite{oli4}. Its emergence as a cosmological cornerstone relies on the
basic consistency of the predictions for the abundances of the light
elements, such as $D$, $^3He$, $^4He$, $^7Li$, with their measured
abundances, which span over more than nine orders of magnitude.
\\ BBN took place in the early Universe when the temperature scale was less
than $1 \, MeV$. The key events leading to the synthesis of the light
nuclides followed from the period when the weak interaction rates were in
equilibrium, thus fixing the ratio of number densities of neutrons to
protons; at temperatures $T \gg 1 \, MeV$ this ratio was $n/p \simeq 1$. As
the temperature fell and approached the point where the weak interaction
rates were no longer fast enough to maintain equilibrium, the neutron to
proton ratio was given approximately by the Boltzmann factor, $(n/p) \simeq
\exp\{- \Delta m / T\}$, where $\Delta m$ is the neutron-proton mass
difference. \\ The nucleosynthesis chain begins with the formation of
deuterium through the process $p \, + \, n \, \rt D \, + \, \gamma$.
However, because of the large number of photons relative to nucleons (of
the order of $10^{10}$), deuterium production is delayed past until the
temperature falls well below the deuterium binding energy of $2.2
\, MeV$ (the average photon energy in a blackbody is $<E_\gamma> \simeq 2.7
\, T$). \\ The dominant product of BBN is $^4He$, resulting in a mass
abundance close to 25 \%. Smaller amounts of other light elements are
produced: $D$ and $^3He$ at the level of about $10^{-5}$, and $^7Li$ at the
level of $10^{-10}$ per hydrogen nucleus. In the standard model (see below)
the abundances depend only on one free parameter, the baryon to photon
ratio $\eta$, and remarkablya single value for $\eta \sim 10^{-10}$ may
accommodate all observed orders of magnitude for $^{4}He$, $D$, $^{3}He$
and $^{7}Li$ data. Furthermore, the primeval yield of $^{4}He$ is also
relatively insensitive to this quantity (pinning down $\eta$ to 20 \% pegs
its value to 1 \% precision).
\\
\begin{figure}
\hspace{0.5truecm}
\epsfysize=5.5truein
\epsffile{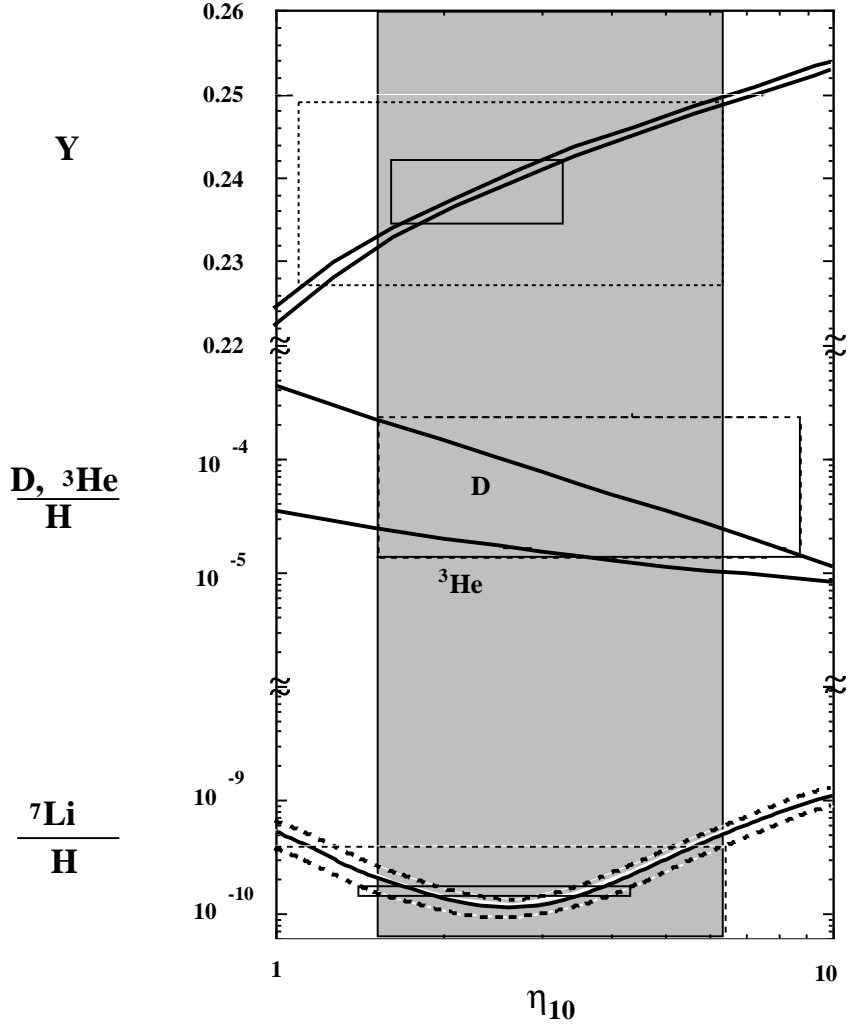}
\caption{{ Predicted abundances (solid curves) of primordial $^4He$, $D$,
$^3He$ and $^7Li$ along with their experimental determination (rectangles)
as a function of $\eta_{10} = 10^{10}\eta$. The two curves for $^4He$
correspond to the $1\sigma$ experimental values for the neutron lifetime,
while the theoretical uncertainties on $^7Li$ are as discussed in Ref.
\protect\cite{oli10}. Uncertainties on $D$ and $^3He$ curves cannot be appreciated
on the scale of this figure. The filled zone is the range of $\eta$
representing the agreement between BBN theory and measurements (taken from
\protect\cite{Olive}). }}
\label{absark}
\end{figure}
The resulting abundances of the light elements are shown in Figure
\ref{absark}, where we report the predictions for the $^4He$ mass fraction
$Y$, and $D$, $^3He$, $^7Li$ abundances relative to $H$ as function of
$\eta$. We also report the present observational situation (a complete
discussion is presented in chapter \ref{pn}). The general agreement between
experimental data and expectations emerging from this figure, ranging over
many orders of magnitude, is the great success of the Big Bang model which
we referred to above. The essential problem in attempting to compare
\cite{Sarkrev, Olive} the theoretical predictions with the observational
data is that the primordial abundances have been significantly altered
during the lifetime of the Universe through nuclear processing in stars and
other galactic chemical evolution effects (we will discuss this subject in
chapter \ref{pn}). The most stable nucleus, $^4He$, grows in abundance with
time since it is always created in stars, while $D$, the most weakly bound,
is always destroyed. The history of $^3He$ and $^7Li$ is more complicated
since these elements may be both destroyed and created during stellar
evolutiom. To avoid corrections which are difficult to treat
quantitatively, it is necessary to measure abundances in the most
primordial available material. Observations of light element abundances
have dramatically improved over the past few years \cite{Olive}. Although
$D$ and $^3He$ abundances have about a 10 \% uncertainty and $^7Li$ data
are even more uncertain, as an example we mention the fact that $^4He$ data
are now reaching a precision \cite{precision} of one per mille. \\ This
perspective, however, the fact that BBN is now entering in his maturity and
precision era, demands similar improvements in the precision of the
theoretical analysis, in order to reduce as much as possible all
uncertainties in the predictions. An increasing precision in the
measurements of $^4He$ mass fraction at the level of $10^{-4}$ requires,
for example, a reliability of neutron-proton conversion rates at the same
level of precision of all other effects which are relevant for the neutron
to proton ratio at the onset of nucleosynthesis \cite{LT, noi}. \\ This is
the main goal of this thesis. \\ The physics of BBN is well understood
\cite{hay50, alp53}: basically, it can be seen as a nuclear reactor in an
expanding box. Therefore nuclear and particle physics and cosmology are the
two basic inputs for studying primordial nucleosynthesis. Since BBN
involves events that occurred at temperatures of order $1 \, MeV$, it
naturally plays a key role in forging the connection between cosmology and
nuclear and particle physics. It is interesting to note how the increasing
interaction between particle physics and cosmology has largely resulted
from the establishments of ``standard models" in both fields which
satisfactorily describe all known phenomena. Regarding cosmology, there are
of course many possibilities for departures from its standard model
\cite{Sarkrev}, e.g. an inhomogeneous nucleon distribution or non zero
neutrino chemical potentials. However, recent developments do not motivate
such non standard scenarios and moreover they are highly constrained by the
observational data. It is therefore reasonable, and we hereafter shall do
so, to adopt the standard picture. In our calculations we will also use the
standard model of strong and electroweak interactions \cite{SM}, but it is
very intriguing to observe how BBN can constrain new physics beyond the
standard $SU(3) {\times} SU(2) {\times} U(1)$ model \cite{Sarkrev}. This is well
illustrated by the BBN limit to the number of light neutrino species
\cite{nnu}. The amount of synthesized $^4He$ strongly depends upon the
expansion rate: a faster expansion rate leads to the production of a larger
amount of $^4He$ since nuclear reactions begin earlier when the neutron
fraction is higher. The expansion rate itself is determined by the energy
density of relativistic particles and the larger the latter, the larger the
former. Now, since the energy density increases by adding new neutrino
species, this can overproduce $^4He$, thus violating the observational
evidences. The BBN limit on the number $N_\nu$ of light neutrino flavours
is close to three, although more work, both observational and theoretical,
is still needed. It is impressive and interesting to compare this result
with the LEP limit based upon the shape of the $Z^0$ resonance: $N_\nu =
2.994 {\pm} 0.012$ \cite{PDG}. While it is unlike that the BBN limit will ever
achieve such precision, the cosmological and laboratory limits are
complementary. The neutrino limit based upon the shape of the $Z^0$ counts
the number of ``active" particle species that have a mass smaller than half
the $Z^0$ mass, weighted by their coupling to the $Z^0$. Differently, BBN
constrains the energy density due to any relativistic particle specie
around the time of primordial nucleosynthesis and thus is sensitive to any
particle species lighter than about $1 \, MeV$. \\ Other several
interesting constraints on physics beyond the standard model of elementary
particles, coming from BBN considerations, can be obtained \cite{Sarkrev},
but this is not our subject. We only stress the fact that accurate BBN
predictions, along with precise experimental observations, greatly helps
also our understanding of fundamental physics.

Summarizing, this thesis is devoted to the study of precision effects
intervening during primordial nucleosynthesis. Our present goal is to give
an estimate of the primeval $^4He$ mass fraction confident, at least, at
the third significant digit (the actual experimental accuracy). The
following step is to write down a new numerical BBN code \cite{progress}
employing our results to further improve light element abundance
predictions.

The thesis is organized as follows. Chapters 1 and 2 review standard
cosmology and, in particular, primordial nucleosynthesis. The subsequent
three chapters are devoted to the calculations of the mentioned precision
effects. In particular we discuss the calculation of QED radiative
corrected Born rates for the weak reactions fixing the neutron to proton
ratio at freeze out, along with the calculation of corrections to this
quantities arising from finite nucleon mass and QED thermal plasma effects.
Finally, in chapter 6 we summarize our results and give our prediction for
the $^4He$ mass fraction. \\ Throughout this thesis, unless otherwise
specificated, we use natural units in which $\hbar = c = k = 1$.

%-------------------------------------------------------------------------
\newpage
\setcounter{page}{-1}
\pagenumbering{arabic}

%------------------------------------------------------------------------
\setcounter{chapter}{0}
\chapter{The standard cosmology: an overview}
\label{bb}

\pagenumbering{arabic}

The standard cosmological model is based on three observational pillars:
\begin{itemize}
\item[{i)}] the uniform distribution of matter in the Universe on large
scales and the isotropic expansion of it that maintains the uniformity;
\item[{ii)}] the existence of a nearly uniform and accurately thermal cosmic
background radiation (CBR);
\item[{iii)}] the abundances (relative to hydrogen) of the light elements
$D$, $^3He$, $^4He$, $^7Li$.
\end{itemize}
As we will see, the validity of the Hubble's expansion law, namely the
proportionality between the observed red-shift $z$ (see below) and the
distance $d$ from the Earth of a given source \footnote{In astronomy, this
distance is defined as $d^2 \, = \, {\cal L}/ (4 \pi {\cal F})$, with
${\cal L}$ the absolute luminosity (energy per unit time produced by the
source in its rest frame) and ${\cal F}$ the measured energy flux (energy
per unit time and surface measured by a detector in the expanding
Universe).}
\begin{equation}\label{9}
  H_0 \, d \; \simeq \; z~~~.
\end{equation}
through the Hubble constant $H_0$, out to red-shifts $z \sim 0.2$, supports
the general notion of an expanding Universe \cite{turn7}. The Hubble
constant $H_0$, defining the present expansion rate, is usually
parametrized as
\begin{equation}\label{11}
  H_0 \; = \; 100 \, h_0 \; Km \, s^{-1} \, Mpc^{-1} ~~~,
\end{equation}
where, observationally \cite{turn101}, \cite{hogan}
\begin{equation}\label{12}
  0.6 \; \leq \; h_0 \; \leq \; 0.8~~~.
\end{equation}
The distribution of matter and radiation in such a Universe is observed to
be homogeneous and isotropic when averaged on scales exceeding a few
hundred $Mpc$  \cite{Peebles}. Instead, the CBR provides a firm evidence of
a hot, dense beginning of the Universe itself (``Big Bang"); the spectrum
of the CBR is a perfect blackbody, at temperature \cite{oli3}
\begin{equation}\label{38}
  T_0 \; = \; 2.728 \, {\pm} \, 0.002 \, K ~~~,
\end{equation}
with deviations that are less than 0.03 \%. \\ Observations of the
primordial light element abundances finally provide the complete success of
the hot Big Bang model, and they will be discussed in the next chapter.

\section{Homogeneity and Isotropy}
\label{bba}

Space-time events in a homogeneous and isotropic Universe are described by
the maximally symmetric Robertson-Walker metric which, in the comoving
reference frame (defined by the property that an observer at rest in this
frame has constant spatial coordinates in time), takes the form
\cite{weinb}
\begin{equation}\label{1}
  d s^2 \; = \; d t^2 \; - \; R^2(t) \left( \frac{d r^2}{1 \, - \, k r^2}
  \; + \; r^2 \, d \theta^2 \; + \; r^2 \, \sin^2 \theta \, d \phi^2
  \right) ~~~,
\end{equation}
where $t \in [0,+ \infty)$ is the proper time measured by an observer at
rest in the comoving frame, and with $r \in [0,1]$, $\theta \in [0,\pi]$,
$\phi \in [0,2\pi]$ the (dimensionless) spherical coordinates in this
frame. The parameter $k \, = \,$+1,0,-1 gives the spatial curvature of the
Universe. The homogeneity and isotropy of space allows us to describe the
dynamics in the space-time in terms of only a quantity (with dimension of a
length), the cosmic scale factor $R(t)$, depending on the time $t$ only.\\
Since, in the Robertson-Walker metric, $R(t)$ is a function of time, the
distance between two space points depends on time; if the space described
by this metric is expanding or contracting, the motion of a particle is
influenced by this expansion/contraction. In fact, let us consider, for
example, a particle in free motion with momentum $\bvec{p}_1$ at time
$t_1$; since the momentum has dimension of an inverse of length, at a later
time $t_2$ the particle momentum will be rescaled by a factor proportional
to $R^{-1}$, that is
\begin{equation}\label{2}
  p_2 \; = \; p_1 \, \frac{R(t_1)}{R(t_2)}~~~.
\end{equation}
Furthermore, due to the finite propagation velocity of light signals, in an
expanding/contracting Universe it is meaningful to introduce the concept of
a ``distance to the horizon" $d_H(t)$: for a comoving observer, this is the
distance at which a light signal emitted at $t=0$ reaches him at (or
before) time $t$. If $d_H(t)$ is finite, then there are sources from which
light has not yet reached us \cite{rind}: a boundary exists (termed
``horizon") between the visible Universe and the part of Universe from
which light signals have not reached us. The distance $d_H(t)$ can be
calculated in the following way: from the isotropy of space, we can
consider an observer at $r=0$ for which $d \theta = d \phi = 0$; hence
\begin{equation}\label{3}
  d_H(t) \; = \; \int_0^{r_H} \, \sqrt{g_{rr}} \, d r \; = \;  R(t) \,
  \int_0^{r_H} \, \frac{d r}{\sqrt{1 \, - \, k r^2}}~~~.
\end{equation}
For a light signal $d s^2 = 0$, thus
\begin{equation}\label{4}
  d t^\prime \; = \; R(t^\prime) \, \frac{d r}{\sqrt{1 \, - \, k r^2}}
  ~~~,
\end{equation}
and correspondingly
\begin{equation}\label{5}
  d_H(t) \; = \; R(t) \, \int_0^t \frac{d t^\prime}{R(t^\prime)}~~~.
\end{equation}
The explicit form of $d_H(t)$ depends on the expression for $R(t)$ which is
determined by the cosmological equations (see next section); however, we
note that the behaviour of $R(t)$ near the initial singularity determines
the finiteness of $d_H(t)$. In the standard cosmology this is indeed finite
and $d_H(t) \sim t$.\\ Another interesting property of an
expanding/contracting Universe is that if at a given point $P_1$ a photon
is emitted with a wavelength $\lambda_1$, at a distant point $P_0$ it will
be detected with a different wavelength $\lambda_0$. In a space described
by the Robertson-Walker metric, on dimensional grounds we have in fact that
\begin{equation}\label{6}
  \frac{\lambda_1}{\lambda_0} \; = \; \frac{R(t_1)}{R(t_0)}~~~.
\end{equation}
The red-shift (or blue-shift) $z$ of a given object is defined as the ratio
between the variation in the detected wavelength and the emitted
wavelength:
\begin{equation}\label{7}
  z \; = \; \frac{\lambda_0 \, - \, \lambda_1}{\lambda_1}~~~.
\end{equation}
From (\ref{6}) we then have
\begin{equation}\label{8}
  1 \,+ \, z \; = \; \frac{R(t_0)}{R(t_1)}~~~.
\end{equation}
An increase (or decrease) in the cosmic scale factor then leads to a
red-shift (or blue-shift) of the light from distant sources. The Hubble law
(\ref{9}), relating the distance to a galaxy with the observed red-shift
through the Hubble parameter at the present epoch
\begin{equation}\label{10}
  H_0 \; = \; \frac{\dot{R}(t_0)}{R(t_0)}~~~,
\end{equation}
$t_0$ being the present time, is a direct consequence of the
Robertson-Walker metric (its derivation is reported in \cite{kolb}, for
example); it is only an approximate relation \cite{harri}, and corrections
are necessary for cosmological large distances. The present age is set by
the Hubble time
\begin{equation}\label{13}
  H_0^{-1} \; \simeq \; 9.778 {\times} 10^9 \, h_0^{-1} \; yr ~~~,
\end{equation}
corresponding to a local spatial scale for the Universe
\begin{equation}\label{14}
  H_0^{-1} \; \simeq \; 3000 \, h_0^{-1} \; Mpc
\end{equation}
(Hubble radius).

\section{Cosmological equations}
\label{bbb}

As anticipated in the previous section, the cosmic scale factor governing
the dynamics of the processes in the Universe is determined by the Einstein
field equations relating the energy-momentum tensor $T_{\mu \nu}$ to the
space-time curvature:
\begin{equation}\label{15}
  R_{\mu \nu} \; - \; \frac{1}{2} \, R_c \, g_{\mu \nu} \; = \;
  8 \pi \, G \, T_{\mu \nu} ~~~.
\end{equation}
Here $G \, = \, M_P^{-2}$ is the Newton gravitational constant, while
$R_{\mu \nu}$ is the Ricci tensor and $R_c$ its trace (the scalar
curvature). For a perfect fluid, as the Universe is assumed to be (see the
next section), the energy-momentum tensor takes the form
\begin{equation}\label{16}
  T_{\mu \nu} \; = \; - \p \, g_{\mu \nu} \; + \; \left( \p \, + \, \rr
  \right) \, u_\mu \, u_\nu ~~~,
\end{equation}
$\p$ and $\rr$ being the pressure and energy density respectively, while
$u_\mu \, = \, d x_\mu / d s$ is the fluid 4-velocity. In a comoving frame
($u_\mu = (1,\bvec{0})$), Eq. (\ref{16}) simplifies to
\begin{equation}\label{17}
  T_{\mu \nu} \; = \; \left(
  \ba{cccc} \rr &  &  &  \\  & \p &  &  \\  &  & \p & \\
   &  &  & \p \ea
  \right) ~~~.
\end{equation}
Note that from homogeneity and isotropy of space it follows that $\rr$ and
$\p$, as well as $R$, only depend on the time $t$. The Ricci tensor for the
Robertson-Walker metric is instead given by
\bea
 R_{00} & = & - \, 3 \frac{\ddot{R}}{R} ~~~~ , \;\;\;\;\;\;\;\;\;\;\;\;
 R_{0i} \; = \; 0 ~~~, \label{18} \\
 R_{ij} & = & - \left\{ \frac{\ddot{R}}{R} \; + \; 2 \, \frac{\dot{R}^2}
 {R^2} \; + \; 3 \, \frac{k}{R^2} \right\} \, g_{ij} ~~~, \label{19}
\eea
and
\begin{equation}\label{20}
  R_c \; = \; - \, 6 \, \left\{ \frac{\ddot{R}}{R} \; + \;
  \frac{\dot{R}^2}{R^2} \; + \;  \frac{k}{R^2} \right\} ~~~.
\end{equation}
Substituting these expressions in the Einstein equations (\ref{15}), we
then obtain the following two independent equations
\footnote{The equation (\ref{21}) can also be cast in the intriguing form
\begin{equation}\nonumber
  \frac{1}{2} \, v^2 \; - \; \frac{G \, M}{R} \; = \; - \, \frac{k}{2} ~~~,
\end{equation}
with $v = \dot{R}$ and $M = \frac{4}{3} \pi R^3 \rr$, showing that the sign
of $k$ is deciding about the final destiny of our Universe.}:
\begin{equation}\label{21}
 \left( \frac{\dot{R}}{R} \right)^2 \; + \; \frac{k}{R^2} \; = \;
 \frac{8 \pi \, G}{3} \, \rr ~~~,
\end{equation}
\begin{equation}\label{22}
  \frac{\ddot{R}}{R} \; = \; - \, \frac{4 \pi \, G}{3} \, \left(
  \rr \, + \, 3 \p \right) ~~~.
\end{equation}
For both matter and radiation $\rr \, + \ 3 \p$ is positive, thus
$\ddot{R}$ is always negative. Since we know that now $\dot{R} > 0$ (the
Universe is now expanding), at a given remote (but finite) time $t_{BB}$ in
the past we have had that $R(t_{BB}) = 0$; this event is called the ``Big
Bang", and it is usually chosen as the starting reference time
($t_{BB}=0$).\\ In the Eqs. (\ref{21}),(\ref{22}) the energy-momentum
conservation equation $D_\nu T^{\mu \nu} = 0$ ($D_\nu$ being the covariant
derivative in the Robertson-Walker metric) is also contained; in fact,
deriving (\ref{21}) with respect to time and substituting (\ref{22}), we
get
\begin{equation}\label{23}
  \dot{\rr} \; = \; - \, 3 \, H \, \left( \rr \, + \, \p \right) ~~~,
\end{equation}
where we have introduced the Hubble parameter (depending on time)
\begin{equation}\label{24}
  H \; = \; \frac{\dot{R}}{R} ~~~.
\end{equation}
An alternative form of (\ref{23}) is as follows:
\begin{equation}\label{25}
  d \left( \rr R^3 \right) \; = \; - \, \p \, d \left( R^3 \right) ~~~,
\end{equation}
expressing just the first law of thermodynamics for the expanding
Universe.\\ Equations (\ref{21}) and (\ref{23})
\bea
H^2 & = & \frac{8 \pi \, G}{3} \, \rr \; - \; \frac{k}{R^2} ~~~,
\label{26} \\
\dot{\rr} & = & - \, 3 \, H \, \left( \rr \, + \, \p \right) ~~~,
\label{27}
\eea
are usually considered as the two basic independent equations governing the
dynamics of the Universe. They are known as the cosmological
Friedmann-Lemaitre equations. Obviously, only two equations are not
sufficient to determine the three unknowns $R(t)$, $\rr(t)$, and $\p(t)$.
The third relation to add to the Friedmann-Lemaitre equations is the
equation of state relating pressure and energy density, that can be written
as
\begin{equation}\label{28}
  \p(t) \; = \; w \, \rr(t) ~~~,
\end{equation}
where, for simplicity, we will assume the coefficient $w$ to be time
independent. From (\ref{23}) we then have
\begin{equation}\label{29}
  \rr \; \sim \; R^{- 3 \, ( 1 + w)} ~~~.
\end{equation}
The interesting case of radiation (relativistic particles), matter (non
relativistic particles) and vacuum are summarized in Table \ref{tb1}.

 \begin{table}
\begin{center}
\begin{tabular}{|l|l|l|} \hline \hline
  & & \\
  Radiation & $\p \, = \, \frac{1}{3} \, \rr$ & $\rr \, \sim \, R^{-4}$ \\
  & & \\
  Matter    & $ \p \, = \, 0$                & $\rr \, \sim \, R^{-3}$ \\
  & & \\
  Vacuum    & $ \p \, = \, - \rr $            & $\rr \, \sim $ constant \\
  & & \\
  \hline \hline
\end{tabular}
\end{center}
\caption{Equations of state for radiation, matter and vacuum.}
 \label{tb1}
 \end{table}

\subsection{Evolution of the Universe}
\label{bbb1}

The evolution of the Universe is determined by the curvature term $k/R^2$
in (\ref{26}) which is positive, zero or negative if the energy density is
greater than, equal to or less than the critical density
\begin{equation}\label{33}
  \rr_c \; = \; \frac{3 \, H^2}{8 \pi \, G} ~~~,
\end{equation}
respectively, since Eq. (\ref{26}) can equivalently be written as
\begin{equation}\label{34}
  \frac{k}{H^2 R^2} \; =  \; \Omega \; - \; 1 ~~~,
\end{equation}
with
\begin{equation}\label{35}
  \Omega \; = \; \frac{\rr}{\rr_c} ~~~.
\end{equation}
For $k = - 1$, $\dot{R}^2$ is always strictly positive and $R \rt t$ as $t
\rt \infty$. Instead, for $k = 0$, $\dot{R}^2$ goes to zero as $R \rt
\infty$ while, for $k = -1$, $\dot{R}^2$ drops to zero at $R_{max} =
\sqrt{3/(8 \pi G \rr)}$ after which $R$ begins decreasing. Thus $\Omega <1$
corresponds to an open Universe which will expand forever, $\Omega =1$ is a
flat Universe which will asymptotically expand to infinity while $\Omega
>1$ corresponds to a closed Universe which will eventually recollapse. \\
The critical density today is (from (\ref{33}))
\begin{equation}\label{36}
  \rr_{c0} \; = \; \left( 2.999 {\times} 10^{-12} \, \sqrt{h_0} \, GeV \right)^4
  \; = \; 1.054 {\times} 10^{-5} \, h_0^2 \, GeV \, cm^{-3} ~~~.
\end{equation}
From dynamical measurements of the present energy density in all
gravitating matter (and excluding the nowadays negligible contribution of
relativistic particles) we deduce \cite{Peebles, Dekel}
\begin{equation}\label{37}
  \Omega_0 \; \approx \; 0.1 \, \div \, 1 ~~~.
\end{equation}
The present energy of photon background alone is known with a very good
accuracy from measurements of the temperature of the cosmic microwave
background radiation (\ref{38}), which gives (see sect. \ref{bbc1})
\begin{equation}\label{39}
  \rr_{\gamma 0} \; = \; \frac{\pi^2 T_0^4}{15} \; \simeq \; 2.02 {\times}10^{-21}
  \, \left( \frac{T_0}{2.73 \, K} \right)^4 \, GeV^4 ~~~,
\end{equation}
and then
\begin{equation}\label{40}
  \Omega_{\gamma 0} \; = \; \frac{\rr_{\gamma 0}}{\rr_{c0}} \; \simeq \;
  2.49 {\times} 10^{-5} \, \left( \frac{T_0}{2.73 \, K} \right)^4 \, h_0^{-2}~~~.
\end{equation}
Including the contribution of a primordial background of three massless
neutrinos (see sect. \ref{bbd1}) we have
\begin{equation}\label{41}
  \Omega_{R0} \; = \; \Omega_{\gamma 0} \; + \; \Omega_{\nu 0} \; \simeq
  \; 1.68 \, \Omega_{\gamma 0} \; \simeq \; 4.18 {\times} 10^{-5}
  \, \left( \frac{T_0}{2.73 \, K} \right)^4 \, h_0^{-2}~~~.
\end{equation}
Comparing (\ref{37}) with (\ref{41}) we then see that the present Universe
is dominated by non relativistic particles; however, as we will see in the
next subsection, the evolution to this situation has been highly not
trivial.

\subsection{Driving the expansion}
\label{bbb2}

The equation (\ref{26}) for the expansion rate of the Universe can be
conveniently normalized to present values of the quantities involved
($R_0=R(t_0)$, $H_0=H(t_0)$ and $\Omega_0 = \rr / \rr_{c0}$, $t_0$ being
the present time), obtaining
\begin{equation}\label{42}
  \left( \frac{H}{H_0} \right)^2 \; = \; \Omega_0(t) \; - \;
  \frac{k}{H_0^2 R_0^2} \, \left( \frac{R_0}{R} \right)^2 ~~~.
\end{equation}
From Table \ref{tb1} we then see that
\begin{equation}\label{43}
  \Omega_0(t) \; = \; \Omega_{R0}(t) \, \left( \frac{R_0}{R} \right)^4
  \; + \; \Omega_{M0}(t) \, \left( \frac{R_0}{R} \right)^3
\end{equation}
(the pedices $i = R,M$ indicating radiation and matter respectively), where
$\Omega_{i0}$ themselves are in general functions of time which can be
assumed, for simplicity, to be step function -like:
\begin{equation}\label{44}
  \Omega_{i0}(t) \; = \; \left\{ \ba{ll} \Omega_{i0} \;\;\;\;\;\;\;\;\;\;
  & t \in [t_i^{in},t_i^{fin}] \\  \left. \right. & \left. \right. \\
  0 \;\;\;\;\;\;\;\;\;\; & t \, {\not \in} \, [t_i^{in},t_i^{fin}]  \ea
  \right.  ~~~.
\end{equation}
Thus we get
\begin{equation}\label{45}
  \left( \frac{H}{H_0} \right)^2 \; = \;
  \Omega_{R0} \, \left( \frac{R_0}{R} \right)^4
  \; + \; \Omega_{M0} \, \left( \frac{R_0}{R} \right)^3 \; - \;
  \frac{k}{H_0^2 R_0^2} \, \left( \frac{R_0}{R} \right)^2 ~~~.
\end{equation}
Given the different powers of $R$ for the terms in (\ref{45}) determining
the expansion rate, we see that the early Universe ($R << R_0$) was
dominated by radiation (RD era), afterwards matter becomes dominating (MD
era). The transition between the RD and MD era can be approximately dated
back at $R_{EQ} \sim 10^{-4} R_0$. Since in the first stages of evolution
of the Universe the curvature term can be neglected, we find that
\bea
\mathrm{RD \; era} \;\;\;\;\;\;\;\;\;\;\;\;\;\;\;\;\;\;\;\;
 H^2 & \sim & R^{-4} \label{46} \\
\mathrm{MD \; era} \;\;\;\;\;\;\;\;\;\;\;\;\;\;\;\;\;\;\;\;
 H^2 & \sim & R^{-3} \label{47}
\eea
and thus, from (\ref{34}),
\begin{equation}\label{48}
  | \Omega \, - \, 1 | \; \sim \; \left\{
  \ba{ll} \frac{R}{R_0} \; = \; (1 + z)^{-1} \;\;\;\;\;\;\;\;\; &
  \mathrm{MD \; era} \\ \left. \right. & \left. \right. \\
  \frac{R_{EQ}}{R_0} \, \left( \frac{R}{R_{EQ}} \right)^2 \; = \;
  10^4 \, (1+z)^{-2} \;\;\;\;\;\;\;\;\; & \mathrm{RD \; era} \ea \right. ~~~.
\end{equation}
Note that, apparently, at very earlier epochs the Universe was very nearly
critical (of the order of one part over $10^4$). However, at these times a
period of ``inflation" (exponential growth of the expansion) is believed to
have occurred \cite{infl}, but a thorough analysis of this issue is beyond
the scope of this thesis.

\section{Thermodynamics of the Universe}
\label{bbc}

In this section we will study the properties of the Universe considered as
a thermodynamic system composed by different species (electrons, photons,
neutrinos, nucleons, etc.) which, in the early phases, were to a good
approximation in thermodynamic equilibrium, established through rapid
interactions. Obviously, coming back to the past, decreasing the cosmic
scale factor we have an increase of the temperature. In this thesis we will
mainly concern ourselves with cosmological processes occurred during the RD
era, thus the discussion is greatly simplified by assuming the Universe as
an ideal gas. This assumption is justified by the fact that, as we will see
in the next chapter, the particle densities do not usually become high
enough for many-body interactions to be important.

\subsection{Equilibrium thermodynamics}
\label{bbc1}

In a gas of a given specie with $\gi$ internal (spin) degrees of freedom
and energy $\ei \, = \, \sqrt{p^2 + \mi^2}$, {\it kinetic equilibrium} is
established by sufficiently rapid elastic scattering processes; in this
case, for an ideal gas, the equilibrium phase-space density is
\begin{equation}\label{58}
  f_i(p) \; = \; \left( \exp\left\{\frac{\ei \, - \, \mu_i}{T_i}\right\}
   \; {\pm} \; 1 \right)^{-1} ~~~,
\end{equation}
where +/- refers to Fermi-Dirac/Bose-Einstein statistics and $\mu_i$ is the
chemical potential. In general each specie has its own equilibrium
temperature $T_i$, and the entire Universe can be represented as a plasma
with different temperatures. However, if several species strongly interact
among them, they will reach a mutual chemical equilibrium and a common
temperature; this is indeed the situation at early times. As the Universe
expands and cools down, some species may start interacting more and more
weakly and eventually decouple. As we will see in sect. \ref{bbd}, we can
consider the photon temperature $T_\gamma$ as the plasma reference
temperature $T$ of the Universe. \\ In {\it chemical equilibrium},
established by processes which can create and destroy particles
(differently from kinetic equilibrium), the chemical potential is
additively conserved. So it is zero for particles such as photons and $Z^0$
which can be emitted and absorbed in any number and consequently opposite
for a particle and its antiparticle which can annihilate into such bosons.
\\ The quantities of interest are the number
density, energy density and pressure of a given specie, defined in general
as:
\bea
\nni & = & \gi \, \int \, \frac{d^3 \bvec{p}}{(2 \pi)^3} \, f_i(p)
~~~, \label{59} \\
\ri & = & \gi \, \int \, \frac{d^3 \bvec{p}}{(2 \pi)^3} \, \ei \,
f_i(p) ~~~, \label{60} \\
\pri & = & \gi \, \int \, \frac{d^3 \bvec{p}}{(2 \pi)^3} \,
\frac{|\bvec{p}|^2}{3 \ei} \, f_i(p) ~~~. \label{61}
\eea
In kinetic equilibrium the phase-space density $f_i(p)$ is given in
(\ref{58}), and these quantities evolve according to temperature. \\ For
non relativistic species ($T_i \ll \mi$) we have (for both Fermi-Dirac and
Bose-Einstein statistics)
\bea
\nni & \simeq & \gi \, \left( \frac{\mi T_i}{2 \pi} \right)^{\frac{3}
{2}} \, e^{- \, \frac{\mi - \mu_i}{T}} ~~~, \label{62a} \\
\ri & \simeq & \nni \, \left( \mi \, + \, \frac{3}{2} \, T_i \right)
~~~, \label{62b} \\
\pri & \simeq & \nni \, T_i \; \ll \; \ri \label{62c}
\eea
(and so we recover Boltzmann statistics). The average energy per particle
$<\ei> \equiv \ri / \nni$ and net number density are instead given by
\bea
<\ei> & \simeq & \mi \; + \; \frac{3}{2} \, T_i ~~~, \label{62d} \\ n_{i+}
\, - \, n_{i-} & \simeq & 2 \gi \, \left( \frac{\mi T_i}{2 \pi}
\right)^{\frac{3}{2}}
\, \sinh \frac{\mu_i}{T} \; e^{- \, \frac{\mi}{T}} \label{62e}
\eea
(assuming $\mu_{i+} = - \mu_{i-} \equiv \mu_i$) . \\ For relativistic
species ($T_i\gg\mi$) we obtain:
\begin{itemize}
\item non degenerate case ($T_i\gg\mu_i$)
\bea
\nni & \simeq & \left\{ \ba{llll} \frac{3}{4} &
\! \frac{1}{\pi^2} \, \zeta(3) \, \gi \, T_i^3 & & \;\;\;\;
 ~~~~~~~~~~~ \, FD \\
 & & & \\
 & \! \frac{1}{\pi^2} \, \zeta(3) \, \gi \, T_i^3  & & \;\;\;\;
 ~~~~~~~~~~~ \, BE
\ea \right. \label{63a} \\
\ri & \simeq & \left\{ \ba{llll} \frac{7}{8} &
\! \frac{\pi^2}{30} \, \gi \, T_i^4  & & \;\;\;\;
 ~~~~~~~~~~~~~~~~~FD \\
 & & & \\
 & \! \frac{\pi^2}{30} \, \gi \, T_i^4  & & \;\;\;\;
 ~~~~~~~~~~~~~~~~~BE
\ea \right. \label{63b} \\
\pri & \simeq & \frac{1}{3} \, \ri  \label{63c} \\
<\ei> & \simeq & \left\{ \ba{llll} \frac{7}{6} &
\! \frac{\pi^4}{30} \, \zeta(3) \, T_i  \; \simeq \; 3.15 \, T_i & &
\;\;\;\; FD \\
 & & & \\
 & \! \frac{\pi^4}{30} \, \zeta(3) \, T_i  \; \simeq \; 2.70 \, T_i
& & \;\;\;\; BE \ea \right. \label{63d} \\ n_{i+} \, - \, n_{i-} & \simeq &
0 \label{63e}
\eea
\item weak degeneracy ($\mu_i < 0$ \footnote{For bosons with $\mu_i >0$
Bose condensation may take place}, $|\mu_i|<T_i$) \footnote{The following
results hold for both FD and BE}
\bea
 \nni & \simeq & \frac{1}{\pi^2} \, \gi \, T_i^3 \, e^{\frac{\mu_i}{T_i}}
 \label{64a} \\
 \ri & \simeq & 3 \, \frac{1}{\pi^2} \, \gi \, T_i^4 \, e^{\frac{\mu_i}{T_i
 }} \label{64b} \\
 \pri & \simeq & \frac{1}{3} \, \ri  \label{64c} \\
 <\ei> & \simeq & 3 \, T_i  \label{64d} \\
 n_{i+} \, - \, n_{i-} & \simeq & 2 \, \frac{1}{\pi^2} \, \gi \, T_i^3 \,
 \sinh \frac{\mu_i}{T_i}  \label{64e}
\eea
\item degenerate case ($T_i \ll \mu_i$)
\bea
 \nni & \simeq & \frac{1}{6 \pi^2} \, \gi \, \mu_i^3
 \label{65a} \\
 \ri & \simeq &  \frac{1}{8 \pi^2} \, \gi \, \mu_i^4
 \label{65b} \\
 \pri & \simeq & \frac{1}{3} \, \ri  \label{65c} \\
 <\ei> & \simeq & \frac{3}{4} \, \mu_i  \label{65d} \\
 n_{i+} \, - \, n_{i-} & \simeq &  \frac{1}{6 \pi^2} \, \gi \, T_i^3 \,
 \left( \left( \frac{\mu_i}{T_i} \right)^3 \, + \, \pi^2
 \left( \frac{\mu_i}{T_i} \right) \right) ~~~.
 \label{65e}
\eea
\end{itemize}
In the relations above $\zeta(x)$ is the Riemann zeta-function, and
$\zeta(3) \simeq 1.2021...$.

Another important quantity for the evolution of a thermodynamic system is
its entropy $S$ which, for thermal equilibrium, is defined by the second
law of thermodynamics:
\begin{equation}\label{66}
  T \, d S \; = \; \delta Q ~~~.
\end{equation}
Assuming zero chemical potentials, for a given comoving volume element
($V=R^3$) we have
\begin{equation}\label{67}
  T_i \, d \sgi \; = \; d (\ri V) \; + \; \pri dV \; = \; d ( (\ri \, + \,
  \pri) \, V) \; - \; V \, d \pri
\end{equation}
(note that $\ri$ and $\pri$ are the energy density and pressure of species
at equilibrium). The entropy function is subject to the integrability
condition
\begin{equation}\label{68}
  \frac{\partial^2 \sgi}{\partial T_i \, \partial V} \; = \;
 \frac{\partial^2 \sgi}{\partial V \, \partial T_i} ~~~,
\end{equation}
which explicitly implies that
\begin{equation}\label{69}
  - \, \frac{\pri \, + \, \ri}{T_i^2} \; + \;
  \frac{1}{T_i} \,  \frac{d \pri}{d T_i} \, + \,  \frac{1}{T_i} \,
  \frac{d \ri}{d T_i} \; = \; \frac{1}{T_i} \,  \frac{d \ri}{d T_i} ~~~,
\end{equation}
or
\begin{equation}\label{70}
  d \pri \; = \; \frac{\pri \, + \, \ri}{T_i} \, d T_i ~~~.
\end{equation}
This relation is in fact verified by (\ref{60}), (\ref{61}) with the
equilibrium distribution (\ref{58}). Substituting in (\ref{67}) we then
have
\begin{equation}\label{71}
  d \sgi \; = \; d \left( V \, \frac{\pri \, + \, \ri}{T_i} \right) ~~~,
\end{equation}
from which we found that the entropy per comoving volume ($V \propto R^3$)
is defined, up to an additive constant, by
\begin{equation}\label{72}
  \sgi \; = \; V \, \frac{\pri \, + \, \ri}{T_i} \; = \; R^3 \,
 \frac{\pri \, + \, \ri}{T_i} ~~~.
\end{equation}
Note that from the first law of thermodynamics (\ref{25}) and (\ref{67})
\begin{equation}\label{73}
  d \sgi \; = \; 0 ~~~,
\end{equation}
i.e. in thermal equilibrium \footnote{During phase transitions entropy is
not conserved, in general.} the entropy in the comoving volume is
conserved.
\\ It is also useful to define the entropy density in the comoving volume,
\begin{equation}\label{74}
  \si \; \equiv \; \frac{\sgi}{V} \; = \;  \frac{\pri \, + \, \ri}{T_i}
  ~~~,
\end{equation}
which, from (\ref{73}), varies as
\begin{equation}\label{75}
  \si \; \propto \; R^{-3} ~~~.
\end{equation}
The total number of particles of a given specie in the comoving volume $N_i
= V \nni = R^3 \nni$ is thus proportional to $\nni / \si$,
\begin{equation}\label{76}
  N_i \; \propto \; \frac{\nni}{\si} ~~~.
\end{equation}
Finally, if chemical potentials are not zero, Eq. (\ref{66}) specializes
into
\begin{equation}\label{77}
  T_i \, d \sgi \; = \; d (\ri V) \; + \; \pri dV  \; - \; \mu_i \,
  d (\nni V) ~~~,
\end{equation}
and then the entropy is now given by
\begin{equation}\label{78}
  \sgi  \; = \; R^3 \, \frac{\pri \, + \, \ri \, - \mu_i \nni}{T_i} ~~~,
\end{equation}
which replaces Eq. (\ref{72}).

\subsection{Energy density and pressure in the Universe}
\label{bbc2}
In the Friedmann-Lemaitre equations (\ref{26}), (\ref{27}), assuming a
perfect fluid, the total energy density $\rr$ and pressure $\p$ of the
cosmological fluid are involved; let us evaluate these quantities
considering all the particles in thermal equilibrium in the Universe and
express them in terms of photon temperature $T_\gamma \equiv T$:
\bea
 \rr & = & T^4 \, \sum_i \, \left( \frac{T_i}{T} \right)^4 \,
 \frac{\gi}{2 \pi^2} \, \int_{x_i}^{\infty} \,
 \frac{(u_i^2 - x_i^2)^{1/2} \, u_i^2 \, d u_i}{\exp \left\{
 u_i - y_i \right\} \, {\pm} \, 1} ~~~, \label{79} \\
 \p & = & T^4 \, \sum_i \, \left( \frac{T_i}{T} \right)^4 \,
 \frac{\gi}{6 \pi^2} \, \int_{x_i}^{\infty} \,
 \frac{(u_i^2 - x_i^2)^{3/2} \, d u_i}{\exp \left\{ u_i - y_i \right\}
 \, {\pm} \, 1} ~~~, \label{80}
\eea
where the sums run over all species and $u_i = \ei / T_i$, $x_i = \mi /
T_i$, $y_i = \mu_i / T_i$. Note that from (\ref{62a}), (\ref{62b}), non
relativistic particles contribute negligibly to the energy density in the
radiation dominated era, since their energy density is exponentially
suppressed with respect to the case of relativistic particles; thus we can
neglect the contribution of non relativistic species in the sums above.
Assuming all species non degenerate, we then get
\bea
 \rr & \simeq & \frac{\pi^2}{30} \, \gs \, T^4  ~~~, \label{81} \\
 \p & \simeq & \frac{1}{3} \, \rr \; = \; \frac{\pi^2}{90} \,
 \gs \, T^4  ~~~, \label{82}
\eea
where in $\gs$ contribute only the relativistic degrees of freedom
\begin{equation}\label{83}
  \gs \; \simeq \; \sum_B \, \gi \left( \frac{T_i}{T} \right)^4 \; + \;
  \frac{7}{8} \, \sum_F \, \gi \left( \frac{T_i}{T} \right)^4
\end{equation}
($B$ = bosons, $F$ = fermions). Note that $\gs$ is in general a function of
$T$, since the number of degrees of freedom becoming relativistic at a
given temperature depends on $T$ itself. Moreover, at a given time, not all
(relativistic) particles in the bath are in equilibrium at a common
temperature $T$. A particle will be in kinetic equilibrium with the
background plasma (that is $T_i = T$) only as long as its interaction with
the plasma is fast enough; although the conditions for this to occur will
be discussed in sect. \ref{bbd}, it is obvious that these involve a
comparison between the particle interaction and the expansion rate $H$.

\subsection{Total entropy density in the comoving volume}
\label{bbc3}

From (\ref{74}) and (\ref{81}), (\ref{82}) we can evaluate the total
entropy density in the comoving volume, obtaining
\begin{equation}\label{84}
  s \; = \; \frac{2 \pi^2}{45} \, \gss \, T^3 ~~~,
\end{equation}
with
\begin{equation}\label{85}
  \gss \; \simeq \; \sum_B \, \gi \left( \frac{T_i}{T} \right)^3 \; + \;
  \frac{7}{8} \, \sum_F \, \gi \left( \frac{T_i}{T} \right)^3 ~~~,
\end{equation}
and the sums again run only over the relativistic degrees of freedom in
equilibrium (in the considered approximation). Note that $s$ in (\ref{84})
can be parametrized in terms of the photon number density as follow
\begin{equation}\label{86}
  s \; = \;  \frac{\pi^4}{45 \zeta(3)} \, \gss \, n_\gamma \; \simeq \;
  1.80 \, \gss \, n_\gamma ~~~.
\end{equation}
From entropy conservation we can now obtain the scaling law relating the
cosmic scale factor with the temperature; in fact from (\ref{73}) and
(\ref{84}) we get
\begin{equation}\label{87}
  T \; \sim \; \gss^{- \, 1/3} \, R^{-1} ~~~.
\end{equation}
We stress that if $\gss$ depends on $T$, then the adiabatic invariant $R T
= const.$ for pure expansion does not hold.

\subsection{Decoupling}
\label{bbc4}

Non equilibrium phenomena are properly described by the Boltzmann equation
for the phase space distribution functions of the given species involved in
them. Such description is reported in appendix \ref{ane}. Here we make a
very simple and general analysis of non equilibrium phenomena occurring
during the evolution of the Universe. \\ A specie $i$ is said to {\it
decouple} at a temperature $\td$ (corresponding to a given $R_D$ and time
$t_D$) when it goes out of equilibrium with the background plasma. The
conditions for this to occur will be considered in sect. \ref{bbd}. Of
course no particle is ever truly decoupled since there are always some
residual interactions with the bath, but it is a convenient assumption to
consider that a specie decouples at a given temperature $\td$. \\ Let us
first consider species that are relativistic at decoupling ($\mi < \td$)
\footnote{For simplicity we consider only the non degenerate case}. In
this case the distribution at decoupling results to be the equilibrium one:
\begin{equation}\label{88}
  f_i(p, \td) \; = \; \left( \exp \left\{ \frac{\ei}{\td} \right\}
  \; {\pm} \; 1 \right)^{-1} ~~~.
\end{equation}
Subsequently the decoupled $i$ particles will expand freely without
interactions, hence their number in a comoving volume is conserved and
their thermodynamical quantities are functions of the scale factor $R$
alone. Although non interacting, after decoupling their phase space
distribution will retain the equilibrium form as long as the particles
remain relativistic, since from (\ref{2}) we have
\begin{equation}\label{89}
  \ei(t_D) \; = \; \ei \, \left( \frac{R}{R_D} \right)~~~,
\end{equation}
and thus for $t \geq t_D$
\begin{equation}\label{90}
  f_i(p, T_i \leq \td) \; = \; \left( \exp \left\{ \frac{\ei R}{\td R_D}
  \right\} \; {\pm} \; 1 \right)^{-1} \; = \;
  \left( \exp \left\{ \frac{\ei}{T_i}  \right\} \; {\pm} \; 1 \right)^{-1}
   ~~~.
\end{equation}
This is because of the adiabatic invariant (for pure expansion)
\begin{equation}\label{91}
  T_i \; = \; \td \left( \frac{R_D}{R} \right) \; \sim R^{-1} ~~~.
\end{equation}
Initially the temperature $T_i$ exactly follows the photon temperature but,
as the Universe cools below some mass thresholds, the corresponding massive
particles will become non relativistic and annihilate. This will heat the
photons and other interacting particles but not the decoupled $i$
particles, so that $T_i$ will drop below $T$ and consequently $n_i /
n_\gamma$ will decrease below its value at decoupling.
\\
For species that are non relativistic at decoupling ($\mi \gg \td$),
(\ref{88}) continues to be valid and after decoupling the phase space
distribution will again retain its equilibrium form, but now $\ei(t_D)
\simeq p_D^2 / 2 \mi$ and from (\ref{2}) we have
\begin{equation}\label{92}
  \ei(t_D) \; = \; \ei \, \left( \frac{R}{R_D} \right)^2~~~,
\end{equation}
thus for $t \geq t_D$
\begin{equation}\label{93}
  f_i(p, T_i \leq \td) \; = \; \left( \exp \left\{
  \frac{\ei R^2}{\td R_D^2} \right\} \; {\pm} \; 1 \right)^{-1} \; = \;
  \left( \exp \left\{ \frac{\ei}{T_i}  \right\} \; {\pm} \; 1 \right)^{-1}~~~,
\end{equation}
where
\begin{equation}\label{94}
  T_i \; = \; \td \left( \frac{R_D}{R} \right)^2 \; \sim R^{-2} ~~~.
\end{equation}
From the discussion above it is then clear that ultrarelativistic and non
relativistic species retain their equilibrium distribution after decoupling
with temperatures varying as in (\ref{91}) or (\ref{94}). This is no longer
true for species with intermediate energy at decoupling, $\td \sim \mi$. In
fact for these we have
\begin{equation}\nonumber
  \frac{\ei(t_D)}{\td} \; = \; \frac{1}{\td} \, \sqrt{p_D^2 \, + \, \mi^2}
  \; = \; \frac{1}{\td} \, \sqrt{\left( \frac{R}{R_D} \right)^2 \, p^2 \,
  + \, \mi^2} \; \equiv \; \frac{\ei}{T_i}~~~,
\end{equation}
which ``defines'' a momentum dependent temperature, indicating thus a non
equilibrium distribution.

\subsection{Time - temperature relationship}
\label{bbc5}

The useful relationship between the time $t$ and the background (photon)
temperature $T$ in the radiation dominated era can be obtained
straightforwardly by integrating the Friedmann-Lemaitre equations by means
of entropy conservation. In fact,
%from (\ref{49})
\begin{equation}\label{95}
  t \; = \; \int_0^{R(t)} \frac{1}{H} \, \frac{dR}{R}~~~,
\end{equation}
by using the Eq. (\ref{26}) (neglecting the curvature term in the RD era)
and (\ref{81}), giving
\begin{equation}\label{96}
  H \; \simeq \; \sqrt{\frac{4 \pi^3}{45 M_P^2}} \, \gs^{1/2} \, T^2 ~~~,
\end{equation}
and entropy conservation (\ref{73}) with Eq. (\ref{84}),
\begin{equation}\label{97}
  - \, 3 \, \frac{dR}{R} \; = \; 3 \, \frac{dT}{T} \; +  \;
  \frac{d \gss}{\gss} ~~~,
\end{equation}
we find
\begin{equation}\label{98}
  t \; = \; - \, \sqrt{\frac{45 M_P^2}{4 \pi^2}} \, \int_0^T \, \gs^{- 1/2}
  \, \left( 1 \, + \, \frac{1}{3} \, \frac{T}{\gss} \, \frac{d \gss}{d T}
  \right) \, \frac{dT}{T^3} ~~~.
\end{equation}
During the periods when both $\gs$ and $\gss$ are approximately constant
(i.e. away from phase transition and mass thresholds where relativistic
degrees of freedom change) the relation (\ref{98}) simplifies to
\begin{equation}\label{99}
  t \; \simeq \; 2.42 \, \gs^{- 1/2} \, \left( \frac{1 \, MeV}{T}
  \right)^2 \; sec
\end{equation}
or, quite approximately, $t(sec) \sim T^{-2}(MeV)$.

\section{Thermal evolution of the Universe}
\label{bbd}

The cosmological model of Friedmann-Lemaitre, as described in the previous
sections, does not allow a Universe in thermal equilibrium since, in this
case, the Universe itself would be stationary, which is not the case.
However, sufficiently away from the Big Bang event, we can nevertheless
approximate the evolution of the Universe as made of several subsequent
phases of (different) thermal equilibrium with temperature $T$ mainly
varying as $R^{-1}$. Thermal equilibrium is realized if the reactions
between the particles in the heat bath take place very rapidly compared
with the expansion rate, set by $H$. Thus denoting with $\Gamma =
\nni <\sigma v>$ the thermal average of the given scattering reaction rate
($\sigma$ and $v$ being the cross section and the particle velocity,
respectively), the (approximate) condition for having equilibrium is given
by \cite{wag80}
\begin{equation}\label{100}
  \Gamma \; \gapproxeq \; H ~~~.
\end{equation}
Of course the opposite condition $\Gamma \lapproxeq H$ is not a sufficient
one for the exit from the equilibrium since, as we have seen, for example,
a relativistic low interacting specie maintain its equilibrium distribution
but with $T \sim R^{-1}$. Nevertheless the decoupling temperature $T_D$, as
used above, is defined by
\begin{equation}\label{101}
  \Gamma(T_D) \; = \; H(T_D) ~~~.
\end{equation}
The correct way of proceeding is to solve the Boltzmann transport equations
for the given specie (see Appendix \ref{ane}), but here we are interested
in a semi-quantitative discussion, and the criteria above fit well. \\ The
scattering rate depends on the particular interaction experienced by the
particles, while (in the RD era) the expansion rate is approximately
\begin{equation}\label{102}
  H \; \sim \; \frac{T^2}{M_P} ~~~.
\end{equation}
For interactions mediated by massless gauge bosons (for example photons or
$W^{\pm}$,$Z^0$ before the electroweak phase transition) there is no particular
mass parameter, so that for reactions of the type $a + b \, \rt \, c + d$
the cross section is approximately given by $< \sigma > \sim \alpha^2
T^{-2}$, $\alpha$ being a dimensionless coupling constant. Thus, for
example, for relativistic particles in the radiation dominated era (and $n
\sim T^3$) we have
\begin{equation}\label{103}
  \Gamma \; \sim \; \alpha^2 \, T ~~~.
\end{equation}
Hence for this type of interactions the decoupling temperature is set by
$T_D \sim \alpha^2 M_P$ and for $T \lapproxeq \alpha^2 M_P$ equilibrium is
established, while for $T \gapproxeq \alpha^2 M_P$ the reaction rates take
place slowly (they are {\it frozen out}). \\ For interactions mediated by
gauge bosons with mass $M_X$ we now have $<\sigma> \sim G_X^2 T^2 \sim
\alpha^2 M_X^{-4} T^2$ and
\begin{equation}\label{104}
  \Gamma \; \sim \; \frac{\alpha^2 \, T^5}{M_X^4} ~~~.
\end{equation}
In this case the decoupling temperature results $T_D \sim (M_X^4 /
\alpha^2 M_P^2)^{1/3}$ and for temperature above this (but below $M_X$)
equilibrium is maintained, while for $T \lapproxeq T_D$ the reactions are
frozen out.

In practice, apart from the massless neutrinos (whose decoupling will be
studied below), all particles in the Standard Model of elementary particles
are strongly coupled to the thermal plasma (of $e^{\pm} , \gamma$) while they
are relativistic. This is why we can reliably describe the evolution only
in terms of the `plasma' (photon) temperature $T$ and neutrino temperature
$T_\nu$. Thus the thermal history of the Universe, as pictorially reported
in table \ref{tb2}, can be fairly reconstructed  reliably back to the Fermi
scale but, with some cautions,  also nearly up to the GUT scale. This
happens because of the dilute RD plasma approximation, in which  non
relativistic particles of plasma are (forced) in equilibrium with the
relativistic species with negligible abundances. Some uncertainties arise,
however, when the ideal gas approximation breaks down, that is at phase
transitions associated with symmetry breaking and at very high
temperatures, where equilibrium is not achieved.

\subsection{Neutrino decoupling}
\label{bbd1}

The above discussion can be interestingly illustrated by the example of the
decoupling of (massless) neutrinos. They are maintained in equilibrium by
reaction such as $\nu \ov{\nu} \rt e^+ e^-$, $\nu e \rt \nu e$ and so on,
whose averaged cross sections are of the order of $<\sigma> \sim G_F^2
T^2$, where $G_F$ is the Fermi coupling constant. From (\ref{101}),
(\ref{102}), (\ref{104}) we then find the neutrino decoupling temperature
\begin{equation}\label{105}
  T_D^\nu \; \sim \; 1 \, MeV
\end{equation}
(a more careful estimate of the cross sections \cite{Dicus82, Enqvist92}
gives $T_D^{\nm , \nt} \simeq 3.5 \, MeV$ and $T_D^{\ne} \simeq 2.3 \,
MeV$, the higher decoupling temperature of $\nm , \nt$ depending on the
fact that they  interact only through weak neutral current, while for $\ne$
there is also a charged current contribution). Thus  above $\sim 1
\, MeV$ neutrinos are in equilibrium with the plasma of photons with $T_\nu
= T$ and (from (\ref{63a})) $n_\nu = (3/4) n_\gamma$, afterwards they
decouple from the plasma and their temperature $T_\nu$ scales approximately
as $R^{-1}$. Subsequently, as the temperature drops below $\sim 0.5 \,
MeV$, $e^+$ and $e^-$ annihilate, heating the photons but not the decoupled
neutrinos. In these three phases we have respectively
\bea
 \gs^{(1)} & = & \gs(T \gapproxeq T_D^\nu) \; = \; 2 \; + \; \frac{7}{8}
 ( 4 \, + \, 3 {\times} 2 ) \; = \; \frac{43}{4}  ~~~, \label{106} \\
 \gs^{(2)} & = & \gs(m_e \lapproxeq T \lapproxeq T_D^\nu) \; = \; 2 \; + \;
  \frac{7}{8} \, 4  \; = \; \frac{11}{2}  ~~~, \label{107} \\
 \gs^{(3)} & = & \gs(T \lapproxeq m_e) \; = \; 2  \label{108} ~~~.
\eea
The evolution of the neutrino temperature through the period of $e^{\pm}$
annihilation can be easily computed by entropy conservation. Note that from
(\ref{73}) the entropy of each specie in equilibrium in the comoving volume
is conserved but, in the present case, also the neutrino entropy after
their decoupling is conserved since, after decoupling, the total neutrino
number in the comoving volume does not change anymore (assuming stable
neutrinos). Thus from the conservation of the total entropy $S = S_I +
S_\nu$ and neutrino entropy $S_\nu$, it follows that also the entropy $S_I$
associated to the species still interacting \cite{alp53}, given
approximately by (see (\ref{86}), (\ref{84}))
\begin{equation}\label{109}
  S_I \; \sim \; \gs \, N_\gamma \; \sim \; \gs \, R^3 \, T^3 ~~~,
\end{equation}
is conserved, i.e.
\begin{equation}\label{110}
  S_I^{(2)} \; = \; S_I^{(3)} ~~~,
\end{equation}
implying, from (\ref{107}), (\ref{108}),
\bea
 N_\gamma^{(3)} & = & \frac{11}{4} \, N_\gamma^{(2)}  ~~~, \label{111} \\
 \left( R \, T \right)^{(3)} & = & \left( \frac{11}{4}
 \right)^{\frac{1}{3}} \, \left( R \, T \right)^{(2)}  \label{112} ~~~.
\eea
However, $\left( R \, T \right)^{(3)} = \left( R \, T_\nu \right)^{(2)} =
const. = \left( R \, T_\nu \right)^{(3)}$ (since $T_\nu \sim R^{-1}$), for
which, after $e^{\pm}$ annihilation
\begin{equation}\label{113}
  \frac{T}{T_\nu} \; \simeq \; \left( \frac{11}{4} \right)^{\frac{1}{3}}
  ~~~.
\end{equation}
Note that after $e^{\pm}$ annihilation there are no other relativistic species
that can become non relativistic (altering the effective degrees of
freedom) so that (\ref{113}) is just the relation between the present
values of  $\gamma$ and $\nu$ temperature, as well as (\ref{111}) relates
the present number of $\gamma$ to that preceding $e^{\pm}$ annihilation.
Furthermore, we also have from (\ref{111})
\begin{equation}\label{114}
  \left( \frac{n_\nu}{n_\gamma} \right)^{(3)} \; = \; \frac{4}{11} \,
  \left( \frac{n_\nu}{n_\gamma} \right)^{(2)} \; = \; \frac{4}{11} \,
  {\times} \, \frac{3}{4} \; = \; \frac{3}{11} ~~~.
\end{equation}
From this value of $T$ till present days, neutrinos  remain relativistic
and therefore continue to retain their equilibrium distribution; hence the
degrees of freedom characterizing the present energy density and entropy
($\gamma$ + 3 $\nu$) are given by
\bea
 \gs^0 & = & 2 \; +  \; \frac{7}{8} \, \left( 3 {\times} 2 \right) \,
 \left( \frac{4}{11} \right)^{\frac{4}{3}} \; \simeq \; 3.36 \label{115} \\
 \gss^0 & = & 2 \; +  \; \frac{7}{8} \, \left( 3 {\times} 2 \right) \,
 \left( \frac{4}{11} \right) \; = \; \frac{43}{11} \label{116}
\eea
from which we deduce, for example, ($T \simeq 2.73 \, K$)
\bea
 n^0_\gamma & = & \frac{2 \zeta(3)}{\pi^2} \, T^3 \; \simeq \; 422 \,
 cm^{-3} \\
 s^0_\gamma & = & \frac{2 \pi^2}{45} \, \gss^0 \, T^3 \; \simeq \;
 2970 \, cm^{-3} \\
 \rr_R^0 & = & \frac{\pi^2}{30} \, \gs^0 \, T^4 \; \simeq \; 8.09 {\times}
 10^{-34} \, g \, cm^{-3} \\
 \Omega_R^0 \, h_0^2 & \simeq & 4.31 {\times} 10^{-5} ~~~.
\eea
The present quantities related to neutrinos are obtained from
\bea
 n_\nu^0 & = & \frac{3}{11} \, n_\gamma^0  \\
 T_\nu^0 & \simeq & 1.96 \, K ~~~.
\eea
A more careful analysis of the $T_\nu$ - $T$ relation is conducted in
appendix \ref{atnu}, where we report the generalization of Eq. (\ref{113}).

 \begin{table}
\caption{History of the Universe.}
\begin{center}
\begin{tabular}{|c|c|c|c|} \hline \hline
  Time & Temperature & Event & Particle content \\
  \hline \hline
  today & $T \simeq 2.73 \, K$ & &  $\gamma$ + 3 decoupled $\nu$  \\
  \hline
  $t \sim 10^{17} \, s$ & $T \sim 10^{-2} \, eV$ & Galaxy formation
  & " \\
  \hline
  $t \sim 10^{13} \, s$ & $T \sim 1 \, eV$ & Matter-Radiation decoupling
  & " \\
  & & Atom formation & \\
  \hline
  $t \sim 10^{11} \, s$ & $T \sim 10 \, eV$ & $\rr_{matter} \sim
  \rr_{radiation}$ & " \\
  & & Structure formation & \\
  \hline
  $t \sim 10^{4} \, s$ & $T \sim 10^{4} \, eV$ & Planck spectrum
  established & " \\
  \hline
  & $T \sim m_e$ & & add $e^{\pm}$ \\
  \hline
  $t \sim 1 \, s$ & $T \sim 1 \, MeV$ & Light element nucleosynthesis &
  " \\
  \hline
  & $T \sim 2 \div 3 \, MeV$ & & $\nu$s become interacting \\
  \hline
  & $T \sim m_\mu$ & & add $\mu^{\pm}$ \\
  \hline
  & $T \sim m_\pi$ & & add $\pi^{\pm}$, $\pi^0$ \\
  \hline
  $t \sim 10^{-6} \, s$ & $T \sim 150 \div 400 \, MeV$ & Quark-hadron
  transition & \\
%  & & (quarks and gluons are confined into hadrons) & \\
  \hline
  & $T \sim m_s$ & & $\gamma , 3 \nu , e^{\pm} , \mu^{\pm}$ \\
  & & &  $u,\ov{u} , d,\ov{d} , s,\ov{s} , gluons$\\
  \hline
  & $T \sim m_c$ & & add $c,\ov{c}$ \\
  \hline
  & $T \sim m_\tau$ & & add $\tau^{\pm}$ \\
  \hline
  & $T \sim m_b$ & & add $b,\ov{b}$ \\
  \hline
  & $T \sim m_{W,Z}$ & & add $W^{\pm}, Z^0$ \\
  \hline
  & $T \sim m_t$ & & add $t,\ov{t}$ \\
  \hline
  $t \sim 10^{-11} \, s$ & $T \sim 300 \, GeV$ & Electroweak phase
  transition & add $H^0$ \\
  \hline
  $t \sim 10^{-34} \, s$ & $T \sim 10^{15} \, GeV$ & GUT symmetry breaking
  & add $X,Y$ \\
  & & Inflation & \\
  \hline
  $t \sim 10^{-43} \, s$ & $T \sim 10^{19} \, GeV$ & Quantum gravity & ? \\
  \hline \hline
\end{tabular}
\end{center}
\label{tb2}
 \end{table}

\chapter{Primordial Nucleosynthesis}
\label{pn}

\section{Preliminaries}
\label{pna}

During about the first 20 minutes in the evolution of the Universe,
conditions were favourable for the synthesis of (significant abundances of)
light nuclides, such as $D$, $^3H$, $^3He$, $^4He$, $^7Li$, ... . The onset
of nuclear reactions which build up the light nuclei takes place slightly
after the decoupling of the weak interactions (see table \ref{tb2}) which
keep neutron and proton in chemical equilibrium. This is a necessary step
since, as long as free nucleons are in equilibrium, the (equilibrium)
abundances of all bound nuclei are quite negligible, due to the very high
value of the entropy per nucleon ($s/n_N\sim10^{10}$). \\ Introducing the
subject, we will start with some definitions and comments. The number
density $n_A$ of a given nuclear specie $(Z,A)$ has been defined in
(\ref{59}) and, since the specie is non relativistic, it is given by the
expression (\ref{62a}):
\begin{equation}\label{n1}
  n_A \; \simeq \; g_A \, \left( \frac{m_A T}{2 \pi} \right)^{\frac{3}
  {2}} \, e^{- \, \frac{m_A - \mu_A}{T}} ~~~.
\end{equation}
The baryon number (density) $n_B$ is the total number of nucleons (bound or
free) in the comoving volume:
\begin{equation}\label{n2}
  n_B \; = \; \sum_A \, A \, n_A ~~~.
\end{equation}
It is also useful to introduce the dimensionless parameter
\begin{equation}\label{n3}
  \eta \; = \; \frac{n_B}{n_\gamma} ~~~,
\end{equation}
which measures the baryon to photon number ratio. \\ The mass fraction
$X_A$ of a given specie is instead defined as \footnote{In many papers, the
mass fraction for the primordial $^4He$ is usually denoted with $Y$.}
\begin{equation}\label{n4}
  X_A \; = \; \frac{A n_A}{n_B}
\end{equation}
and verifies the normalization condition
\begin{equation}\label{n5}
  \sum_A \, X_A \; = \; 1 ~~~.
\end{equation}
Finally, the abundance $Y_A$ of a nuclide relative to hydrogen is
\begin{equation}\label{n6}
  Y_A \; = \; \frac{n_A}{n_H} ~~~,
\end{equation}
and, in terms of this, the mass fraction can be also expressed as
\begin{equation}\label{n6b}
  X_A \; = \; \frac{A Y_A}{\sum_A \, A Y_A} ~~~.
\end{equation}
In chemical equilibrium, the parameter $\mu_A$ in (\ref{n1}) is given by
\begin{equation}\label{n7}
  \mu_A \; = \; Z \, \mu_p \; + \; \left( A \, - \, Z \right) \, \mu_n ~~~,
\end{equation}
where $\mu_p , \mu_n$ are the proton and neutron chemical potential,
respectively. In terms of the quantities of the constituent nucleons, the
number density $n_A$ can therefore be written as
\begin{equation}\label{n8}
  n_A \; = \; g_A \, \frac{A^{\frac{3}{2}}}{2^A} \, \left( \frac{2 \pi}
  {m_N T} \right)^{\frac{3}{2} (A - 1)} \, n_p^Z \, n_n^{A - Z} \,
  e^{\frac{B_A}{T}} ~~~,
\end{equation}
where $m_N$ is the nucleon mass and
\begin{equation}\label{n9}
  B_A \; = \; Z \, m_p \; + \; \left( A \, - \, Z \right) \, m_n \; - \;
  m_A
\end{equation}
is the binding energy of the given nuclear specie $(Z,A)$. From (\ref{n8})
and (\ref{63a}) (applied to photons) we can then deduce the following
formula for the mass fraction in terms of the nucleon quantities and the
$\eta$ parameter:
\begin{equation}\label{n10}
  X_A \; = \; \left( \frac{\zeta(3)}{\sqrt{8 \pi}} \right)^{A-1} \,
  \frac{g_A}{2} \, A^{\frac{5}{2}} \, \left( \frac{T}{m_N}
  \right)^{\frac{3}{2} (A-1)} \, \eta^{A-1} \, X_p^{Z} \, X_n^{A-Z} \,
  e^{\frac{B_A}{T}} ~~~.
\end{equation}
From (\ref{n8}) or (\ref{n10}) we then see that the greater is the binding
energy of a given specie, the larger is its abundance (though, of course,
this is not the only condition for having an appreciable amount of a
nuclide). In table \ref{tpn1} we report the values of the binding energy of
light nuclides relevant for nucleosynthesis.

\begin{table}
\caption{Binding energies (in KeV) of light nuclides (data taken from
\protect\cite{ndata}).}
\begin{center}
\begin{tabular}{|rl|rl|} \hline \hline
$n$ & 0.0 & $^{11}B$ & 76204.800 ${\pm}$ 0.421   \\
\hline
$p$ & 0.0 & $^{11}C$ & 73439.899 ${\pm}$ 0.952  \\
\hline
$D$ & 2224.573 ${\pm}$ 0.002 & $^{12}B$ &  79575.205 ${\pm}$ 1.400  \\
\hline
$^3H$ & 8481.821 ${\pm}$ 0.004 & $^{12}C$ & 92161.753 ${\pm}$ 0.014 \\
\hline
$^3He$ & 7718.058 ${\pm}$ 0.002 & $^{12}N$ & 74041.317 ${\pm}$ 1.000  \\
\hline
$^4He$ & 28295.673 ${\pm}$ 0.005 & $^{13}C$ & 97108.065 ${\pm}$ 0.016 \\
\hline
$^6Li$ & 31994.564 ${\pm}$ 0.475 & $^{13}N$ & 94105.267 ${\pm}$ 0.270  \\
\hline
$^7Li$ & 39244.526 ${\pm}$ 0.473 & $^{14}C$ & 105284.507 ${\pm}$ 0.019 \\
\hline
$^7Be$ & 37600.358 ${\pm}$ 0.472 & $^{14}N$ & 104658.628 ${\pm}$ 0.016 \\
\hline
$^8Li$ & 41277.328 ${\pm}$ 0.488 & $^{14}O$ & 98733.236 ${\pm}$ 0.076 \\
\hline
$^8B$  & 37737.813 ${\pm}$ 1.107 & $^{15}N$ & 115491.930 ${\pm}$ 0.019 \\
\hline
$^9Be$ & 30258.837 ${\pm}$ 62.471 & $^{15}O$ & 115955.627 ${\pm}$ 0.503 \\
\hline
$^{10}B$ & 64750.700 ${\pm}$ 0.370 & $^{16}O$ & 127619.336 ${\pm}$ 0.019
 \\  \hline \hline
\end{tabular}
\end{center}
\label{tpn1}
 \end{table}

\section{Observed primordial abundances}
\label{obs}

Primordial abundances of light elements are, in general, significantly
altered by nuclear processing in stars; hence, the deduction of the
primordial abundances from the ones observed today is very difficult and,
in some sense, is always dependent by the chemical evolution model.  \\
Primordial $D$ is easily destroyed by $(p,\gamma)$ reactions in stars where
the temperature is greater than $\sim 6 {\times} 10^5 \, ^oK$; it can be,
furthermore, converted in $^3He$ which is, however, also burned. Instead
$^4He$ grows in abundance with time: throughout stellar evolution, a star
becomes more and more enriched in $^4He$ and, through stellar mixing and
mass loss, the interstellar gas can increase its content in $^4He$.
Finally, $^7Li$ is destroyed by nuclear reactions with protons in low mass
stars, but it is nevertheless produced in supernov{\ae}, red giant interiors,
supermassive objects, etc. \\ We will shortly review the actual
experimental situation for the light element abundances.

\subsection{$D$}
\label{obsa}

Deuterium present in stellar interior is destroyed by the energy-generating
reactions; in fact, it has not been detected in any star . Nevertheless,
there are accurate data from giant planets in the solar system \cite{oli34}
and from the local interstellar medium (ISM) \cite{oli31}.
\\ Since there are no known astrophysical sources of $D$ \cite{sark34}, a firm lower
bound to the primordial abundance from the lowest observed one can be
obtained
\begin{equation}\label{o1}
   \frac{D}{H} \; > \; 1.1 {\times} 10^{-5}
  ~~~.
\end{equation}
Other estimates involve models of galactic chemical evolution to infer the
pre-galactic abundance in $D$ from those actually present in the ISM. \\
Deuterium has also been observed by many authors in the spectra of high
red-shift quasar absorption systems (QAS) \cite{oli3678}, \cite{oli39},
\cite{oli401}. In principle these measurements would be capable of
determining the primordial value of $D/H$. However, at present, this is not
the case. In fact, while in several measurements a rather high value of
$D/H \sim 2.0 {\times} 10^{-4}$ has been reported \cite{oli3678}, for other
observed QAS a significantly lower value ($\sim 3.4 {\times} 10^{-5}$) has been
quoted \cite{oli39}, \cite{oli401}. The present situation, then, does not
lead to a conclusive value for $D/H$ \cite{oli401}; a ``reasonable"  bound
from these observations is the following \cite{kern}:
\begin{equation}\label{o2}
 \frac{D}{H} \; \gapproxeq \; 2.5 {\times} 10^{-5} ~~~.
\end{equation}

\subsection{$^3He$}
\label{obsb}

$^3He$ is the product of incomplete $H$ burning in stars comparable in mass
to the Sun, but it is burned away in the interiors of heavier stars. \\
$^3He$ has been detected in meteoritic extractions \cite{oli323} as well as
in several interstellar medium (ISM) \cite{oli424} measurements of $^3He$
in galactic HII regions
\footnote{These are regions in which interstellar gas is heated and ionized
by the radiation from hot young stars. In astrophysical notations AI means
neutral A element, AII stands for A ionized once, and so on.} and in
planetary nebulae \cite{oli45}. Observations show a wide dispersion which
may be indicative of pollution or a bias \cite{oli43}, and large
evolutionary uncertainties make very difficult to extrapolate the
primordial abundance. However, noting that $D$ is burnt in stars to $^3He$
a (fairly known) fraction of which survives stellar processing, a more
reliable result can be obtained by considering the sum of $D$ and $^3He$
and requiring that these two isotopes not be overabundant at the time of
formation of the solar system \cite{yang}. This yields the limit
\begin{equation}\label{o3}
  \frac{D \, + \,^3He}{H} \; \gapproxeq \; 10^{-4} ~~~.
\end{equation}

\subsection{$^4He$}
\label{obsc}

\begin{figure}
\hspace{0.5truecm}
\epsfysize=7truecm
\epsffile{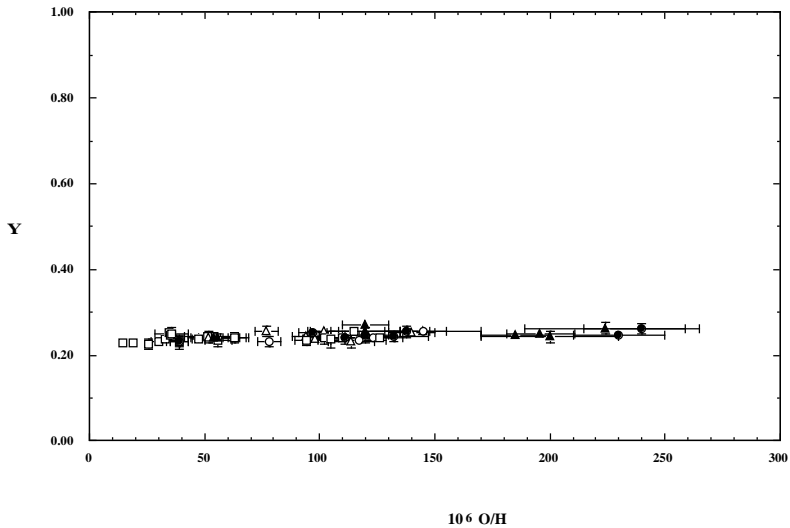}
\caption { The $^4He$ (Y) vs oxygen (O/H) abundances
in extragalactic HII regions (taken from Ref. \protect\cite{Olive}).  }
\label{he4con}
\end{figure}

The abundance of $^4He$ has been determined in a variety of astrophysical
sources \cite{oli112}, \cite{oli13}, such as atmospheres of young and old
stars, planetary nebulae, HII regions (galactic and extragalactic) and so
on. The almost, quite amazing constancy (within a $\sim 20 \%$ factor) of
the observed $^4He$ abundance by mass of about 25 \% in all these objects,
as can be seen from Figure \ref{he4con}, points to an uniform origin, which
is the primordial nucleosynthesis. However, although this abundance is
measured with a much higher accuracy than for the other light elements, one
has to be careful to extrapolate the primordial abundance from the today
observed one because of stellar processing. In fact, stars generate their
energy by burning hydrogen into helium and, at their death, they return
this processed material to the ISM to be incorporated into subsequent
generation of stars.
\\ To determine the primordial abundance of $^4He$, one must allow for the
stellar helium component through its correlation with some other element
which is made only in stars (such as $N , O$). The primordial value is then
obtained with an extrapolation to zero metallicity. For this reason, it is
better to concentrate  on data from  those regions least contaminated by
the products of stellar evolution: the $^4He$ abundance is, in fact, best
determined by observations of HeII $\rt$ HeI recombination lines in
extragalactic HII regions. Measuring both $^4He$ and $O/H$ abundances, it
has been found that the data are well fitted by a linear correlation
\cite{oli14} for $X_{^4He}$ ($\equiv Y$) versus $O/H$, and then the
primordial value can be determined from the intercept of that relation.
Moreover, the primordial $^4He$ abundance can also be determined by
studying the correlation between $X_{^4He}$ and $N/H$ (in almost all HII
regions both $O/H$ and $N/H$ data are available). However, in this case, it
is not still clear if a linear fit should be appropriate (at least in the
situations in which $N/H$ is not proportional to $O/H$) \cite{oli14},
\cite{oli15}, \cite{oli19}. Nevertheless, the quoted difference between the
intercepts of $X_{^4He}$ versus $N/H$ and $X_{^4He}$ versus $O/H$ is small
($\lapproxeq 0.003$). \\ At present, there are two different quoted
abundances \cite{oli112}, \cite{oli13}
\begin{eqnarray}
X_{^4He} & = &  0.234 {\pm} 0.0054
\label{o4}
\\
X_{^4He} &= & 0.243 {\pm} 0.003
\label{o5} ~~~.
\end{eqnarray}
The Izotov et al. \cite{oli13} result (\ref{n5}) is higher because of
dropping the lowest metallicity galaxy from their sample due to its
anomalously low HeI line intensities (which seems underlying stellar and
interstellar absorption). Also, different ways to extract the primordial
value of $X_{^4He}$ are employed. Although we do not enter into the details
of the discussion for the quoted values, we limit ourselves to observe that
the systematic error may be significantly larger than the ones included
above \cite{oli15}. In particular, the atomic physics inputs seem to be
uncertain (regarding mainly the HeI line intensities) and more accurately
known physical conditions in the HII regions are auspicable. This can be
done by measuring several different line intensities. \\ We finally note
that the error on the $^4He$ mass fraction is on the third significant
digit, and consider the upper bound
\begin{equation}\label{o6}
  X_{^4He} \; < \; 0.24 ~~~,
\end{equation}
as a reasonable one and
\begin{equation}\label{o7}
  X_{^4He} \; < \; 0.25 ~~~,
\end{equation}
as a reliable bound \cite{kern}.

\subsection{$^7Li$}
\label{obsd}

\begin{figure}
\hspace{0.5truecm}
\epsfysize=7truecm
\epsffile{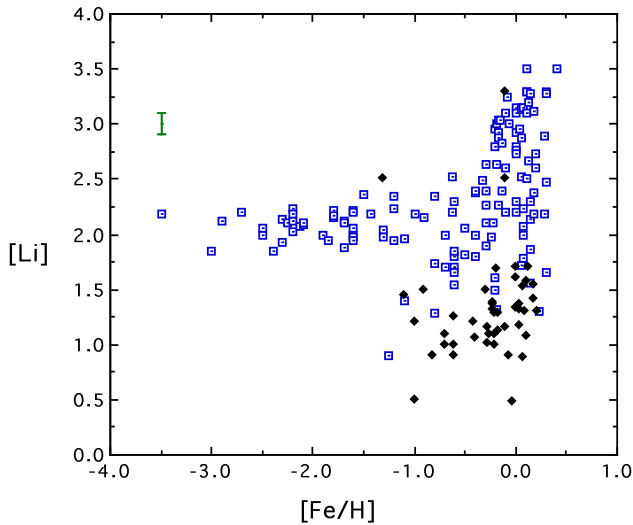}
\caption{The $^7Li$ abundance in halo stars with $T_{surf} > 5500 \, K$
as a function of metallicity. Here $[Li] = \log ( ^7Li/H) + 12$ and $[Fe/H]
= (\log(Fe/H))/(\log(Fe/H)_{\odot})$ (taken form Ref. \protect\cite{Olive}).}
\label{li7}
\end{figure}

As a star evolves, its surface lithium is subject to destruction and
dilution. In fact $^7Li$, like $D$, is easily destroyed (at temperatures
above  $\sim 2 {\times} 10^6 \, K$) by ($p, \alpha$) reactions, so that only the
lithium remaining on the stellar surface survives. This, however, will be
further diluted through mixing of the outer layers with the interior. To
avoid this, stars with a surface temperature $T > 5500 \, K$ and a
metallicity less than 1/20th solar are mainly observed, so that effects
such as stellar convection may be not important. \\ $^7Li$ is observed in
the atmospheres of both very old stars (Population II) and young ones
(Population I) \cite{oli20}, \cite{oli21}. When we plot the $^7Li$
abundance as a function of metallicity (for stars with $T_{surf} > 5500 \,
K$) a plateau region is observed, as can be seen in Figure \ref{li7},
indicating a primordial (or very close to it) value for the abundance. \\
The best estimate for the mean $^7Li$ abundance (including statistical
uncertainties) is the following \cite{oli21}:
\begin{equation}\label{o8}
   \frac{^7Li}{H}  \; = \; \left( 1.6 \,  {\pm}  \, 0.1 \right) \, {\times} \,
   10^{-10}  ~~~.
\end{equation}
However, we stress that $^7Li$ abundance determination is sensitive to
several stellar parameters (surface temperature, metallicity and so on) and
an important source of systematic error is due to the possibility that
$^7Li$ has been depleted in stars from their initial abundances. These
uncertainties are limited by the observation of $^6Li$ \cite{oli26} (with
$^6Li / ^7Li \, = \, 0.05 {\pm} 0.02$), a very fragile isotope; in fact
standard stellar models predict that any depletion of $^7Li$ would be
accompanied by a very severe depletion of $^6Li$ \cite{oli2789}. \\
Finally, $Li$ is also produced, along with $Be$ and $B$, in cosmic ray
spallation of $C$, $N$, $O$ by $p$, $\alpha$ (and also by $\alpha - \alpha$
fusion). Hence abundances of $Be$ and $B$ have been served as a consistency
check \cite{oli30} on primordial $Li$, concluding that no more than 10 - 20
\% of the $^7Li$ is due to cosmic ray nucleosynthesis.
\\ Of course, for $^7Li$ the uncertainties are dominated by systematic
effects.

\vspace{1cm}

Abundances of other elements have also been observed, such as those of the
intermediate mass isotopes $^9Be$, $^{10}B$ and $^{11}B$. However, for
these, large uncertainties are introduced, since it is believed that these
isotopes are formed in cosmic ray nucleosynthesis. In fact, the observed
abundances are far above the BBN predictions and a comparison between
theory and experiments is extremely difficult.

\section{Production of the light elements}
\label{pnb}

Before going into the details of the primordial nucleosynthesis, in this
section we will discuss (mainly qualitatively) the main features and
results of Big Bang Nucleosynthesis.

\subsection{$T \gapproxeq 1 \, MeV$ ($t \lapproxeq 1 \, s$)}
\label{pnb1}

At sufficiently high temperatures, neutrons and protons are maintained both
in kinetic \footnote{Kinetic equilibrium is maintained also and primarily
by nuclear and electromagnetic elastic scattering reactions} and chemical
equilibrium by charged current weak interactions:
\bea
\ne \, + \, n & \lrt & e^- \, + \, p   \label{n11} \\
e^+ \, + \, n & \lrt & p \, + \, \neb  \label{n12} \\ n & \lrt & p \, + \,
e^- \, + \, \neb \label{n13} ~~~.
\eea
Because of chemical equilibrium, the following relation between the
chemical potentials holds
\begin{equation}\label{n14}
  \mu_n \; + \; \mu_\nu \; = \; \mu_p \; + \; \mu_e ~~~.
\end{equation}
The neutron to proton density ratio is then given by
\begin{equation}\label{n15}
  \left( \frac{n_n}{n_p} \right)_{eq} \; = \; e^{- \frac{Q}{T}}
  \, e^{\frac{\mu_e - \mu_\nu}{T}}
\end{equation}
(detailed balance), where $Q = m_n - m_p \simeq 1.29 \, MeV$ is the
difference between the neutron and proton mass. As discussed in sect.
\ref{bbc1}, the electron chemical potential is small compared to the
temperature $T$ in the region relevant for BBN. Furthermore, assuming zero
neutrino chemical potential, we have then
\begin{equation}\label{n16}
  \left( \frac{n_n}{n_p} \right)_{eq} \; \simeq \; e^{- \frac{Q}{T}} ~~~.
\end{equation}
The nucleons abundances track their values in equilibrium,
\begin{equation}\label{n17}
  X_n(T) \; \simeq \; X_p(T) \; = \; \frac{1}{e^{\frac{Q}{T}} + 1} ~~~,
\end{equation}
as long as the rates for the reactions (\ref{n11}),(\ref{n12}),(\ref{n13})
decrease sufficiently and become comparable with the expansion rate $H$.
This happens, as we have seen in sect. \ref{bbd}, at about $T \sim 1
\, MeV$. At this time, neutrons ``freeze out'', i.e. go out of chemical
equilibrium, and  $X_n$ (and $X_p$) relaxes to the final constant value
\begin{equation}\label{n18}
  X_n^F \; \simeq \; X_p^F \; \simeq \; \frac{1}{e^{\frac{Q}{T_F}} + 1} ~~~,
\end{equation}
($T_F \sim 1 \, MeV$ is the freeze out temperature) rather than following
the exponential falling of (\ref{n17}). Since $Q/T_F \sim O(1)$, a
substantial fraction of nucleons survives when chemical equilibrium between
them is broken, while if $T_F$ would be much lower than $Q$ then $n_n
\simeq 0$. \\ The abundances of the other nuclides are very small for
$T \gg 1 \, MeV$, and can be calculated from (\ref{n10}). They become
appreciable when the temperature goes down the binding energy  per nucleon
of the given specie, which is typically of the order of $1 \div 8 \, MeV$.
However, there are two facts that delay the onset of nucleosynthesis: the
low binding energy of the first nuclide in the nucleosynthesis chain,
deuterium, and the high number of photons (or high entropy) present at the
nucleosynthesis era. In fact, for $T \lapproxeq 1 \div 8 \, MeV$ deuterium
(and hence the other nuclides to be formed from this) synthesis is
energetically favoured, so the reaction $n + p \rt D + \gamma$ takes place
but, due to the high number of photons, the inverse reaction takes place as
well and more efficiently than other reactions such as $n + D \rt \,^3H +
\gamma$ or $p + D \rt \,^3He + \gamma$. Hence the abundances of these light elements are still very small
for these temperatures (deuterium ``bottleneck"). For reference, we report
on the typical abundances of some light nuclei for $T\simeq10 \, MeV$, as
calculated from (\ref{n10}) with $\eta \sim 10^{-9}$:
\bea
 X_n & \simeq & X_p \; \simeq \; 0.5 ~~~, \\
 X_D & \simeq & 6 {\times} 10^{-12} ~~~, \\
 X_{^3He} & \simeq & 2 {\times} 10^{-23} ~~~, \\
 X_{^4He} & \simeq & 2 {\times} 10^{-34} ~~~, \\
 X_{^{12}C} & \simeq & 2 {\times} 10^{-126} ~~~.
\eea
Instead, as neutrons freeze out, the $n/p$ ratio is given by (\ref{n16})
with $T\simeq T_F$ ($n_n / n_p \simeq 1/6$) and for $T \sim 1 \, MeV$ the
abundances of light nuclides are
\bea
 X_n & \simeq &  \frac{1}{7} ~~~, \\
 X_p & \simeq &  \frac{6}{7} ~~~, \\
 X_D & \simeq & 10^{-12} ~~~, \\
 X_{^3He} & \simeq & 10^{-23} ~~~, \\
 X_{^4He} & \simeq & 10^{-28} ~~~, \\
 X_{^{12}C} & \simeq & 10^{-108} ~~~.
\eea

\subsection{$T \simeq 0.3 \div 0.1 \, MeV$ ($t \simeq 1 \div 3 \, min$)}
\label{pnb2}

\begin{figure}
%\hspace{0.5truecm}
%\vskip 2.0cm
\epsfysize=5.0truein
\epsffile{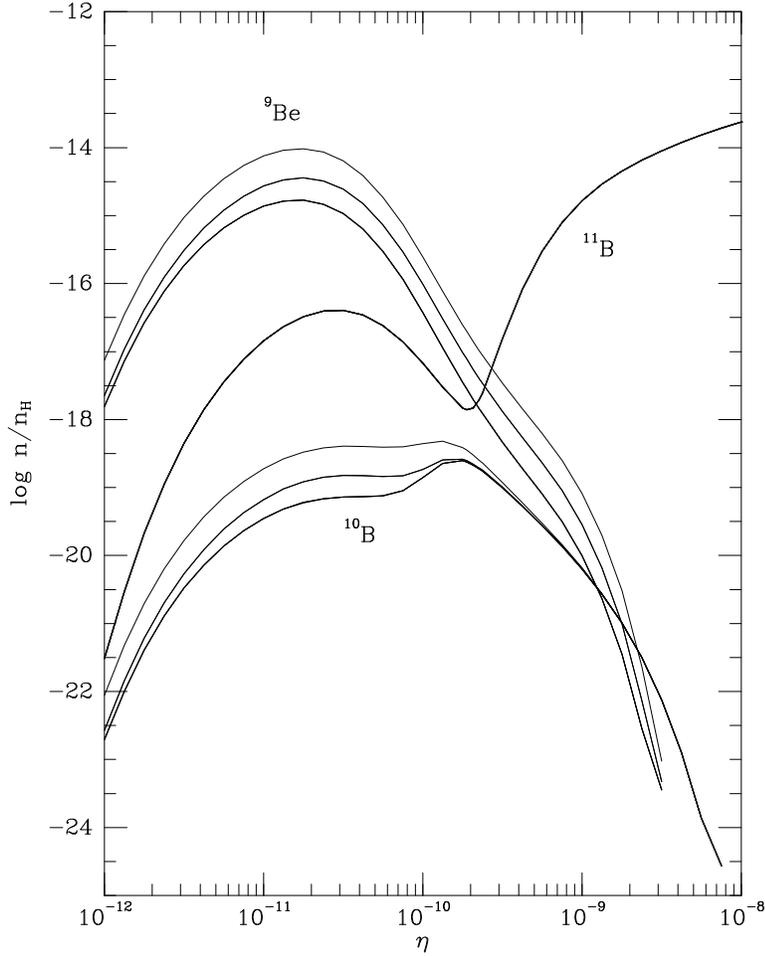}
\caption{{The intermediate mass element abundances from big bang
nucleosynthesis as a function of $\eta$ (taken from Ref.
\protect\cite{oli8}).}}
\label{oli2}
\end{figure}

As the temperature cools below $0.5 \div 0.3 \, MeV$ the synthesis of
complex nuclei becomes thermodynamically favourable. Thus, neutrons and
protons react with each other to build up light nuclides through the
following sequence of two-body reactions:
\begin{equation}\label{n19}
  \ba{lll}
  p \, (n,\gamma) \, D & & \\
  D \, (p,\gamma) \, ^3He & D \, (D,n) \, ^3He & D \, (D,p) \, ^3H \\
  ^3H \, (D,n) \, ^4He & ^3H \, ( ^4He,\gamma) \, ^7Li & \\
  ^3He \, (n,p) \, ^3H & ^3He \, (D,p) \, ^4He & ^3He \, ( ^4He,\gamma) \,
  ^7Be \\
  ^7Li \, ( p.^4He) \, ^4He & ^7Be \, (n,p) \, ^7Li & \\
  ... & & ~~~.
  \ea
\end{equation}
(a complete list of nuclear reactions is reported in appendix \ref{netw}).
The first reaction is, of course, crucial since deuterium has to be formed
in appreciable amount, before the other reactions can start. In fact, many
body interactions such as $2p + 2n \rt ^4He$ are in general ineffective,
due to the small number densities of nucleons. After deuterium, appreciable
quantities of $^3H$ and $^3He$ may form. The synthesis of $^4He$ is,
however, delayed until enough tritium has been built up, since the main
process for making $^4He$ involves this hydrogen isotope. \\ Nearly all
neutrons surviving down  to freeze out are captured in $^4He$ because of
its large binding energy per nucleon; its mass fraction is then
approximately given by
\begin{equation}\label{n20}
  X_{^4He} \; = \; \frac{4 n_{^4He}}{n_B} \; \simeq \; \frac{4 \left(
  n_n / 2 \right)}{n_n + n_p} \; = \; \frac{2 (n_n / n_p)}{1 + (n_n /
  n_p)} ~~~.
\end{equation}
For $T \simeq 0.1 \, MeV$, the $n/p$ ratio is about 1/7 (actually, it is a
bit smaller than the freeze out value since neutrons have been depleted by
$\beta$-decay), so that
\begin{equation}\label{n21}
  X_{^4He} \; \simeq \; \frac{1}{4} ~~~.
\end{equation}
Heavier nuclei do not form in any significant quantity both because of
absence of stable nuclei with $A = 5,6$ which forbids nucleosynthesis
through $(n,^4He)$, $(p,^4He)$ or $(^4He, ^4He)$ reactions) and the large
Coulomb barriers for reactions such as $^3H( ^4He, \gamma) ^7Li$ and $^3He(
^4He,
\gamma) ^7Be$. \\
For $t \gapproxeq 10^3 \, s$ BBN is over, since the low temperature and low
density in the Universe suppress nuclear reactions. Hence the most abundant
nuclides with which we are left are hydrogen ($p$) and $^4He$
(incorporating almost all neutrons), followed by trace amounts of $D$,
$^3He$ and $^7Li$. Note that the $^3He$ abundance includes that of survived
$^3H$ which subsequently $\beta$-decays and similarly the $^7Li$ abundance
includes that of $^7Be$.\\ Observe that the $^4He$ abundance, as
approximately given by (\ref{n21}), is quite insensitive to the nucleon
density (or $\eta$) present in the Universe, and almost only depends on the
weak reaction rates determining neutron freeze out. This is not true for
the other elements, for which the nucleon density directly determines the
two-body nuclear reaction rates. It is expected that $D$ and $^3He$
abundances decrease with increasing $\eta$, since for higher values of
$\eta$  we have a more efficient burning into $^4He$. In fact, from
(\ref{n10}), abundances increase with $\eta$ so that BBN can start earlier
when at higher temperature and the $n/p$ ratio is  larger. As far as $^7Li$
abundance, it is expected to decrease with increasing $\eta$ when it is
determined by the competition between $^4He(^3H,\gamma)
^7Li$ and $^7Li(p, ^4He) ^3He$, while for sufficiently high $\eta$ it
starts increasing again with $\eta$ due to larger production of $^7Be$
through $^4He( ^3He,\gamma) ^7Be(e^-,\nu_e) ^7Li$.
\begin{figure}
%\hspace{0.5truecm}
%\vskip 2cm
\epsfysize=5.0truein
\epsffile{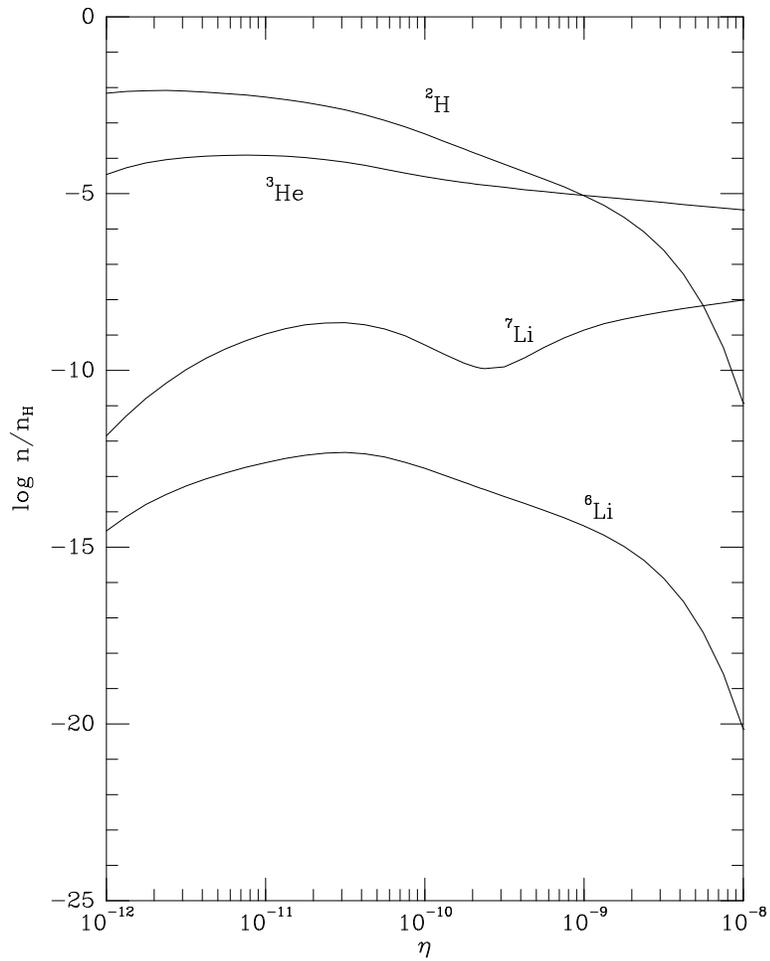}
\caption{{The light element abundances from big bang
nucleosynthesis as a function of $\eta$ (taken from Ref.
\protect\cite{oli8}).}}
\label{oli1}
\end{figure}
\\ In Figures \ref{oli1}, \ref{oli2} we report \cite{Olive}
the predicted primordial abundances of some light elements versus the
baryon to photon ratio $\eta$ as calculated with the standard numerical
code \cite{Kawano}. The discussed features come evident from these plots.

\section{Calculation of the primordial abundances}
\label{pnc}

In this section we will write down the equations governing the time
evolution, during nucleosynthesis, of the abundances of the produced light
elements.\\ The variables describing the Universe as a thermodynamic system
during BBN are the following:
\begin{equation}
  \ba{l}
  R \\
  T \, , \, T_\nu \\
  \mu_e \, , \, \mu_\nu \, , \, \mu_n \, , \, \mu_p \\
  Y_i ~~~.
  \ea \nonumber
\end{equation}
The first is associated to the dynamical aspects of the system, i.e.
expansion of the Universe, while the others describe the primordial plasma
\footnote{Note that it is explicitly assumed that species interacting
electromagnetically or strongly are always in kinetic equilibrium, though
chemical equilibrium may be not hold. For neutrinos, it is instead supposed
that their decoupling is ``instantaneous'' (see below and chapter
\ref{sum})} of $e_{\pm} , \nu , n , p$ and $Y_i$ are the nuclide abundances. \\
The relevant part of nucleosynthesis takes place after neutrino decoupling,
so that the quantities describing the neutrino plasma evolve independently
from the plasma in equilibrium and, for our purposes, may be disregarded.
Hence, $T_\nu$ and $\mu_\nu$ may be dropped from the list above. \\
Moreover, only two of the remaining three chemical potentials are
independent, since the relation (\ref{n14}) holds at chemical equilibrium.
The first independent variable we choose is the electron chemical potential
or, equivalently, the parameter $\phi_e = \mu_e /T$. Instead, the second
one, which is related to the nucleon chemical potential, is the total
baryon number density $n_B$. Thus the independent variables we have to
consider are the following
\begin{equation}
  R \, , \, n_B \, , \, \phi_e \, , \, T \, , \, Y_i
  \nonumber ~~~,
\end{equation}
for which it is necessary to write down the time evolution equations;
however, let us note that in many cases is more convenient to follow the
temperature evolution rather the time one, and this can be done once the
relation between $T$ and $t$ is established \cite{wag67, wag69, wag73,
Kawano}.

\subsection{Equation for $R$}
\label{pnc1}

As seen in the previous chapter, the equation satisfied by the cosmic scale
factor is the following:
\begin{equation}\label{n22}
  \frac{1}{R} \, \frac{dR}{dt} \; = \; H \; \equiv \;
  \sqrt{\frac{8 \pi}{3 M_P^2} \, \rho \; - \; \frac{k}{R^2} } ~~~,
\end{equation}
where $\rho$ is the total energy density. During the nucleosynthesis era,
the curvature term $k / R^2$ is completely negligible, so that we can
safely neglect its contribution to expansion.

\subsection{Equation for $n_B$}
\label{pnc2}

The total number $N_B$ of baryons in the comoving volume is constant
\footnote{At the energy scales relevant for BBN there are no baryon
number violating processes.} and, assuming homogeneity, it is also
independent of $R$. Hence the baryon number density $n_B = N_B / V$ varies
as $R^{-3}$, so that
\begin{equation}\label{n23}
  \frac{1}{n_B} \, \frac{d n_B}{dt} \; =  \; - \, 3 \,
  \frac{1}{R} \, \frac{dR}{dt} ~~~,
\end{equation}
which is the equation for the evolution of $n_B$.

\subsection{Equation for $\phi_e$}
\label{pnc3}

The electron chemical potential (or $\phi_e$) is fixed by the conservation
of the electric charge and the total neutrality of the Universe
\cite{neutro}
\begin{equation}\label{n24}
  n_{e^-} \; - \; n_{e^+} \; = \; n_p ~~~,
\end{equation}
neglecting the possible presence of antiprotons. The proton number density
is
\begin{equation}\label{n25}
  n_p \; = \; \sum_i \, Z_i \, n_i \; = \; n_B \, \sum_i \, Z_i \, Y_i
  \; \equiv \; n_B \, q_B ~~~,
\end{equation}
while, for $\phi_e \ll 1$, we have
\bea
n_{e^-} \; - \; n_{e^+} & \simeq & \frac{2}{\pi^2} \, \phi_e \, T^3
\, f(z) ~~~, \label{n26} \\
f(z) & \equiv & \int_0^\infty dx \, \frac{x^2 \, e^\epsilon} {(e^\epsilon +
1)^2} ~~~, \nonumber
\eea
with $\epsilon = \sqrt{x^2 + z^2}$ and $z = m_e /T$. From these we then
obtain the equation for $\phi_e$:
\begin{equation}\label{n27}
  \phi_e \; \simeq \; \frac{\pi^2}{2} \, \frac{n_B \, q_B}{T^3 \, f(z)} ~~~,
\end{equation}
with $q_B \equiv \sum_i Z_i Y_i$ defined in (\ref{n25}).

\subsection{Equation for $T$}
\label{pnc4}

The important equation relating temperature with time is deduced from
energy conservation, which  gives
\begin{equation}\label{n28}
  \frac{d}{dt} \left( \rr V \right) \; + \; \p \, \frac{dV}{dt} \; = \; 0 ~~~,
\end{equation}
where $V \sim R^3$ is the comoving volume and $\rr , \p$ are the total
energy density and pressure of the particles involved, namely:
\bea
 \rr & = & \rr_\gamma \; + \; \rr_e \; + \; \rr_B  ~~~, \label{n29} \\
 \p & = & \p_\gamma \; + \; \p_e \; + \; \p_B  \label{n30} ~~~.
\eea
(assuming that neutrinos are completely decoupled
\footnote{In this approximation, neutrino plasma energy is separately
conserved,}) The photon contributions $\rr_\gamma ,\p_\gamma$ depend only
on the temperature and are given by equations (\ref{63b}), (\ref{63c}),
while $\rr_e = \rr_{e^-} + \rr_{e^+}$, $\p_e = \p_{e^-} + \p_{e^+}$ depend
in general on both $T$ and $\phi_e$:
\begin{equation}\label{n31}
  \rr_\gamma \; = \; \frac{\pi^2}{15} \, T^4 \;\;\;\;\;\;\;\;\;\;\;\;\;\;\;
  \p_\gamma \; = \; \frac{1}{3} \, \rr_\gamma
\end{equation}
\begin{equation}\label{n32}
  \rr_e \; \simeq \; \frac{2}{\pi^2} \, T^4 \, g(z) \; + \; O(\phi_e^2)
  \;\;\;\;\;\;\;\;\;\;\;\;\;\;\;
  \p_e \; \simeq \; \frac{2}{3 \pi^2} \, T^4 \,  h(z) \; + \; O(\phi_e^2)
\end{equation}
\begin{equation}
  g(z) \; \equiv \; \int_0^\infty dx \, \frac{x^2 \, \epsilon}
  {e^{\epsilon} + 1} \;\;\;\;\;\;\;\;\;\;\;\;\;\;\;
  h(z) \; \equiv \; \int_0^\infty dx \, \frac{x^4}{\epsilon} \,
  \frac{1}{e^{\epsilon} + 1}
  \nonumber ~~~.
\end{equation}
The baryon terms can be written, in the non relativistic approximation, as
(see (\ref{62b}), (\ref{62c}))
\bea
 \rr_B & = & \sum_i \, \ri \; \simeq \sum_i \, \left( m_i \nni \, +
 \, \frac{3}{2} \, T \, \nni \right) \; = \nonumber \\
 & = & M_u \, n_B \left( 1 \, + \, \sum_i \left( \frac{\Delta \mi}{M_u}
 \, + \, \frac{3}{2} \, \frac{T}{M_u} \right) Y_i \right) \label{n33} \\
 \p_B & = & \sum_i \, \pri \; \simeq \sum_i \, T \, \nni \; = \; T \, n_B
 \, \sum_i \, Y_i ~~~, \label{n34}
\eea
where $M_u$ is the atomic mass unit (referred to $^{12}C$) while $\Delta
\mi$ is the mass excess for the nuclide $i$
(mass excesses for the light nuclides relevant for BBN are reported in
appendix \ref{mex}). \\ From (\ref{n28}) after some simple algebra we
arrive at the formal equation for $T$
\begin{equation}\label{n35}
  \frac{dT}{dt} \; = \; - \, 3 \frac{1}{R} \, \frac{dR}{dt} \, \left( \rr \, +
  \, \p \right) \, \left( \frac{d \rr}{dT} \right)^{-1} ~~~.
\end{equation}
First of all, let us note that the energy density derivative with respect
to $T$ has to be evaluated along the time direction, that is
\begin{equation}
  \frac{d\rr}{dT} \; = \; \left( \tpunto \right)^{-1} \, \frac{d\rr}{dt}
  \; = \; \left( \tpunto \right)^{-1} \, \left( \frac{d\rr_\gamma}{dt}
  \, + \, \frac{d\rr_e}{dt} \, + \, \frac{d\rr_B}{dt} \right) \nonumber
\end{equation}
\bea
 \frac{d\rr_\gamma}{dt} & = & \frac{d\rr_\gamma}{dT} \, \tpunto
 \nonumber \\
 \frac{d\rr_e}{dt} & = & \left( \aphie \right) \, \tpunto \; + \;
 \left( \bphie \right) \nonumber \\
 \frac{d\rr_B}{dt}
 & \simeq & \rr_B \left( - \, \tet \; + \; \frac{3}{2} \frac{1}{M_u} \tpunto \, \sum_i \, Y_i \; + \;
 \sum_i \, Y_i \; + \;
 \sum_i \, \left( \frac{\Delta \mi}{M_u} \, + \, \frac{3}{2} \frac{T}{M_u}
 \right) \frac{d Y_i}{dt} \right) \nonumber ~~~.
\eea
Hence we have
\bea
 \frac{d\rr}{dT} & = & \left( \aphie \; + \; \frac{d\rr_\gamma}{dT} \; + \;
 \frac{3}{2} \frac{1}{M_u} \, \rr_B \, \sum_i \, Y_i \right) \; + \nonumber
 \\
 & - & \tet \, \left( \tpunto \right)^{-1} \left\{ \rr_B \left( 1 \, -
 \left( \tet \right)^{-1} \sum_i \,
 \left( \frac{\Delta \mi}{M_u} \, + \, \frac{3}{2} \frac{T}{M_u}
 \right) \frac{d Y_i}{dt} \right) \; + \right. \nonumber \\
 & - & \left.  \left( \tet \right)^{-1}
 \left( \bphie \right) \right\} \nonumber
\eea
and substituting Eq. (\ref{n35}), solving with respect to $dT/dt$, we find
\bea
  \frac{dT}{dt} & = & - \left\{  \tet \left(\rr_\gamma \, + \, \rr_e \, +
  \, \p_\gamma \, + \, \p_e \, + \, \p_B \right) \; + \right. \nonumber \\
  & + & \left. \bphie \; + \; \sum_i
  \, \left( \frac{\Delta \mi}{M_u} \, + \, \frac{3}{2} \frac{T}{M_u}
  \right) \frac{d Y_i}{dt} \right\} {\cdot} \nonumber \\
  & {\cdot} & \left\{
  \aphie \; + \; \frac{d\rr_\gamma}{dT} \; + \; \frac{3}{2}
  \frac{1}{M_u} \, \rr_B \, \sum_i \, Y_i \right\}^{-1} ~~~.
\label{n36}
\eea
This is the equation relating temperature with time we looked for. Note
that it is now written in terms of known quantities, once Eqs. (\ref{n22})
and (\ref{n23}) are taken into account.

\subsection{Equations for $Y_i$}
\label{pnc5}

The change in time of the abundance $Y_i$ of a given nuclide $i$ is driven
by the rates for the nucleon reactions producing or destroying it which we
will consider to be of the form
\begin{equation}\label{n37}
  i \; + \; j \; \longleftrightarrow \; k \; + \; l ~~~.
\end{equation}
If $\Gamma_{ij \rt kl}$ is the reaction rate for the direct process and
$\Gamma_{kl \rt ij}$ that for the inverse one, the time evolution of $Y_i$
is given by the following equation:
\begin{equation}\label{n38}
  \frac{d Y_i}{dt} \; = \; \Gamma_{kl \rt ij} \, Y_l \, Y_k \; - \;
  \Gamma_{ij \rt kl} \, Y_i \, Y_j ~~~.
\end{equation}
In general, one can consider also more complex reactions, in which more
than one nuclide for each specie is present:
\begin{equation}\label{n39}
  N_i \left( i \right) \; + \; N_j \left( j \right) \; \longleftrightarrow
  \; N_k \left( k \right) \; + \; N_l \left( l \right) ~~~,
\end{equation}
where $N_a$ is the (integer) number of a given nuclide taking part to the
considered reaction. In this case the evolution equations are of the
following general form:
\begin{equation}\label{n40}
  \frac{dY_i}{dt} \; = \; \sum_{j,k,l} \, N_i \left(
  \Gamma_{kl \rt ij} \, \frac{Y_l^{N_l} \, Y_k^{N_k}}{N_l! \, N_k !}
  \; - \; \Gamma_{ij \rt kl} \, \frac{Y_i^{N_i} \, Y_j^{N_j}}{N_i ! \, N_j
  !} \right) ~~~.
\end{equation}
The complete nuclear reaction network used in BBN calculations is reported
in appendix \ref{netw}.

\section{Theory versus observations. Constraints from BBN}
\label{pncon}

\begin{figure}
\vspace*{-2cm}
\epsfysize=20cm
\epsfxsize=16cm
\epsffile{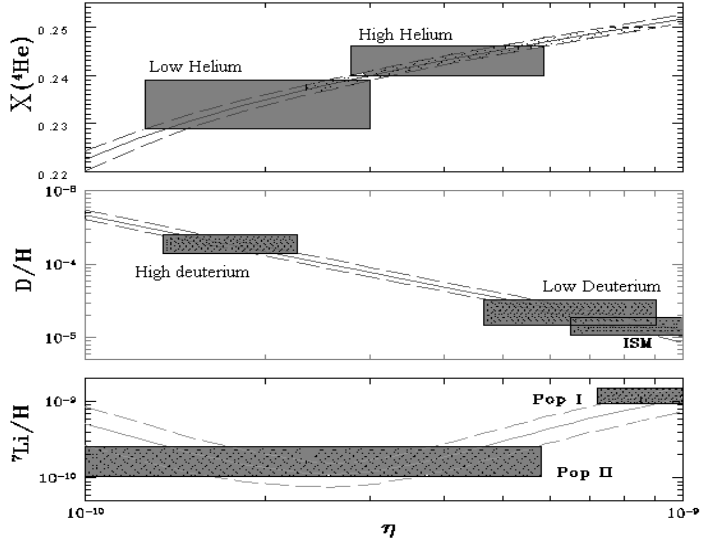}
\vspace{-9cm}
\caption { $^4He$ mass fraction and $D$, $^7Li$ abundances
predicted in the standard BBN scenario. Also shown are the observed values
of these isotopes (adapted from Ref. \protect\cite{Sarkrev}).  }
\label{sfig}
\end{figure}

We can now compare the inferred bounds on the abundances of light elements
discussed in section \ref{obs} with their values computed within the
standard BBN scenario. \\ The two conflicting $^4He$ mass fraction results
reflect into the QSO determinations of deuterium abundance, and two
different sets of measurements mutually incompatible
\begin{eqnarray}
X_{^4He} & = &  0.234 {\pm} 0.0054 ~~~,~~~~~ D/H  \; = \; ( 1.9 {\pm} 0.4 ) {\cdot}
10^{-4} \label{star1}
\\ X_{^4He} & = &  0.243 {\pm} 0.003  ~~~~,~~~~~ D/H \; = \;
(3.40 {\pm} 0.25) {\cdot} 10^{-5} \label{star2}
\end{eqnarray}
arise. Obviously, large helium mass fractions require low values for
deuterium abundance and vice versa. The situation is well illustrated in
Figure \ref{sfig} where both theoretical predictions and experimental
observations are reported; from this the ranges of the $\eta$ parameter
corresponding to the different sets of data (\ref{star1}), (\ref{star2})
can be directly read off. However, a very recent (re-)analysis \cite{Olive}
of observational data has superseded the discussed dichotomy, and a unique
range for $\eta$ emerges, as can be seen from Figure \ref{absark}.
Nevertheless, adopting the reliable bounds of  Eqs. (\ref{o2}), (\ref{o7}),
(\ref{o8}) we can derive the conservative limit on $\eta$ \cite{Sarkrev}
\begin{equation}\label{star3}
  4.1 {\times} 10^{-11} \; < \; \eta \; < \; 9.1 {\times} 10^{-10} ~~~,
\end{equation}
overwhelming all the quoted difficulties in abundance determinations. These
values of $\eta$ translate into the following range for the nucleon density
(normalized to the critical density)
\begin{equation}\label{star4}
  0.0015 \; < \; \Omega_B \, h^2 \; < \; 0.033 ~~~,
\end{equation}
which is then an independent estimate of the baryon content of the
Universe. Confronting with Eq. (\ref{37}), it seems evident that non
baryonic (dark) matter exists in the Universe. For an account of dark
matter problems, see for example \cite{dark}.

Having established the consistency of standard BBN, we can use it to
constrain new physics beyond the Standard Model of elementary particles.
Here we don't perform an exhaustive analysis of these constraints, which is
far from the scope of this thesis and for which we remind to excellent
reviews (such as \cite{Sarkrev}); our aim is to show how precise BBN
results can be used to extract important informations on particle physics.
\\ Limits on this are mostly sensistive to the bounds imposed on the $^4He$
abundance, which is predominantly determined by the neutron to proton ratio
at freeze out. This is determined by the competition between the weak
interaction rates and the expansion rate of the Universe, according to Eq.
(\ref{101}). The presence of additional neutrino flavours (or any other
relativistic particle species) at the time of BBN increases the overall
energy density and hence the expansion rate, leading to a larger value of
$T_F$, $n/p$ and ultimately $X_{^4He}$. The dependence of $X_{^4He}$ on the
effective number $N_\nu$ of neutrinos can be parametrized as follows
\cite{Olive}:
\begin{equation}\label{star5}
  X_{^4He} \; = \; 0.2262 \; + \; 0.0131 \, \left( N_\nu \, - \, 3 \right)
  \; + \; 0.0135 \, \ln \, \left( \frac{\eta}{10^{-10}} \right)
\end{equation}
(this fit holds near the central value of (\ref{star3})). Note the weak
logarithmic dependence of $X_{^4He}$ on the $\eta$ parameter. By requiring
that all constraints on elemental yields be simultaneously satisfied, a
Monte Carlo analysis has given the following conservative limit on $N_\nu$
\cite{kern}
\begin{equation}\label{star6}
  N_\nu \; \leq \; 3.75 \; + \; 78 \, \left( X_{^4He}^{max} \, - \, 0.24
  \right) ~~~.
\end{equation}

A number of other constraints on known or hypothetical particles, both
relativistic and non relativistic at the time of BBN, as well as limits on
the strength of new forces, can be deduced. These also have important
implications for the problem of dark matter in the Universe. While these
constraints are much more stringent than the ones obtained in laboratory
experiments, pointing out the relevance of BBN studies, they are (particle
physics) model dependent and cannot be discussed here. We remind to the
recent reviews \cite{Olive, Sarkrev} and references therein.

\chapter{The Born rates for $n \lrt p$ reactions}
\label{c1}

The initial conditions for the primordial nucleosynthesis are settled out
by the ratio of the relative abundances of neutrons and protons. When BBN
is starting, the primordial plasma consists of nucleons, electrons,
positrons, neutrinos, antineutrinos and photons in thermal equilibrium. The
relative abundance of neutrons and protons is determined by the following
charged-current weak interactions:
\bea
(a)~~~ \ne \, + \, n \; \rt \; e^- \, + \, p  &~~~,~~~~~~& (b)~~~ e^- \, +
\, p \; \rt \; \ne \, + \, n  \label{cb1} \\
(c)~~~ e^+ \, + \, n \;  \rt \; \neb \, + \, p &~~~,~~~~~~& (d)~~~
\neb \, + \, p \;  \rt \; e^+ \, + \, n \label{cb2} \\
(e)~~~ n \; \rt \; p \, + \, e^- \, + \, \neb &~~~,~~~~~~& (f)~~~ p \, + \,
e^- \, + \, \neb \; \rt \; n \label{cb3}
\eea
In this chapter, the leading contribution for the rates (per nucleon and
per time) for these six reactions is calculated and the relevant
approximations are particularly pointed out.

For a generic process in \reac, the reaction rate can be written in the
form
\begin{equation} \label{cb4}
d \Gamma (i \rt f) \, = \, \mod \, \four \, d \Phi ~~~,
\end{equation}
where $|M|^2$ is the squared matrix element, which has to be summed over
all spin degrees of freedom. The $\delta$-function gives the
energy-momentum conservation, ${\mathrm p}_a \, = \, (E_a, \bvec{p}_a)$
being the 4-momenta of the particles entering in the process (each $p_a$ is
intended with a positive sign if the particles involved in the reaction is
in and with a negative sign if it is out). The quantity $d \Phi \, = \,
\prod_a d \Phi_a$ is the phase space factor for the given reaction.
Since the processes \reac occur in a thermal bath, the phase space factor
associated with each initial (Fermi) particle is
\begin{equation}\label{cb5}
  d \Phi_a \, = \, \frac{d^3 \bvec{p}_a}{(2 \pi)^3} \,
  \frac{1}{2 E_a} \, F_a ~~~,
\end{equation}
while for final (Fermi) particles we have
\begin{equation}\label{cb6}
  d \Phi_a \, = \, \frac{d^3 \bvec{p}_a}{(2 \pi)^3} \,
  \frac{1}{2 E_a} \, (1 \, - \, F_a) ~~~,
\end{equation}
where $F_a$ is the phase-space density of the $a$-th particle at
temperature $T$, which takes the thermal equilibrium value (in the rest
frame of the thermal radiation) given by the Fermi function
\begin{equation}\label{cb7}
  F_a(E_a) \, = \, \frac{1}{e^{\frac{E_a}{T_a}} \, + \, 1}
\end{equation}
only for those species actually in equilibrium. Note that in writing
(\ref{cb4}) we have assumed that $n_a$ does not depend on spin, so that the
summation over this degree of freedom is limited to $|M|^2$.

The quantity of interest in BBN is the reaction rate per incident nucleon,
so that $\Gamma$ in (\ref{cb4}) has to be divided by the number density
\begin{equation}
  n_b \, = \, 2 \, \int \frac{d^3 \bvec{p}_b}{(2 \pi)^3} \, F_b \nonumber
\end{equation}
of the incident nucleon (say $b$). Assuming that $\mod$ does not depend on
the initial nucleon momentum (this assumption, which is justified in the
infinite nucleon mass approximation, will be discussed later), the net
effect on the reaction rate is the following
\begin{equation} \label{cb8}
d \Gamma (i \rt f) \, = \, \frac{1}{2} \mod \, \four \, \prod_{a \neq b}
d\Phi_a \, \frac{1}{2 M_b}
\end{equation}
(from now on, we indicate with $\Gamma$ the reaction rate per incident
nucleon). Furthermore, since at the epoch of BBN the baryon density is very
low ($\eta \sim 10^{-9}$), we can also neglect the occupation number of the
final nucleon (say $c$), so the final expression for $\Gamma (i \rt f)$ to
be considered is
\begin{equation} \label{cb9}
d \Gamma (i \rt f) \, = \, \frac{1}{2} \mod \, \four \,
\left( \prod_{leptons} d\Phi_a \right) \, \frac{d^3 \bvec{p}_c}{(2 \pi)^3} \,
 \, \frac{1}{2 M_b} \, \frac{1}{2 M_c} ~~~.
\end{equation}
We now proceed to explicitly calculate the reaction rates for the processes
in \reac in the Born approximation.

\section{The reaction $\ne \, + \, n \, \rt \, e^- \, + \, p$ and the
other processes}
\label{c1a}

\begin{figure}
\epsfysize=6.0cm
\epsfxsize=6.0cm
\centerline{\epsffile{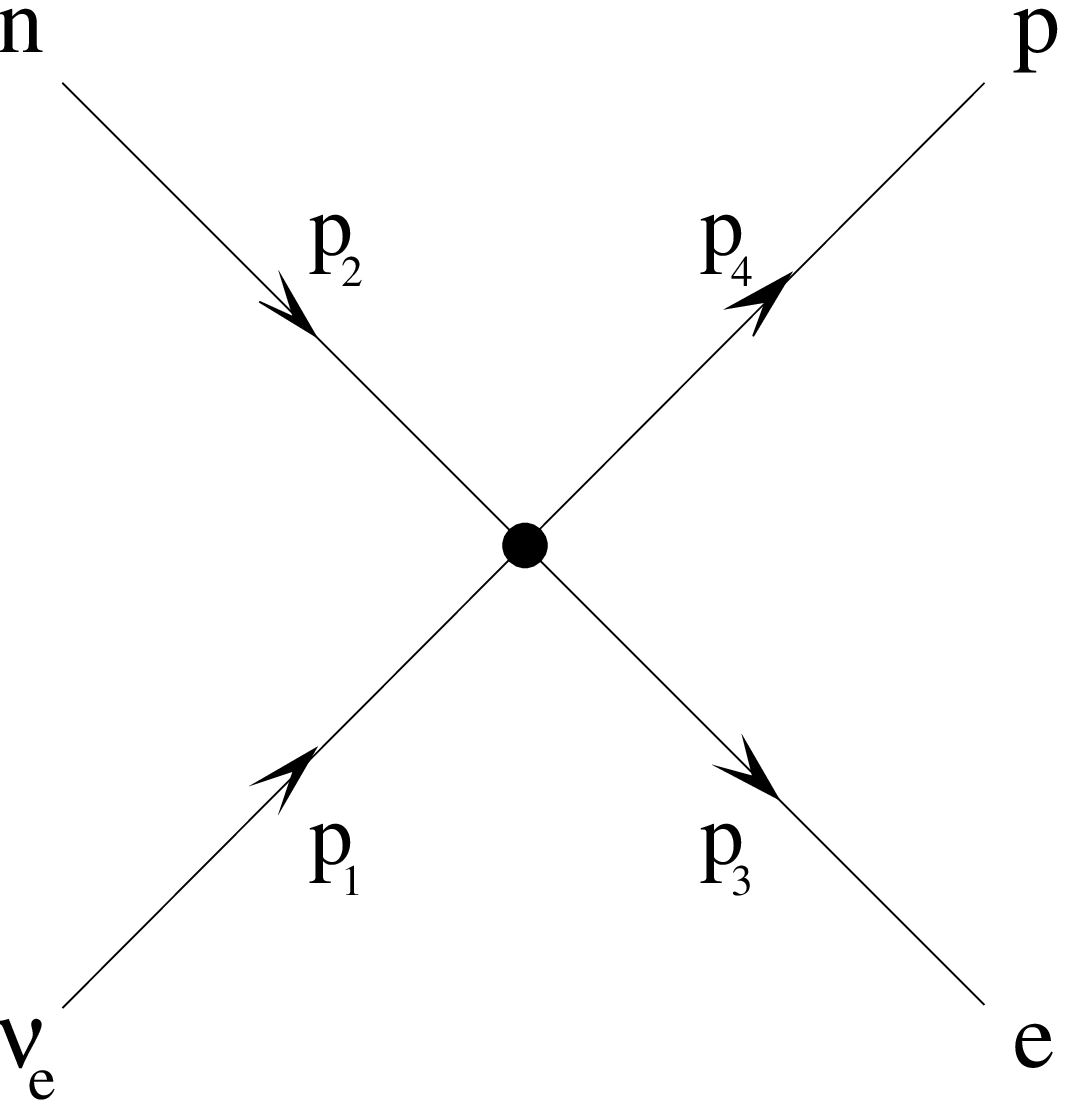}}
\caption{The Feynman diagram for the reaction
 $\ne \,+ \, n \, \rt \, e^- \, + \, p$.}
 \label{c1.1}
\end{figure}

Let us start with considering the direct reaction in (\ref{cb1}); in the
Born approximation the Feynman diagram for this process is given in Figure
\ref{c1.1}. The amplitude is
\begin{equation}\label{cb10}
  M \, = \, \frac{G_F}{\sqrt{2}} \, \ov{u}_p({\mathrm p}_4) \gamma_\mu
  (\cv - \ca \gamma_5)
  u_n({\mathrm p}_2) \ov{u}_e({\mathrm p}_3) \gamma^\mu (1 - \gamma_5)
  u_\nu({\mathrm p}_1) ~~~,
\end{equation}
where $G_F$ is the Fermi coupling constant and $\cv$, $\ca$ are the vector
and axial coupling of the nucleon. In the neutron rest frame (which, in the
limit $m_N \rt \infty$, coincides with the plasma rest frame), neglecting
proton motion ($\bvec{p}_4 \sim 0$, $E_4 \sim M_4$), we then have
\begin{equation}\label{cb11}
  \mod \, = \, 32  G_F^2 \,  M_2  M_4 \, \left( \tre E_1 E_3 \, + \,
  \left( \cv^2 - \ca^2 \right) p_2 p_3 \cos \,
  \theta_{e \nu} \right)
\end{equation}
(hereafter $p_\alpha \equiv |{\bvec{p}}_\alpha|$)> In $\Gamma ( \ne  +  n
\rt e^- + p )$, after integration over angles, the last term does not
contribute, thus we can disregard it, and obtain the well known result
\begin{equation}\label{cb12}
  \mod \, = \, 32 \,  G_F^2 \, \tre \, M_2 M_4 E_1 E_3 ~~~.
\end{equation}
The reaction rate in (\ref{cb9}) for the present process then takes the
form
\newpage
\begin{eqnarray}
  d \Gamma ( \ne  +  n  \rt  e^-  +  p ) & = & \frac{1}{2} \,
   32 \, G_F^2 \, \tre \, M_2 M_4 E_1 E_3 \, {\cdot} \nonumber \\
   & {\cdot} &  (2 \pi)^4 \, \delta^4 (p_1 + p_2 - p_3 - p_4 ) \,
   \frac{1}{2 M_2} \frac{1}{2 M_4} \frac{1}{2 E_1} \frac{1}{2 E_3}
   \, {\cdot} \nonumber \\
   & {\cdot} & \frac{d^3 \bvec{p}_1}{(2 \pi)^3} \frac{d^3 \bvec{p}_3}{(2 \pi)^3}
   \frac{d^3 \bvec{p}_4}{(2 \pi)^3} \, F_\nu (p_1) \left( 1 - F_e(p_3)
   \right)
   \label{cb13}
\end{eqnarray}
(we have distinguished the neutrino Fermi function from that of electrons
since, in general, neutrino temperature $T_\nu$ is different from the $e^{\pm}
, \gamma$ bath temperature $T$, as discussed in the previous chapter). The
integration over the proton 3-momentum can be eliminated by using
 $\delta^3(\bvec{p}_1 + \bvec{p}_2 - \bvec{p}_3 - \bvec{p}_4 )$.
From the isotropy of space, also the integration over the incident neutrino
solid angle $\int d \Omega_{\ne}$ can be easily performed, leading to an
overall factor $4\pi$:
\begin{equation}\label{cb14}
 d \Gamma ( \ne  +  n  \rt  e^-  +  p ) \, = \,
 \frac{2 G_F^2 \tre}{(2 \pi)^4} \, \delta (E_1 - Q) \, p_1^2 d p_1
 \, p_3^2 d p_3 \, d \Omega_3  \, F_\nu (p_1) \left( 1 - F_e(p_3) \right)
 ~~~,
\end{equation}
where $Q \, = \, M_4 - M_2 + E_3$. The energy $\delta$-function can be used
to perform the integration in $d p_1$, assuming massless (or almost
massless) neutrinos, i.e.
 $E_1 \sim|\bvec{p}_1|$. Finally, also the integration over the electron
angles $\int d \Omega_{3}$ is easily performed since the differential rate
does not depend (in the present approximations) on these quantities. We are
then left with the following result;
\begin{equation}\label{cb15}
 \Gamma ( \ne  +  n  \rt  e^-  +  p ) \, = \,
 \frac{G_F^2 \tre}{2 \pi^3} \, \int \, d p_3 \, p_3^2 \, Q^2 \, \theta(Q)
 \, F_\nu (Q) \left( 1 - F_e(p_3) \right) ~~~.
\end{equation}
The integration limits are imposed by the condition $Q \geq 0$; in the
present case we have $p_3 \in \left[ \sqrt{\Delta^2 - m^2} , + \infty
\right)$,
with $\Delta = M_2 - M_4$ (in the following, with $\Delta$ we always denote
the neutron - proton mass difference).

For the other five processes in \reac one simply has to observe that the
form of $\mod$ in (\ref{cb12}) remains unchanged due to crossing symmetry.
Then in the final result (\ref{cb15}) we have just to replace $Q$ (which is
determined by the energy conservation for each reaction) and the thermal
factors (which depend on the initial and final lepton states for each
process). In Table \ref{tc1.1} we report the values for $Q$ and the thermal
factors for each reaction in \reac, using the same notation adopted for
$\ne  +  n  \rt  e^-  +  p$ (i.e. $1 = \ne$ or $\neb$, $2 = n$, $3 = e^-$
or $e^+$, $4 = p$).
\begin{table}
\caption{The $Q$ parameter and thermal factors for the reactions in
\protect\reac .}
\begin{center}
\begin{tabular}{|c|c|c|} \hline \hline
  Reactions & $Q$ & Thermal factors \\ \hline \hline
  $\ne \, n \, \rt \, e^- \, p$ & $- \, \Delta \, + \, E_3$ &
  $F_\nu(Q) \; (1  -  F_e(E_3))$ \\
  $e^- \, p \, \rt \, \ne \, n$ & $- \, \Delta \, + \, E_3$ &
  $F_e(E_3) \; (1  -  F_\nu(Q))$ \\
  $e^+ \, n \, \rt \, \neb \, p$ & $\Delta \, + \, E_3$ &
  $F_e(E_3) \; (1  -  F_\nu(Q))$ \\
  $\neb \, p \, \rt \, e^+ \, n$ & $\Delta \, + \, E_3$ &
  $F_\nu(Q) \; (1  -  F_e(E_3))$ \\
  $n \, \rt \, p \, e^- \, \neb$ & $\Delta \, - \, E_3$ &
  $(1  -  F_\nu(Q)) \; (1  -  F_e(E_3))$ \\
  $p \, e^- \, \neb \, \rt \, n$ & $\Delta \, - \, E_3$ &
  $F_\nu(Q) \; F_e(E_3)$ \\ \hline \hline
\end{tabular}
\end{center}
\label{tc1.1}
\end{table}
The Born rates for each reaction in \reac, as functions of photon
temperature in the range relevant for nucleosynthesis, are plotted in
Figure \ref{fborn}. The relation between neutrino and photon temperature we
adopt is reported in the Appendix \ref{atnu}.
\begin{figure}
\epsfysize=8.4cm
\epsfxsize=7.0cm
\centerline{\epsffile{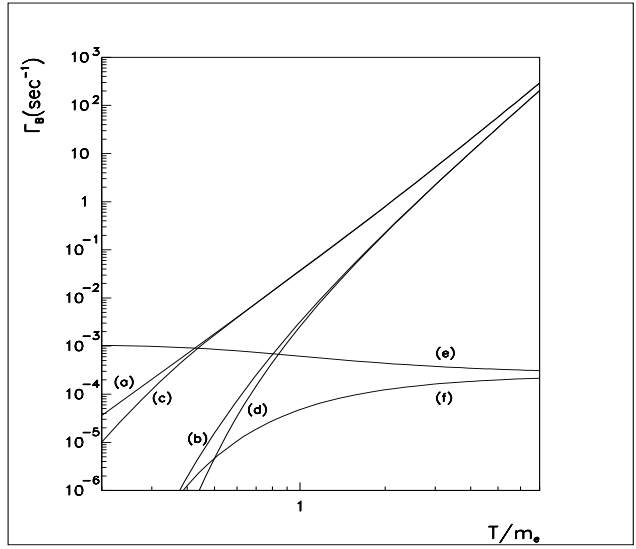}}
\caption{The Born rates $\Gamma_B$ versus photon temperature divided by
electron mass $m_e$. Hereafter reaction labels are the same as in 
\protect\reac.}
\label{fborn}
\end{figure}

\section{Radiative electromagnetic corrections: the neutron lifetime}
\label{c1b}

The reaction rates for the six weak processes in \reac are proportional to
the quantity
\begin{equation}\label{cb16}
  G_F^2 \, \tre \, = \, G_V^2 \, \left( 1 \, + \, 3 \frac{\ca^2}{\cv^2}
  \right) ~~~,
\end{equation}
as it can be seen from Eq. (\ref{cb15}), for example. The value of $G_F^2$
is very well known from the measurements of the muon decay rate \cite{PDG},
while $G_V^2$ is obtained from the observations of
 $0^+ \rt 0^+$ nuclear $\beta$ decays after applying radiative and
isospin-mixing corrections \cite{PDG}. Instead, the empirical value of the
ratio $\ca / \cv$ is deduced, for example, from experiments measuring the
angular distribution of decay products of polarized neutrons \cite{PDG}. It
is very common in the literature on BBN (see for example \cite{kolb}) to
rewrite the effective coupling constant in (\ref{cb16}) in terms of the
neutron lifetime (free neutron decay in vacuum) since, as it can be
immediately deduced from (\ref{cb15}) and Table \ref{tc1.1} dropping out
the thermal factors, it is given by the following relation
\begin{equation}\label{cb17}
 \tau_n^{-1} \, = \,
 \frac{G_F^2 \tre}{2 \pi^3} \, m_e^5 \,  \int_1^{\frac{\Delta}{m_e}} \,
 d \epsilon \, \epsilon \, \left( \epsilon \, - \, \frac{\Delta}{m_e}
 \right)^2 \, \left( \epsilon^2 \, - \, 1 \right)^{\frac{1}{2}} ~~~,
\end{equation}
where $m_e$ is the electron mass. \\ Rewriting the reaction rates in terms
of the neutron lifetime, it is then customary to substitute $\tau_n$ with
its experimental value. This way of reasoning does not bring to accurate
BBN theoretical predictions at a certain degree of approximation, but
introduces an uncertainty in the BBN calculations due to the experimental
uncertainty on the neutron lifetime. This procedure is usually adopted to
circumvent the problem of an accurate estimation of radiative
(electromagnetic) corrections to $G_V$ which is, nevertheless, a very
important one. To give an idea of the relevance of this problem, it
suffices to observe that if we calculate the neutron lifetime from the tree
level formula (\ref{cb17}) (not containing radiative corrections),
inserting the value of $G_V$ and $\ca / \cv$ reported in \cite{PDG}, we
obtain for $\tau_n$ the value $\sim 961 \, s$ which is not at all
consistent with the experimental value $886.7 \, {\pm} \, 1.9 \, s$: we need a
correction factor of about 8 \%.

This simple exercise shows that the inclusion of radiative corrections is a
very important step, which we are going to discuss.

Let us consider order $\alpha$ corrections to the neutron lifetime
($\alpha$ is the fine structure constant). These can be separated into
''outer" corrections, involving the nucleon as a whole, and ''inner"
corrections, depending on nucleon structure. Obviously, inner corrections
are sensible to the details of the strong interactions inside the nucleon,
while it can be shown \cite{Sirlin} that outer corrections, at least up to
terms of order $\alpha$, are independent of these. \\
\begin{figure}
\epsfysize=15.0cm
\epsfxsize=12.0cm
\centerline{\epsffile{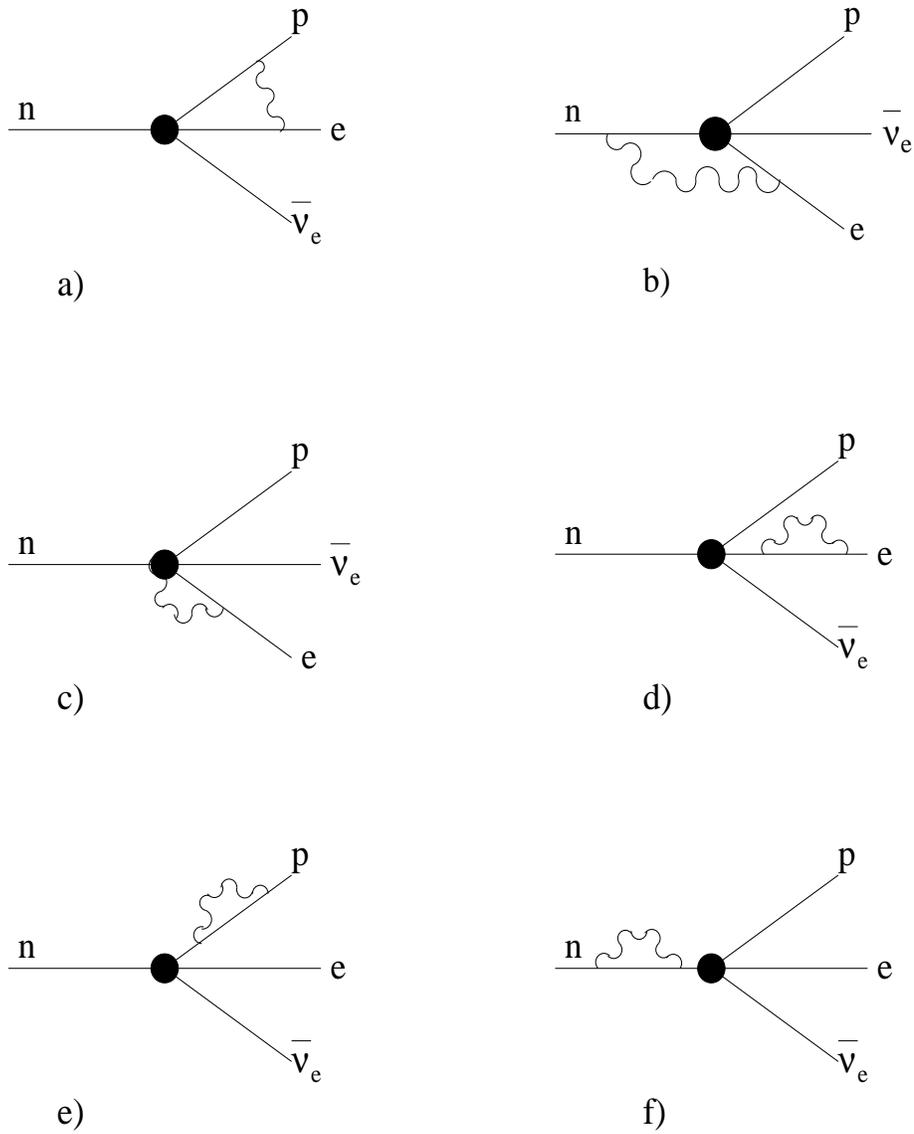}}
\caption{The Feynman diagrams for order $\alpha$ radiative
corrections to neutron decay.}
\label{c1.2}
\end{figure}
The main Feynman diagrams contributing to order $\alpha$ corrections to
neutron lifetime are sketched in Figure \ref{c1.2}. The electromagnetic
interactions of the proton are through its charge and magnetic moment,
while those of the neutron are only through its magnetic moment. All
diagrams reported in Figure \ref{c1.2} are of order $\alpha$ but the ones
involving magnetic moment interactions of the nucleons are suppressed with
respect to the others by the inverse of nucleon mass. Notice that since we
use the value of $G_F$ as measured from muon decay, we have also to
consider radiative corrections for the muon decay rate (at the same order
in $\alpha$). The relevant diagrams are similar to those pictured in Figure
\ref{c1.2}, but with $n,p$ replaced by $\mu , \nm$; obviously for muon
decay there are only outer corrections, since we deal with pointlike
particles. In the following we consistently take into account also order
$\alpha$ corrections to muon decay and report the global correction factor
for the neutron decay rate. \\
 The outer correction to the (non integrated) decay rate can be written as
\begin{equation}\label{cb18}
  \frac{\alpha}{2 \pi} \, g(E,E_m) ~~~,
\end{equation}
where $g(E,E_m)$ is a function of electron energy $E$ and end-point energy
$E_m$, reported in \cite{Sirlin}. This function, which is the same for
electron and positron capture, describes the deviations from the allowed
electron spectrum arising from the radiative corrections of order $\alpha$
\footnote{It is common to substitute the function $g(E,E_m)$ in (\ref{cb18}) with its
mean value $\ov{g} (E_m)$ obtained averaging over the allowed electron
spectrum. However, since we will consider not only neutron decay but also
the other five reactions in \reac, we do not employ this approximation.}.
It increases the decay probability for neutron $\beta$ decay (and then
decreases neutron lifetime) by about 1.5 \%. \\
 Inner corrections are much more difficult to handle, since they strongly
depend on nucleon structure. In general, one can follow two different
approaches. On one side one can directly consider radiative corrections to
the effective nucleon weak current
 $\ov{u}_p \gamma_\mu (1 - \frac{\ca}{\cv} \gamma_5) u_n$ \cite{Kubo} while
on the other side, one can study corrections for the weak quark current
$\ov{q} \gamma_\mu (1 - \gamma_5) q$ and then translate the quark-based
description to hadronic description \cite{Marciano}. Here we adopt the
second point of view, and report the results obtained by Marciano and
Sirlin \cite{Marciano}
\begin{equation}\label{cb19}
  \frac{\alpha}{2 \pi} \left( 4 \, \ln \frac{M_Z}{M_P} \, + \, \ln
  \frac{M_P}{M_A} \, + \, 2 C \, + A_g \, \right) ~~~.
\end{equation}
The first term represent the dominant model-independent short-distance
contribution ($M_P$ is the proton mass, while $M_Z$ is the $Z$ boson mass).
The second and third terms are axial-current induced contributions, where
$M_A$ is a low energy cutoff applied to the short-distance part of the
$\gamma W$ box diagram and $2 C$ is the remaining long-distance (low
energy) correction. These terms depend on the details of the strong
interaction structure and are the main sources of uncertainty in the
radiative corrections. The allowed range for the cutoff $M_A$ is
 $400 \div 1600 \, MeV$, while $C \, = \, 3 \ca {\cdot} 0.266 {\cdot} ( \mu_p + \mu_n)$,
$\mu_p + \mu_n \, \simeq \, 0.88$ being the nucleon isoscalar magnetic
moment. We then have
\begin{equation}\label{cb20}
  \frac{\alpha}{2 \pi} \left( \ln
  \frac{M_P}{M_A} \, + \, 2 C  \right) \, \simeq \, 0.0012 \, {\pm} \, 0.0018
  ~~~.
\end{equation}
The last term in (\ref{cb19}) is a perturbative QCD correction whose
calculation is rather reliable and gives
\begin{equation}\label{cb21}
  A_g \, \simeq \, - \, 0.34 ~~~.
\end{equation}
Let us observe that the largest correction comes from the first term in
(\ref{cb19}). Thus, it seems appropriate to approximate the effects of
higher orders by summing all leading-logarithmic corrections of the type
$\alpha^n
\ln^n M_Z$ ($n \, = \, 1,2,3, ...$) via a renormalization group analysis.
This has been done in \cite{MS}; the corrected rate then acquires the
factor
\begin{equation}\label{cb22}
  {\cal G}(E) \; = \; \left( 1 \, + \,  \frac{\alpha}{2 \pi} \left( \ln
  \frac{M_P}{M_A} \, + \, 2 C  \right) \, + \,
  \frac{\alpha (M_P)}{2 \pi} \, \left( g(E,E_m) \, + \, A_g \right) \right)
  S(M_P , M_Z) ~~~.
\end{equation}
where $\alpha (\mu)$ is the QED running coupling constant defined in the
$\ov{MS}$ scheme satisfying the equation
\begin{equation}\label{cb23}
  \mu \frac{d}{d \mu} \, \alpha (\mu) \, \simeq \, b_0 (\mu) \, \alpha^2
  (\mu) ~~~,
\end{equation}
with
\begin{equation}\label{cb24}
  b_0 (\mu) \, = \, \frac{2}{2 \pi} \, \sum_f \, Q_f^2 \, \theta (\mu - m_f)
  \, - \, \frac{7}{2 \pi} \, \theta (\mu - M_W) ~~~.
\end{equation}
The sum runs over all elementary fermions with mass $m_f$ and charge $Q_f$,
$M_W$ is the $W$ boson mass. The short-distance enhancement factor
$S(M_P,M_Z)$ is given by \cite{MS}, \cite{Marciano} \footnote{The
expression reported in \cite{Marciano} is here corrected allowing the (now
observed) top quark mass to be higher than the $W,Z$ boson masses}
\begin{equation}\label{cb25}
 S(M_P,M_Z) \, = \, \left( \frac{\alpha(m_c)}{\alpha(M_p)} \right)^
 {\frac{3}{4}}  \left( \frac{\alpha(m_\tau)}{\alpha(m_c)} \right)^
 {\frac{9}{16}}  \left( \frac{\alpha(m_b)}{\alpha(m_\tau)} \right)^
 {\frac{9}{19}}  \left( \frac{\alpha(M_W)}{\alpha(m_b)} \right)^
 {\frac{9}{20}}  \left( \frac{\alpha(M_Z)}{\alpha(M_W)} \right)^
 {\frac{36}{17}} ~~~.
\end{equation}
In the $\ov{MS}$ scheme one has $\alpha^{-1} (M_Z) \simeq 127.90$
\cite{PDG}. The running coupling constants at different scales are reported
in Table \ref{tc1.2} where we use the {\it central} values for the quark
masses as quoted in \cite{PDG}. From this one easily finds
\begin{equation}\label{cb26}
 S(M_P,M_Z) \, \simeq \, 1.02254 ~~~.
\end{equation}

 \begin{table}
 \caption{QED running coupling constant for several energy scales.}
\begin{center}
\begin{tabular}{|lcccr|} \hline \hline
 $ \alpha^{-1} (M_W)$ & = & $\alpha^{-1} (M_Z) \, + \, \frac{17}{18 \pi}
  \, \ln \frac{M_Z}{M_W}$ & $\simeq$ & $ 127.94
 $ \\
 $ \alpha^{-1} (m_b)$ & = & $\alpha^{-1} (M_W) \, + \, \frac{40}{9 \pi}
  \, \ln \frac{M_W}{m_b}$ & $\simeq$ & $ 132.08
 $ \\
 $ \alpha^{-1} (m_\tau)$ & = & $\alpha^{-1} (m_b) \, + \, \frac{38}{9 \pi}
  \, \ln \frac{m_b}{m_\tau}$ & $\simeq$ & $ 133.27
 $ \\
 $ \alpha^{-1} (m_c)$ & = & $\alpha^{-1} (m_\tau) \, + \, \frac{32}{9 \pi}
  \, \ln \frac{m_\tau}{m_c}$ & $\simeq$ & $ 133.62
 $ \\
 $ \alpha^{-1} (M_p)$ & = & $\alpha^{-1} (m_c) \, + \, \frac{8}{3 \pi}
  \, \ln \frac{m_c}{M_p}$ & $\simeq$ & $ 133.90
 $ \\ \hline \hline
\end{tabular}
\end{center}
\label{tc1.2}
 \end{table}

The radiative corrections considered here, which are described by the
factor in (\ref{cb22}), are not, however, the only corrections to be taken
into account for the neutron decay (they contribute for about 4 \% to the
lifetime, against the required global correction of about 8 \%). The
remaining leading corrections are usually viewed as corrections to the
phase space factor \cite{Wilk}, namely to the integrand function in
(\ref{cb17}) since they are (electron) energy dependent. We will express
the contribution to the phase space factor as multiplicative (energy
dependent) terms to the integrand function.

Let us first examine the correction due to the distortion of the outgoing
electron wave by the Coulomb field of the proton. This can be calculated by
solving the Dirac equation for an electron under the influence of a
spatially finite proton charge distribution of radius $R \simeq 1 \, fm$;
the wave function is then evaluated at the centre of the proton. The
correction factor is then given by \cite{Wilk}
\begin{equation}\label{cb27}
 {\cal F}(E) \, {\cal L}(E) \, \simeq \, \left( 1 \, + \, \alpha \pi \,
 \frac{E}{\sqrt{E^2 \, - \, m^2_e}} \right) \left( 1 \, - \,
 \alpha \, R \, E \left( \frac{m_e^2}{2 E^2} \, + \, 1 \right) \right)
 ~~~.
\end{equation}
Note that the first term in (\ref{cb27}) is usually denoted as the Fermi
function for the Coulomb scattering.

Another relevant correction comes from the fact that neither the electron
wavefunction evaluated at the centre of the proton through
 ${\cal F}(E) \, {\cal L}(E)$) nor the (anti-)neutrino one are constant
through the proton volume. Thus the decay rate has to be found by an
appropriate convolution of the electron, (anti-)neutrino and proton
wavefunctions through the proton volume. This brings to the following
correction factor \cite{Wilk}
\begin{equation}\label{cb28}
  {\cal C} \, \simeq \, 1 \, + \, c_0 \, + \, \frac{c_1}{E} \, + \,
  c_2 \, E \, + \, c_3 \, E^2 ~~~,
\end{equation}
\begin{eqnarray*}
  c_0 & = & \frac{1}{5} \left( R^2 \, m_e^2 \left( 1 \, - \frac{2}{3} B
  \, - ,\ \frac{E_m^2}{m_e^2} \right) \, - \, \alpha \, B \, R \, E_m
  \right) ~~~, \\
  c_1 & = & \frac{2B}{15} \, R^2 \, m_e^2 \, E_m ~~~, \\
  c_2 & = & \frac{1}{5} \, (3 \, - \, B) \, R \, \left( \frac{2}{3} \,
  R \, E_m \, - \, \alpha \right) ~~~, \\
  c_3 & = & \frac{2}{15} \, (3 \, - \, B) \, R^2 ~~~, \\
  B   & = & \frac{\cv^2 \, - \, \ca^2}{\tre} ~~~.
\end{eqnarray*}

Finally, there are also small corrections related to the fact that the
outgoing proton is not at rest, since it has a finite mass. This type of
effect will be considered in detail in the following chapter. Here, we only
want to point out that if the proton recoils, then the Coulomb field that
distorts the electron wave comes from a moving source. The correction
associated to this effect is contained in the factor \cite{Wilk}
\begin{equation}\label{cb29}
  {\cal Q} (E) \, \simeq \, 1 \, - \, \frac{\alpha \, \pi \, m_e^2}
  {M_p \, \sqrt{E^2 \, - \, m_e^2}} \, \left( 1 \, + \, B \,
  \frac{E_m \, - \, E}{3 E} \right) ~~~.
\end{equation}

From previous results, the neutron lifetime corrected at order $\alpha$
against electromagnetic interactions reads
\begin{eqnarray}
 \tau_n^{-1} \; = \;
 \frac{G_F^2 \tre}{2 \pi^3}  \,  \int_{m_e}^{\Delta} \!\!\!\!\!\!
 & \left. \right. &
 \!\!\!\! d E \, E \, \left( E \, - \, \Delta
 \right)^2 \, \left( E^2 \, - \, m_e^2 \right)^{\frac{1}{2}}
 {\cdot} \nonumber \\ & {\cdot} &
 {\cal G}(E) \, {\cal F}(E) \, {\cal L}(E) \, {\cal C}(E) \, {\cal Q}(E)
 \label{cb30} ~~~,
\end{eqnarray}
where ${\cal G}(E)$ is the correction function in (\ref{cb22}) and the
other terms are reported in (\ref{cb27}) - (\ref{cb29}). Concerning the
relevance of the different corrections just considered, a comment is in
turn. For seek of completeness we have reported an exhaustive description
of all corrections at order $\alpha$, but as clearly appears from the
explicit expressions of (\ref{cb22})-(\ref{cb29}), the main contributions
come from (\ref{cb22}) and ${\cal F}(E)$ term in (\ref{cb27}). All other
contributions, contained in ${\cal L}(E)$, ${\cal C}(E)$ and ${\cal Q}
(E)$, are in fact much smaller and thus they can be safely neglected, since
they are suppressed by a factors of the order $\Delta/\Lambda_{QCD}$.
Actually the $\alpha^2$ contributions coming from radiative and Coulomb
effects are even larger, or of the same order of magnitude, than these
terms, so they should be included for consistency if ${\cal L}(E)$, ${\cal
C}(E)$ and ${\cal Q} (E)$ are taken into account \cite{Wilk}. For the level
of accuracy of our analysis it will be sufficient to include ${\cal F}(E)$
and ${\cal G}(E)$ at order $\alpha$ only. \\ Evaluating numerically the
integral in (\ref{cb30}) we then obtain for the neutron lifetime the value
\footnote{For completeness we have included also the corrections coming
from the finite nucleon mass changing the squared matrix element, which are
considered in the next chapter. However, these corrections are very small.}
$893.8 \, s$ which is now quite compatible with the experimental value.
Note that the theoretical prediction can be further refined (leading to a
slightly better agreement with the experiments) by considering other small
effects (again of order $\alpha$) such as magnetic moment (normal and
anomalous) interactions, residual average proton polarization due to parity
non-conservation and so on. While these effects are briefly discussed in
\cite{Wilk}, for our purposes we do not consider them, since their
contributions are energy independent and then account only for an overall
factor.

\section{Results for the reaction rates}
\label{c1c}

The analysis of radiative corrections for $\beta$-decay shows that their
contribution to the rates for the processes in \reac$\!\!$, relevant for
BBN, is expected to be as large as few percent. To reach accuracy of the
order of one percent in rate evaluations it is therefore necessary to
correct Born rates for both radiative and Coulomb effects. This last
contribution, however, is only present when both electron and proton are
present in the initial or final state, since it can be viewed as the
electromagnetic rescattering of the two charged particles. In this way it
is straightforward to show that no Coulomb corrections ${\cal F}(E)$ are
present for the channels $e^+ + n \leftrightarrow \neb + p$. The results
reported in (\ref{cb22}), (\ref{cb27}) can be applied to the general
formula for the decay rates (\ref{cb9}) by rewriting the dependence on the
electron spectrum end-point $E_m$ in terms of neutrino energy. In fact,
with the same notation of Figure \ref{c1.1} we have that $E_m \, = \, E_1
\, + \, E_3$. Then, in general the correction factors depend on both the
electron energy $E_3$ and neutrino energy $E_1$, which are both integration
variables.

Indicating with $d \Gamma_B$ the uncorrected decay rate in (\ref{cb9}), the
corrected one is given by
\begin{equation}\label{cb31}
  d \Gamma_r \, \simeq \, {\cal G}(E_1 , E_3) \, {\cal F}(E_3)
  \, d \Gamma_B ~~~,
\end{equation}
where the factors are the same ones reported in (\ref{cb22}), (\ref{cb27})
but with $E_m$ substituted by $E_1 \, + \, E_3$, and the factor ${\cal
F}(E_3)$ does not apply to reactions (c), (d).
\begin{figure}
\epsfysize=8.4cm
\epsfxsize=7.0cm
\centerline{\epsffile{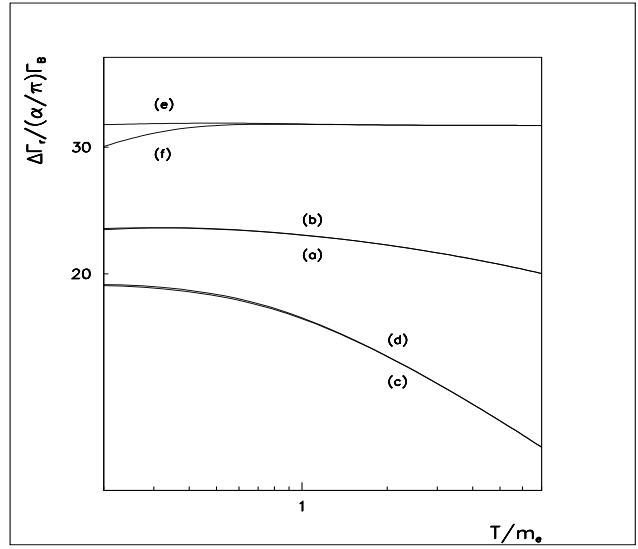}}
\caption{Radiative corrections to the Born rates for the processes in
\protect\reac.}
\label{frad}
\end{figure}

Equipped with the corrected formula in (\ref{cb31}), we may now numerically
evaluate the Born rates for the reactions in \reac. In Figure \ref{frad} we
show the relative difference between the Born rates and the radiatively
corrected ones, namely the quantity
\begin{equation}\label{cb32}
  \Delta \Gamma_r \; = \; \frac{\Gamma_r \, - \, \Gamma_B}{(\alpha / \pi)
  \, \Gamma_B} ~~~.
\end{equation}
Note that at sufficiently high temperatures the corrections for a given
process and the inverse one coincide. Apart from the $\beta$-decay
reactions, whose corrections are practically constant with temperature, for
the scattering reactions the considered radiative corrections are relevant
especially at low temperatures.

In our BBN calculations reported in chapter \ref{sum} we have also included
an overall rescaling factor $\tau_n^{th} / \tau_n^{exp}$ for the reaction
rates allowing the theoretical prediction for the neutron lifetime to be
fully compatible with the experimental value. This is a standard procedure,
which allows to overcome the problem of a precise determination of the
coupling constant for weak interactions involved in BBN, expressing this
overall factor in terms of the experimental value of neutron lifetime. We
stress, however, that it is worth rescaling the rate only after all known
corrections have been included, since this increases the accuracy of the
prediction. In this way the ansatz that the residual correction be an
overall factor and not, as it is reasonable to expect, a function of
leptons energies, may introduce errors less than $1 \%$. Although not
included in Figure \ref{frad} (it would cause a trivial overall shift of
the curves), we have also rescaled the corrected rates in (\ref{cb31}) by
the factor
\begin{equation}\label{cb33}
  1 \; + \; \delta_\tau \; \equiv \; \frac{\tau_n^{th}}{\tau_n^{exp}} \;
  \simeq \; 1.008 ~~~,
\end{equation}
representing an energy independent constant correction which should be
included to reproduce the experimental value for neutron lifetime at zero
temperature and density.

\chapter{Finite nucleon mass corrections}
\label{c2}

In this chapter we calculate the corrections to the Born rates for the $n
\lrt p$ reactions \reac  induced by relaxing the assumption of infinite
mass nucleons \cite{finb}. Such an assumption was made (implicitly or
explicitly) in several points of the calculations done in the previous
chapter, which will be now subject to critical examination.

The main effect of considering finite values for the nucleon masses are due
to a non vanishing nucleon velocity (and then recoil effect) and to the
presence to effective extra nucleon interactions such as weak magnetism.

Recoil effects have been neglected during the calculation of Born rates in
the evaluation of the spin summed squared matrix element (Eq. (\ref{cb11})
is valid only for fixed nucleons) and in the calculation of phase space
integrals. The last point (kinematical changes) comes out from the fact
that a finite nucleon velocity modifies the energy-momentum conservation
relations and then the phase space for the given process changes. In
particular, both the statistical distributions (Fermi functions) and the
integration limits change with respect to the Born case (note also that Eq.
(\ref{cb8}) is valid only if one neglects recoil effect in $\mod$). The
most practical implication of considering recoil effects is that it is no
more possible to write the reaction rate as a one-dimensional integral, as
in Eq. (\ref{cb15}).

However, recoil effects are not the only ones involved in the finite
nucleon mass corrections. Up to the same order of approximation we have
also to consider modifications of the effective nucleon weak current due to
weak magnetism, and, in general, to interaction terms due to the scalar and
pseudoscalar couplings of the nucleons.

In the following section we firstly calculate the dynamical changes in the
squared matrix element induced by weak magnetism, scalar and pseudoscalar
interactions. Then, in the subsequent section, we will examine kinematical
changes induced in the phase space structure of the reaction rates. At
first order in $T/M_N$ these two corrections can be treated independently.

\section{Corrections to the transition amplitude}
\label{c2a}

For definiteness let us consider the process shown in Figure (\ref{c1.1}).
At first order in $1/M_N$, the transition amplitude can be written as
\begin{equation}\label{cm1}
  M \, = \, \frac{G_F}{\sqrt{2}} \, \ov{u}_p(p_4) O_\mu
  u_n(p_2) \, \ov{u}_e(p_3) \gamma^\mu (1 - \gamma_5) u_\nu(p_1) ~~~,
\end{equation}
where the effective nucleon-nucleon weak coupling is given in general by
\cite{weak}
\begin{equation}\label{cm2}
  O_\mu \; = \; \gamma_\mu (\cv - \ca \gamma_5) \; + \; i \,
  \frac{f_2}{M_N} \, \sigma_{\mu \nu} \, q^\nu \; + \; f_3 \, q_\mu \; + \;
  f_{ps} \, \gamma_5 \, q_\mu ~~~.
\end{equation}
Here $q_\mu$ denotes the momentum transfer to the final nucleon and $f_2$,
$f_3$, $f_{ps}$ are the anomalous weak charged-current magnetic moment,
scalar and pseudoscalar couplings of the nucleon, respectively
\cite{f2fps}. In general, the couplings $\cv$, $\ca$, $f_2$, $f_3$,
$f_{ps}$ are form-factors, which all depend on $q^2$. However, at the
relevant energy scales for the considered processes, this $q^2$-dependence
can be neglected, being of higher order than $1/M_N$.

With some algebra, one can evaluate the squared modulus of (\ref{cm1})
summed over all spins. It is very useful, for references to the other
processes, to write the resultant expression in terms of the relativistic
invariants $s$, $t$
\footnote{The third invariant $u$ has been eliminated through the relation
 $u \, = \, - \, s \, - t \, + \, M_1^2 \, + \, M_2^2 \, + \, M_3^2 \, +
 M_4^2$
which follows from energy-momentum conservation.}. With the same notation
of section (\ref{c1a}) we obtain the result reported in appendix
\ref{matrix}. The expression in (\ref{cm3}) holds for all the six processes
in \reac.
%;however for reactions with charged anti-leptons (namely (\ref{cb2})) one
%has to replace $\ca$ with $ - \, \ca$.

Let us note that in the formula (\ref{cm3}) we have retained only the
leading terms, namely those corresponding to first order in the couplings
$f_2$, $f_3$ and $f_{ps}$. Aa a further simplification we can drop out all
the terms coming from scalar and pseudoscalar interactions, which are
indeed very weak (in the following we then use $f_3
\, = \, f_{ps} \, = 0$)

\begin{figure}
\epsfysize=8.4cm
\epsfxsize=7.0cm
\centerline{\epsffile{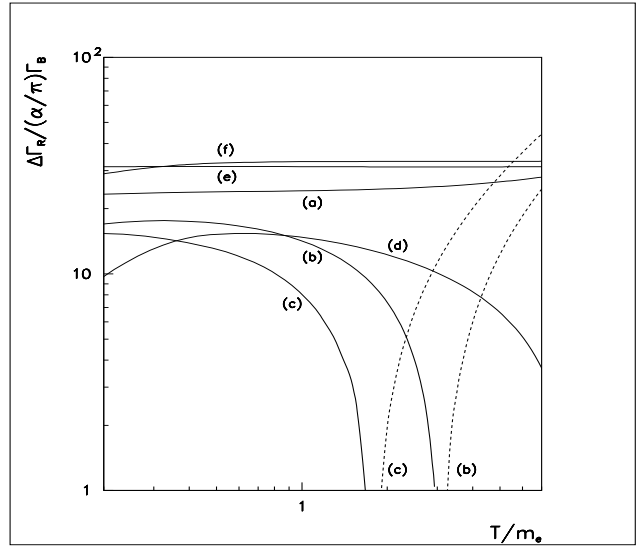}}
\caption{``Zero temperature" radiative corrections to the Born rates
for the processes in \protect\reac (see text). 
We use the same notations of chapter \protect\ref{c1}.}
\label{fzerot}
\end{figure}

In Figure \ref{fzerot} we plot the ``zero temperature" cumulative
corrections, i.e. the corrections to the transition amplitude considered
here plus the radiative QED corrections evaluated in the previous chapter;
for the process $\nu_e + n \rightarrow e^- + p$ we have $\Delta \Gamma_R \,
= \, \Gamma_R  - \Gamma_B$ with
\bea
\Gamma_R( \ne  +  n  \rightarrow  e^-  +  p ) &=&
\frac{1 + \delta_\tau}{128 \pi^3 M_n}
\int_0^\infty d E_1 \int^{E_{sup}}_{E_{inf}} d E_3 ~\mod(E_1,E_3)
\nonumber\\
&{\times}&{\cal G}(E_1,E_3)~{\cal F}(E_3) ~F_\nu(E_1)~\left[ 1-F_e(E_3)
\right]~~~, \label{mm1}
\eea
where
\bea
E_{inf}&=&\frac{\left[(M_n +E_1)(M_n^2  - M_p^2 + m_e^2 + 2 E_1 M_n) - 2E_1
\xi \right]} {2M_n(M_n + 2 E_1)} ~~~,\label{mm3} \\
E_{sup}&=&\frac{\left[(M_n + E_1) (M_n^2  - M_p^2 + m_e^2 + 2 E_1 M_n) +
2E_1 \xi \right]}{2M_n(M_n + 2 E_1)}~~~,\label{mm4} \\
\xi&=&\frac 12 \sqrt{\left[M_n^2  - M_p^2 - m_e^2 + 2 E_1 M_n\right]^2 - 4
m_e^2 M_p^2} ~~~. \label{mm2}
\eea
The reason for this name is that in the limit of zero temperature these
corrections do not vanish, differently from  what occurs to the other
corrections considered in this thesis. Note that the constant shift
correction $\pi
\delta_\tau /\alpha$ (see Eq. (\ref{cb33})) has been subtracted in order to
show the pure radiative and Coulomb effects and finite mass corrections. As
we will see, these corrections turn to be the most relevant ones for the
Born rates.

\section{Kinematical corrections}
\label{c2b}

Apart from the dynamical corrections to the squared matrix element
calculated in the previous section, there are also other corrections to the
reaction rates which are of kinematical nature, arising properly from the
finite nucleon velocity. While dynamical corrections are due to the fact
that for a slightly moving nucleon the collision probability is higher than
for a fixed one, kinematical corrections involve changes in the energy
carried off by leptons due to nucleon motion.

It is important to note that the reaction rates (per incident nucleon)
intervening in the Boltzmann equations for the calculations of the
elemental abundances have to be evaluated in the reference frame of the
comoving volume (radiation rest frame).

\begin{figure}
\epsfysize=12cm
\epsfxsize=10cm
\centerline{\epsffile{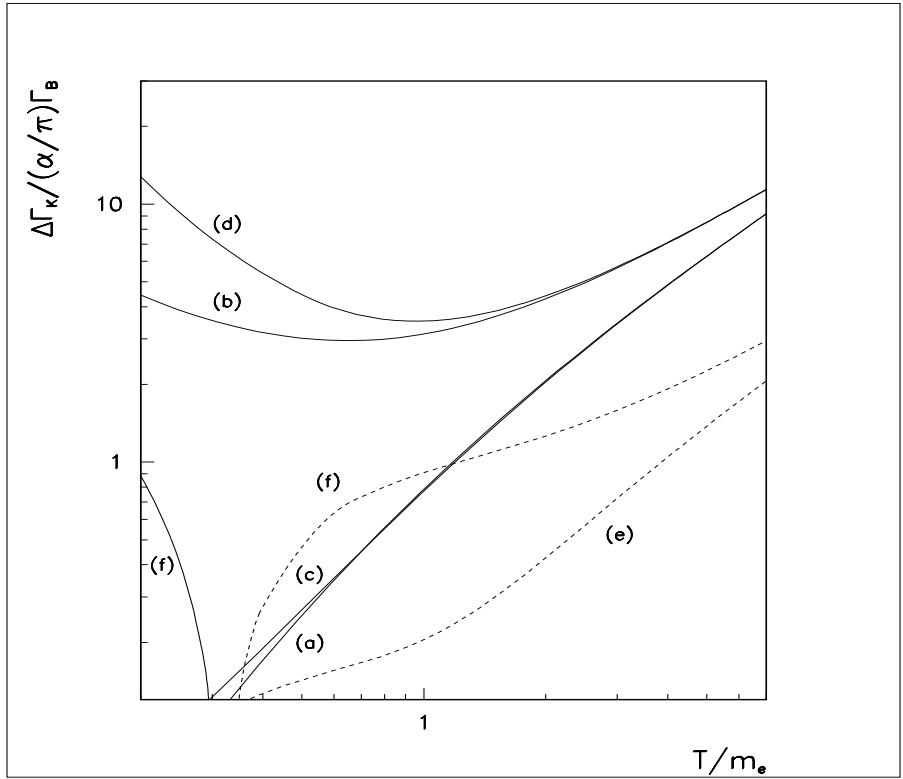}}
\caption{The ratio $\Delta \Gamma_K / (\alpha / \pi) \Gamma_B$. Dashed
lines correspond to negative values and labels refer to reactions in
\protect\reac .}
\label{fkin}
\end{figure}

For definiteness let us consider the reaction $\ne \,+ \, n \rt e^- \, +
\, p$ whose exact expression for the rate is given by
\begin{eqnarray}
  \Gamma ( \ne  \, + \,  n  \rt  e^-  \, + \,  p ) \!\!\!\! & \left. \right. &
  \!\! = \; \frac{1}{n} \, \int
  \frac{d^3 \bvec{p}_2}{(2 \pi)^3} \, \frac{1}{2 E_2} \, F_n(E_2) \,
  \left\{
  \int \frac{d^3 \bvec{p}_4}{(2 \pi)^3} \, \frac{1}{2 E_4} \,
  \frac{d^3 \bvec{p}_1}{(2 \pi)^3} \, \frac{1}{2 E_1} \,
  \frac{d^3 \bvec{p}_3}{(2 \pi)^3} \, \frac{1}{2 E_3} \, \right.
  \nonumber \\
  & \left. \right. & \left. (2 \pi)^4 \, \delta^4 (p_1 + p_2 - p_3 - p_4 ) \,
  \mod \right\} \, F_\nu (E_1) \left( 1 - F_e(E_3) \right) \label{cm4}
\end{eqnarray}
(as already discussed, we approximate $1 \, - \, F_p(E_4) \, \simeq \, 1$),
where
\begin{equation}\label{cm5}
  n \; = \; 2 \, \int \frac{d^3 \bvec{p}_2}{(2 \pi)^3} \, F_n(E_2)
\end{equation}
is the incident neutron number density. All the quantities in (\ref{cm4})
are evaluated in the comoving volume reference frame; we denote with a
prime the corresponding values in the neutron rest frame (in the infinite
nucleon mass limit the two reference frames coincide). If the neutron
motion lies along the $z$-axis, we have
\begin{equation}\label{cm6}
  \left( \begin{array}{c} E_2 \\  p_{2 z} \end{array} \right)
  \; = \; \Lambda \,
  \left( \begin{array}{c} E_2^{\prime} \\  p_{2 z}^{\prime} \end{array}
  \right) \; = \; \left( \begin{array}{cc} \gamma & \gamma \, u \\
  \gamma \, u & \gamma \end{array} \right) \,
   \left( \begin{array}{c} M_2 \\  0 \end{array} \right) ~~~ ,
\end{equation}
where $u \, = \, p_{2 z} / M_2$ is the neutron velocity and
\begin{eqnarray}
  \gamma & = & \frac{1}{\sqrt{1 \, - \, u^2}} \; \simeq \; 1 \; + \;
  \frac{v^2}{2} \, = \; 1 \; + \; \frac{p^2_{2 z}}{2 M_2^2}
  ~~~ , \label{cm7} \\
  \gamma \, u & = & \frac{u}{\sqrt{1 \, - \, u^2}} \; \simeq \; u \; = \;
  \frac{p_{2 z}}{M_2} ~~~ . \label{cm8}
\end{eqnarray}
For the leptons we then obtain
\begin{equation}\label{cm9}
  \left( \begin{array}{c} E_{1,3} \\  p_{1,3}^n \end{array} \right)
  \; = \; \Lambda \,
  \left( \begin{array}{c} E_{1,3}^{\prime} \\  p_{1,3}^{n \, \prime}
  \end{array} \right) \; = \;
  \left( \begin{array}{cc}
  1 \; + \; \frac{p^2_{2 z}}{2 M_2^2} & \frac{p_{2 z}}{M_2} \\
  \frac{p_{2 z}}{M_2} & 1 \; + \; \frac{p^2_{2 z}}{2 M_2^2} \end{array}
  \right) \,
  \left( \begin{array}{c} E_{1,3}^{\prime} \\  p_{1,3}^{n \, \prime}
  \end{array} \right) ~~~ ,
\end{equation}
where $p^n_{1,3}$ is the lepton momentum component along neutron motion. In
general for the lepton energies we can write
\begin{equation}\label{cm10}
  E_{1,3} \; \simeq \; E^\prime_{1,3} \; + \; \frac{\bvec{p_2} {\cdot}
  \bvec{p_{1,3}}}{M_2} \; + \; \frac{\bvec{p_2}^2}{2 M_2^2} \,
  E^\prime_{1,3} ~~~ ,
\end{equation}
Now, let us note that even if $\Gamma$ in (\ref{cm4}) has to be evaluated
in the comoving volume reference frame, the quantity in
 $\left\{ ... \right\}$ is a Lorentz invariant which takes the same value
in any reference frame. We choose to evaluate this quantity in the nucleon
rest frame. In this way, the kinematical corrections involve only the
quantities outside the braces \footnote{Note that even if the quantity in
$\left\{ ... \right\}$ is an invariant, nevertheless the integration
limits, in general, change their values allowing changes in the phase space
due to neutron motion. However, we only want to calculate the leading
kinematical correction contribution, thus we have to consider only
corrections coming from  the terms outside braces, while the quantities
inside them (including the integration limits) have to be calculated in the
infinite nucleon mass approximation.}. The correction to the phase space
term is then easily calculated
\begin{eqnarray}
  \delta \left( F_\nu \, (1 \, - \, F_e) \right) & = & F_\nu(E_1) \,
  \left( 1 \, - \, F(E_3) \right) \; - \; F_\nu(E_1^\prime) \,
  \left( 1 \, - \, F(E_3^\prime) \right) \; \simeq \nonumber \\
  & \simeq & F_\nu(E_1^\prime) \,
  \left( 1 \, - \, F(E_3^\prime) \right) \, \left\{
  \frac{\bvec{p_2} {\cdot}  \bvec{p_{3}}}{M_2 \, T} \, F(E_3^\prime) \; - \;
  \frac{\bvec{p_2} {\cdot}  \bvec{p_{1}}}{M_2 \, T_\nu} \,
  \left( 1 \, - \, F_\nu(E_1^\prime) \right) \; + \right. \nonumber \\
  & - & \frac{1}{2} \, \left(
  \frac{\bvec{p_2} {\cdot}  \bvec{p_{3}}}{M_2 \, T} \right)^2 \,
  F(E_3^\prime) \, \left( 1 \, - \, 2 F(E_3^\prime) \right) \; +
  \nonumber \\
  & - &
  \frac{\bvec{p_2} {\cdot}  \bvec{p_{3}} \, \bvec{p_2} {\cdot}  \bvec{p_{1}}
  }{M_2^2 \, T \, T_\nu} \, F(E_3^\prime) \,
  \left( 1 \, - \, F_\nu(E_1^\prime) \right) \; + \nonumber \\
  & - & \frac{1}{2} \, \left(
  \frac{\bvec{p_2} {\cdot}  \bvec{p_{1}}}{M_2 \, T} \right)^2 \,
  \left( 1 \, - \, F_\nu(E_1^\prime) \right) \,
  \left( 1 \, - \, 2 F_\nu(E_1^\prime) \right) \; + \nonumber \\
  & + &
  \frac{\bvec{p_2}^2}{2 M_2^2 T} \, E^\prime_3 \, F(E_3^\prime) \;
  - \: \left. \frac{\bvec{p_2}^2}{2 M_2^2 T} \, E^\prime_1 \,
  \left( 1 \, - \, F_\nu(E_1^\prime) \right) \right\}
  ~~~ . \label{cm11}
\end{eqnarray}
It is here appropriate to approximate the neutron distribution function
with a Boltzmann function
\begin{equation}\label{cm12}
  F_n(E_2) \; = \; \frac{1}{e^{\frac{E_2}{T}} \, + \, 1} \; \simeq \;
  e^{- \frac{E_2}{T}} \; \simeq \; e^{- \frac{M_2}{T}} \,
  e^{- \frac{\bvec{p_2}^2}{2 M_2 T}} ~~~ ,
\end{equation}
thus the correction to the rate for the considered reaction is
\begin{eqnarray}
  \delta \Gamma \!\!\!\! & ( & \!\!\!\! \ne  \,  n  \rt  e^-  \,  p ) \;
  \simeq \; \frac{1}{n} \, \int
  \frac{d^3 \bvec{p}_2}{(2 \pi)^3} \, \frac{1}{2 E_2} \,
  e^{- \frac{M_2}{T}} \, e^{- \frac{\bvec{p_2}^2}{2 M_2 T}} \,
  \left\{
  \int \frac{d^3 \bvec{p}_4}{(2 \pi)^3} \, \frac{1}{2 E_4} \,
  \frac{d^3 \bvec{p}_1}{(2 \pi)^3} \, \frac{1}{2 E_1} \, \right.
  \nonumber \\
  & \left. \right. & \!\!\!\! \left. \!\!\!\!
  \frac{d^3 \bvec{p}_3}{(2 \pi)^3} \, \frac{1}{2 E_3} \,
  (2 \pi)^4 \, \delta^4 (p_1 + p_2 - p_3 - p_4 ) \,
  \mod \right\} \, \delta \left( F_\nu \, (1 \, - \, F_e) \right)
  ~~~ . \label{cm13}
\end{eqnarray}
Since the expression in $\left\{ ... \right\}$ does not depend on neutron
momentum (see the above discussion), it is very convenient to perform first
the integration over this quantity. First order terms in (\ref{cm11}) do
not contribute, since
\begin{equation}\label{cm14}
  \int \frac{d^3 \bvec{p}_2}{(2 \pi)^3} \, \frac{1}{2 M_2} \,
  e^{- \frac{M_2}{T}} \, e^{- \frac{\bvec{p_2}^2}{2 M_2 T}} \, p_{2 i} \;
  = \; 0 ~~~ ,
\end{equation}
for symmetry reasons. Instead for second order terms we have
\begin{eqnarray}
  \frac{1}{n} \int \frac{d^3 \bvec{p}_2}{(2 \pi)^3} \, \frac{1}{2 M_2}
  \!\!\!\!\!\! & \left. \right. & \!\!\!\!\!\!
  e^{- \frac{M_2}{T}} \, e^{- \frac{\bvec{p_2}^2}{2 M_2 T}} \, p_{2 i} \,
  p_{2 j} \; \simeq \; \frac{\int \frac{d^3 \bvec{p}_2}{(2 \pi)^3} \,
  \frac{p_{2 i} \, p_{2 j}}{2 M_2} \, e^{- \frac{\bvec{p_2}^2}{2 M_2 T}}}
  {2 \, \int \frac{d^3 \bvec{p}_2}{(2 \pi)^3} \, e^{- \frac{\bvec{p_2}^2}{2
  M_2 T}}}  \; = \nonumber \\
  & = & \frac{1}{2} \, \delta_{i j} \,
  \frac{\int \frac{d^3 \bvec{p}_2}{(2 \pi)^3} \,
  \frac{p_{2 i}^2}{2 M_2} \, e^{- \frac{\bvec{p_2}^2}{2 M_2 T}}}
  {2 \, \int \frac{d^3 \bvec{p}_2}{(2 \pi)^3} \, e^{- \frac{\bvec{p_2}^2}{2
  M_2 T}}}  \; = \; \frac{T}{4} \, \delta_{i j } ~~~ , \label{cm15}
\end{eqnarray}
hence Eq. (\ref{cm13}) now becomes
\begin{eqnarray}
  \delta \Gamma ( \ne  \,  n  \rt  e^-  \,  p ) & \simeq &
  \int \frac{d^3 \bvec{p}_4}{(2 \pi)^3} \, \frac{1}{2 E_4} \,
  \frac{d^3 \bvec{p}_1}{(2 \pi)^3} \, \frac{1}{2 E_1} \,
  \frac{d^3 \bvec{p}_3}{(2 \pi)^3} \, \frac{1}{2 E_3}
  \, {\cdot} \nonumber \\
  & {\cdot} &  (2 \pi)^4 \, \delta^4 (p_1 + p_2 - p_3 - p_4 ) \,
  \mod  \; \delta \Phi ~~~ , \label{cm16}
\end{eqnarray}
with
\begin{eqnarray}
  \delta \Phi & = & \frac{T}{4 M_2^2} \left( - \frac{1}{2} \,
  \frac{\bvec{p_3}^2}{T^2} \,
  F(E_3^\prime) \, \left( 1 \, - \, 2 F(E_3^\prime) \right) \; - \;
  \frac{\bvec{p_1^\prime} {\cdot} \bvec{p_3^\prime}}{T T_\nu} \,
  F(E_3^\prime) \, \left( 1 \, - \, F_\nu(E_1^\prime) \right) \; +
  \right. \nonumber \\
  & + &  \frac{1}{2} \,  \frac{\bvec{p_1}^2}{T_\nu^2} \,
  \left( 1 \, - \,  F_\nu(E_1^\prime) \right)
  \, \left( 1 \, - \, 2 F_\nu(E_1^\prime) \right) \; + \;
  \frac{3}{2} \, \frac{E^\prime_3}{T} \, F(E_3^\prime) \: +
  \nonumber \\
  & - & \left.
  \frac{3}{2} \, \frac{E^\prime_1}{T_\nu} \,
  \left( 1 \, - \,  F_\nu(E_1^\prime) \right) \right) ~~~ , \label{cm17}
\end{eqnarray}
and $\mod$ is given by the expression in (\ref{cb11}). As in section
(\ref{c1a}), the integration over the proton 3-momentum can be eliminated
by using the
 $\delta^3 ( \bvec{p_1} \, + \, \bvec{p_2} \, - \, \bvec{p_3} \, - \,
 \bvec{p_4})$,
whereas the neutrino energy integration (for approximatively massless
neutrinos) through the energy $\delta$-function. Also integration over
angles can be analytically performed (only the one over electron-neutrino
angle is now non trivial). The result is
\begin{eqnarray}
\Delta \Gamma_K ( \ne  \,  n  \rt  e^-  \,  p )
& = & \frac{G_F^2 (c_V^2 + 3 c_A^2)}{2 \pi^3}
\left( \frac{T}{2 M_n} \right) \int dp_3 \; p_3^2 \, Q^2 \, \theta(Q) \,
F_\nu(Q)~(1-F(E_3)) \; {\cdot} \nonumber\\ & {\cdot} & \left[
\frac{Q^2}{T_\nu^2}(1-F_\nu(Q))(1-2 F_\nu(Q))
-\frac{p_3^2}{T^2}F(E_3) (1-2 F(E_3)) + \right.
\nonumber\\
& + & 3 \frac{E_3}{T}F(E_3) -  3 \frac{Q}{T_\nu}(1-F_\nu(Q)) - \, 3 \, +
\nonumber\\
& + &
\left. \left(\frac{c_A^2 - c_V^2}{c_V^2 + 3 c_A^2}\right)
\left(\frac{2 p_3^2 Q}{3 T T_\nu E_3}\right) F(E_3)(1-F_\nu(Q))\right]
~~~ , \label{cm18}
\end{eqnarray}
where we have used the same notation of section (\ref{c1a}). The above
expression for $\Delta \Gamma_K( \ne  +  n \rt  e^-  +  p )$ cannot be
extended to all reactions of \reac by simply using the substitution rules
of Table \ref{tc1.1}. This is due to the different form of the statistical
factors involved in the corresponding expressions of (\ref{cm4}). However,
it is easy to find that the factor in square brackets must be replaced as
follows. First of all, note that for the reaction $\neb + p \rt e^+ + n$
this factor is the same as in (\ref{cm18}). Differently, for $e^- + p \,
\rt \, \ne + n$ and $e^+ + n \, \rt \, \neb + p$, one has
\begin{eqnarray}
\left[-3 -\frac{Q^2}{T_\nu^2} F_\nu(Q) \left[ 1-2 F_\nu(Q) \right]
+\frac{{p_3}^2}{T^2} \left[ 1-F(E_3) \right]
\left[1-2 F(E_3) \right]\right.
+  3 \frac{Q}{T_\nu} F_\nu(Q)  \nonumber \\ \left. -  3
\frac{E_3}{T} \left[ 1-F(E_3) \right] + \left(\frac{\ca^2 -
\cv^2}{\tre}\right)
\left(\frac{2 {p_3}^2 Q}{3 T T_\nu E_3}\right) F_\nu(Q)
\left[ 1-F(E_3) \right] \right] ~~~,
\end{eqnarray}
while for $n \, \rt \,  e^- + \neb + p$
\begin{eqnarray}&&
\left[ -3 -\frac{Q^2}{T_\nu^2} F_\nu(Q)  \left[1-2 F_\nu(Q)
\right]
-\frac{{p_3}^2}{T^2} F(E_3) \left[ 1-2 F(E_3) \right]
+ 3 \frac{q}{T_\nu} F_\nu(Q)
\right. \nonumber \\  &+& \left. 3 \frac{E_3}{T} F(E_3)-
\left(\frac{\ca^2 - \cv^2}{\tre}\right)
\left(\frac{2 {p_3}^2 Q}{3 T T_\nu E_3}\right) F(E_3)
F_\nu(Q) \right] ~~~.
\end{eqnarray}
Finally for $e^- + \neb + p \, \rt \, n$
\begin{eqnarray}
&& \left[ - 3 + \frac{Q^2}{T_\nu^2} \left[ 1-F_\nu(Q) \right]
\left[1-2
F_\nu(Q) \right] +\frac{{p_3}^2}{T^2} \left[ 1 - F(E_3)
\right] \left[ 1-2 F(E_3) \right] \right.
 -3 \frac{Q}{T_\nu} \left[ 1-F_\nu(Q)\right]
 \nonumber \\ &-&\left.3 \frac{E_3}{T} \left[ 1 - F(E_3)\right]
- \left(\frac{\ca^2 - \cv^2}{\tre}\right)
\left(\frac{2 {p_3}^2 Q}{3 T T_\nu E_3}\right) \left[ 1 - F(E_3)
\right] \left[ 1-F_\nu(Q) \right] \right] ~~~.
\end{eqnarray}
We report the results in Figure \ref{fkin} in the form of the ratio $\Delta
\Gamma_K / (\alpha /\pi) \Gamma_B$ for comparison with the other corrections
which are of order $\alpha$. Note that in the entire temperature range
relevant for nucleosynthesis the dominant corrections are positive and
apply to the scattering reactions (b), (d).

\chapter{QED thermal radiative corrections}
\label{c3}

The reaction rates for the processes in \reac strongly depend whether they
take place in vacuum or in a heat bath at given temperature $T$. The most
important contribution comes from the density of states in the integration
over phase space. This has been already  taken into account in the previous
chapters for the calculation of the rates in the Born approximation and of
the finite nucleon mass corrections. However, there are also a number of
effects \cite{Dicus82}, \cite{Cambier} - \cite{EMMP} induced by a $T\neq 0$
background which have to be considered; they are usually referred to as
thermal radiative corrections.

\begin{figure}
\epsfysize=13.0cm
\epsfxsize=12.0cm
\centerline{\epsffile{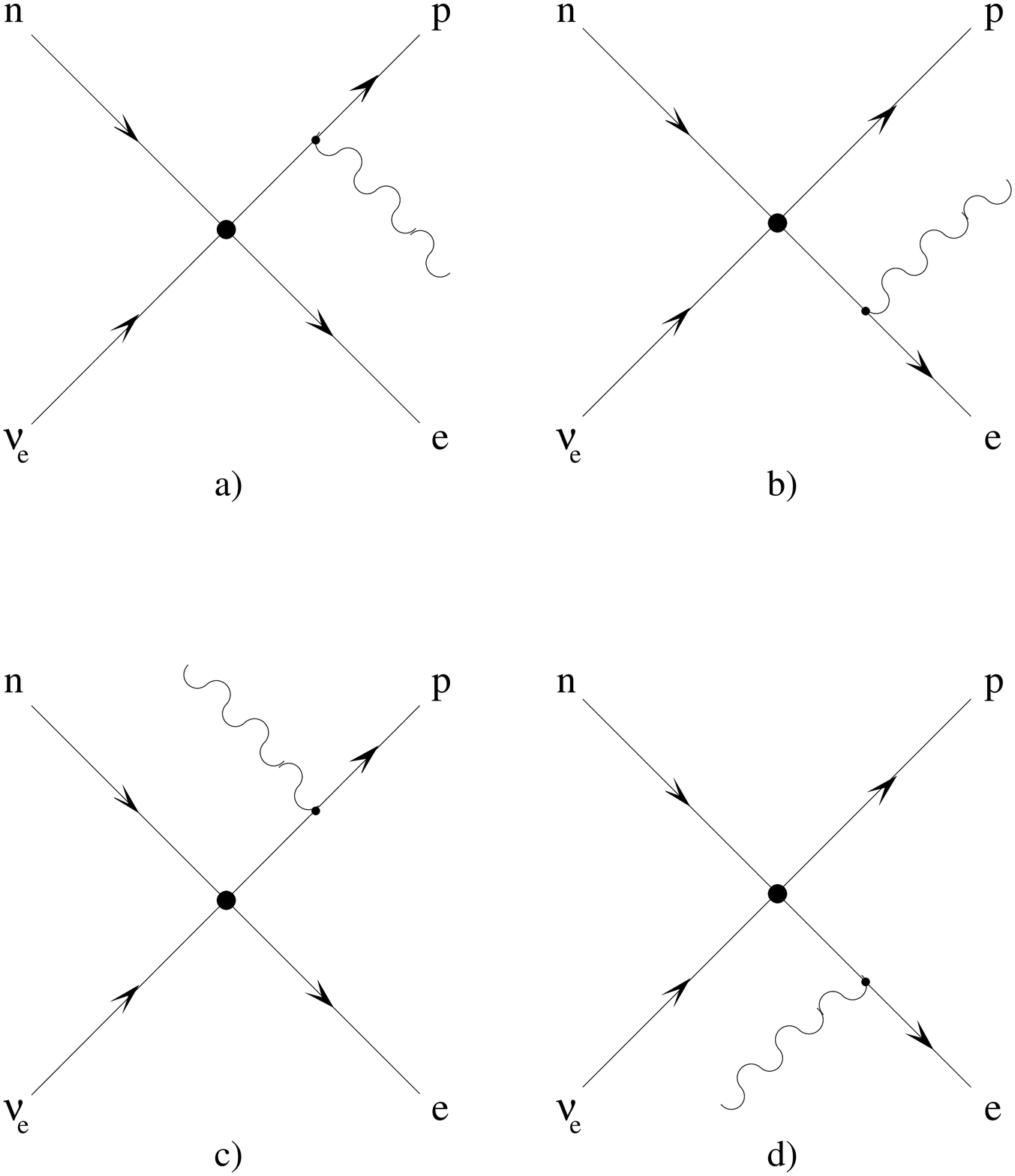}}
\caption{Photon emission and absorption diagrams
for the process in (\protect\ref{cb1}).}
\label{c3.3}
\end{figure}

Since the temperature range of interest for BBN is around $1 \, MeV$, the
most significative thermal radiative corrections to the processes in
\reac arise in the context of QED.
\begin{figure}
\epsfysize=6.0cm
\epsfxsize=12.0cm
\centerline{\epsffile{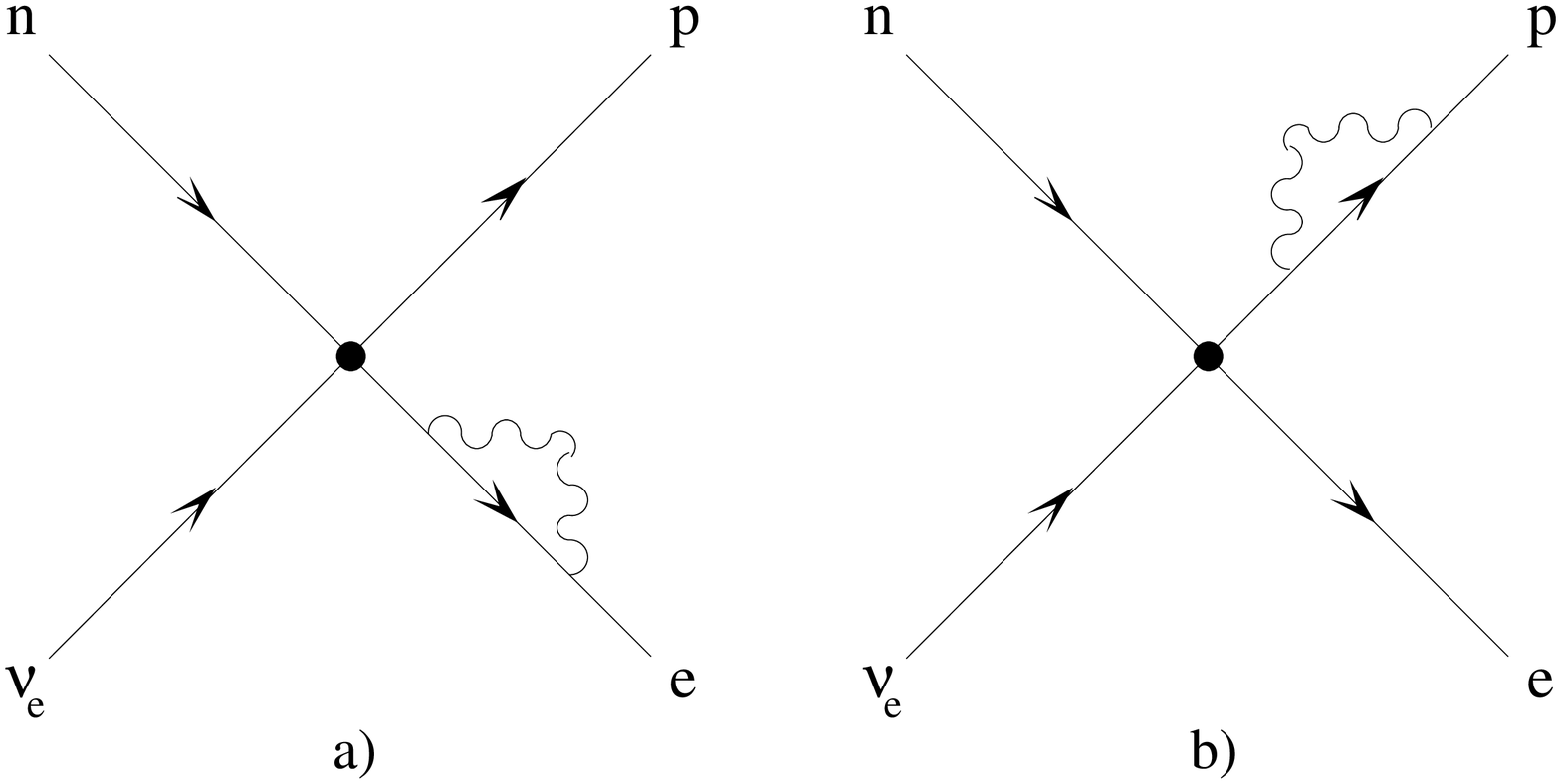}}
\caption{The Feynman diagrams for mass shift and wavefunction
renormalization for the process in (\protect\ref{cb1}).}
\label{c3.1}
\end{figure}
First of all, if the particles entering in the given process propagate in a
heat bath, they in general acquire an effective thermal mass induced by
coherent interactions with the medium (in particular with the photons in
the bath). The Feynman diagrams to be considered in this case are reported
in Figure \ref{c3.1} for the process (\ref{cb1}).\\ Moreover, wavefunction
renormalization caused by $T \neq 0$ background has to be taken into
account as well.
\begin{figure}
\epsfysize=6.0cm
\epsfxsize=6.0cm
\centerline{\epsffile{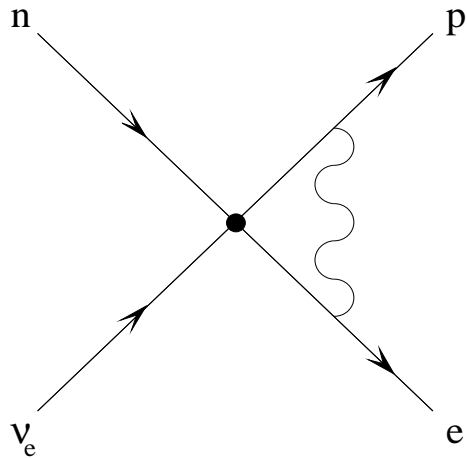}}
\caption{The Feynman diagram for vertex correction for the
process in (\protect\ref{cb1}).}
\label{c3.2}
\end{figure}
Finally, there are vertex corrections mediated by the exchange of a photon
in the bath; they are sketched in Figure \ref{c3.2}. These corrections
contain infrared divergences, which have to be eliminated via the inclusion
of corrections coming from the spontaneous and induced emission and
absorption of thermal photons, as described by the diagrams in Figure
\ref{c3.3}.

All these effects will be here considered at order $\alpha$. While we have
to evaluate the square of the diagrams in Figure \ref{c3.3}, for the mass
shift and vertex corrections only the interference of the diagrams in
Figure \ref{c3.1} and \ref{c3.2} with the Born graph in Figure \ref{c1.1}
will give contribution of order $\alpha$. Note that we have considered only
electromagnetic interactions of electrons and protons interacting with the
particles in the plasma through their electric charge; this gives indeed
the main contribution to the effect we are considering.\\ Moreover, since
these effects already contribute as corrections to the Born rates, they
will be calculated in the infinite nucleon mass approximation. In
particular we will adopt the nucleon rest frame.

Finally, the modified dispersion relation for $e^{\pm}$ and photons also call
for corrections to the thermodynamic quantities (energy density, pressure
and so on) \cite{heck} calculated in chapter \ref{bb}, which are relevant
for the BBN itself.

All these effects will be considered in this chapter; in particular in the
following four sections we focus on the thermal radiative corrections to
the Born rates for the reactions in \reac, while in the last section we
consider finite temperature QED corrections to the relevant thermodynamic
quantities. Note, however, that since we are looking only for the main
contributions coming from these radiative corrections, we can safely
assume, throughout this chapter, a negligible electron chemical potential,
which is usually, in fact, a small quantity. \\ Our calculations will be
carried out in the framework of the real time formalism (RTF) \cite{dolan}
of the finite temperature and density quantum field theory \cite{ftft1,
ftft2}, an account of which is given in appendix \ref{aftd}.

\section{Mass shift correction}
\label{c3a}

\begin{figure}
\epsfysize=8.4cm
\epsfxsize=7.0cm
\centerline{\epsffile{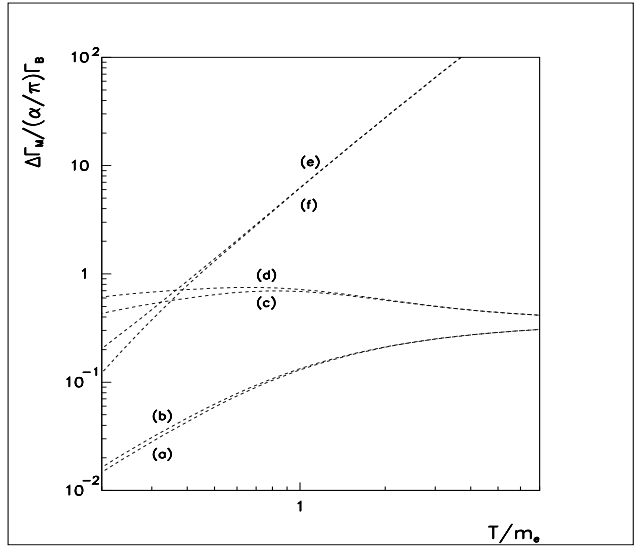}}
\caption{The ratio $\Delta \Gamma_M / (\alpha / \pi) \Gamma_B$. Dashed
lines correspond to negative values and labels refer to reactions in
\protect\reac .}
\label{fmass}
\end{figure}

Due to interactions with the particles in the plasma, an electron (or a
proton \footnote{Since electrons and protons, at a first approximation,
have equal electromagnetic interactions, the self-energy terms for these
two particles are equal. However, the mass shift correction for protons is
much less important than that for electrons because of their large (vacuum)
mass. We are only interested in leading terms and so we disregard proton
mass shift corrections.}) acquires an effective mass depending on the
temperature of the bath. Therefore, the mass parameter entering in the
expressions for the decay rates should be substituted with its effective
value, which has been explicitly calculated in the appendix \ref{aftd}. \\
We focus on the reaction $\ne \, + \,  n \, \rt \, e^- \, + \,  p$ (see
Figure \ref{c3.1}), whose Born rate is given in Eq. (\ref{cb15}); in this
expression, according to (\ref{cr24}) (see appendix \ref{aftd}), we have to
substitute
 $E_3 \, \rt \, E_3 \, + \, \mu$ and, correspondingly,
\begin{equation}\label{cr25}
  Q \; \rt \; Q \; + \; \mu ~~~,
\end{equation}
in all quantities appearing in (\ref{cb15}). With the same notation of the
appendix \ref{aftd}, the parameter $\mu$ is given by
\begin{equation}
  \mu \; = \; \frac{\alpha}{\pi} \left\{ \frac{\pi^2}{3} \frac{T^2}{E_3}
  \, + \, \frac{2}{E_3} \, \int_0^\infty \, d k \, \frac{k}{E} \, F(E)
  \; + \; \frac{m_e^2}{2 E_3 p_3} \, \int_0^\infty \, d k \frac{k}{E}
  \, F(E) \, \left( \log \, A \, - \, \log \, B \right) \right\}
\end{equation}
with $E \equiv \sqrt{k^2 + m_e^2}$, $E_3 \equiv \sqrt{p_3^2 + m_e^2}$ and
$A,B$ are reported in (\ref{cicci}). The mass shift correction we then
obtain is
\bea
  \left. \right. & \Delta \Gamma_M \!\!\!\! &
  ( \ne  \, + \,   n  \rt  e^-  \, + \,   p ) \; = \;
  \frac{G_F^2 \tre}{2 \pi^3} \, \int \, d p_3 \, p_3^2 \, \theta(Q) \,
  {\cdot} \nonumber \\
  & {\cdot} & \!\!\!\!\!\! \left\{  \left( Q + \mu \right)^2 \, F_\nu (Q  + \mu) \,
  \left( 1 -  F(E_3 + \mu) \right) \; - \;
  Q^2 \, F_\nu (Q) \, \left( 1 - F(E_3) \right) \right\}
  \;\;\;\; ~~~,  \label{cr26}
\eea
Expanding in the variable $\mu$, at first order in $\alpha$ we finally get
\bea
  \Delta \Gamma_M ( \ne  \,  n  \rt  e^-  \,  p ) \; = \;
  \frac{G_F^2 \tre}{2 \pi^3} \!\!\!\! & {\cdot} & \!\!\!\! \int \, d p_3 \,
  p_3^2 \, Q^2 \, \mu \,  \theta(Q) \, F_\nu (Q) \, \left( 1 \, - \,
  F(E_3) \right) \, {\cdot} \nonumber \\
  & {\cdot} & \left( \frac{2}{Q} \; + \; \frac{F(E_3)}{T} \; - \;
  \frac{F_\nu(-Q)}{T_\nu} \right)  \label{cr27}
\eea
Similar expressions for the other reactions in \reac can be easily found by
using the substitutions of Table \ref{tc1.1} and replacing the factor
\begin{equation}
   \left( \frac{2}{Q} \; + \; \frac{F(E_3)}{T} \; - \;
  \frac{F_\nu(-Q)}{T_\nu} \right) ~~~,
\end{equation}
in the above expression as follows. This factor applies to the reaction
$\neb + p \rt e^+ + n$ as well. For $e^- + p \, \rt \, \ne + n$ and $e^+ +n
\, \rt \, \neb + p$, we have instead
\begin{equation}
\left(\frac{2}{Q} - \frac{1-F(E_3)}{T} +
\frac{F_\nu(Q)}{T_\nu}\right) ~~~,
\end{equation}
while for $n \, \rt \, e^- + \neb + p$
\begin{equation}
\left(- \, \frac{2}{Q} + \frac{F(E_3)}{T} -
\frac{F_\nu(Q)}{T_\nu}\right) ~~~,
\end{equation}
and $e^- + \neb + p \rt \, n$
\begin{equation}
\left(- \, \frac{2}{Q} - \frac{1-F(E_3)}{T} +
\frac{1-F_\nu(Q)}{T_\nu}\right) ~~~.
\end{equation}
In Figure \ref{fmass} the ratio $\Delta \Gamma_M / (\alpha / \pi) \Gamma_B$
is reported as a function of the (photon) temperature $T$ for the six
reactions in \reac. Note that mass shift corrections are negative for all
these reactions and decreasing (in absolute value) with decreasing $T$ for
(a), (b), (e), (f) while they are almost constant for (c), (d) in the
temperature range relevant for nucleosynthesis.

\section{Wavefunction renormalization correction}
\label{c3b}

\begin{figure}
\epsfysize=8.4cm
\epsfxsize=7.0cm
\centerline{\epsffile{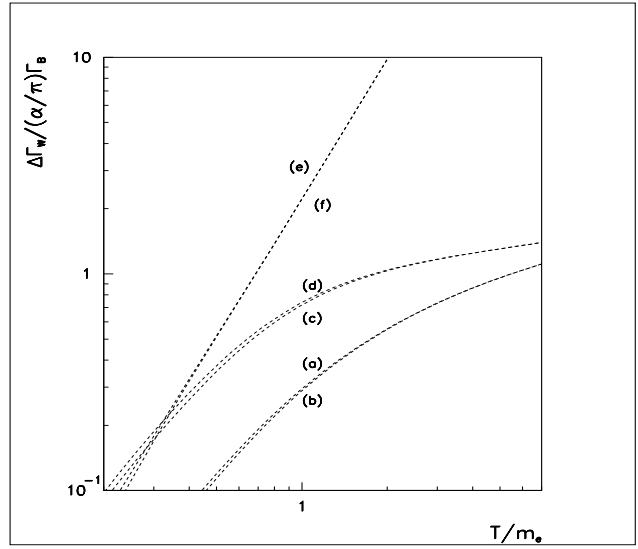}}
\caption{The ratio $\Delta \Gamma_W / (\alpha / \pi) \Gamma_B$. Dashed
lines correspond to negative values and labels refer to reactions in
\protect\reac .}
\label{fwavefun}
\end{figure}

The same diagrams in Figure \ref{c3.1} leading to mass shift corrections
also contribute to wavefunction renormalization corrections.  For
definiteness we consider in detail the reaction (\ref{cb1}). As studied in
appendix \ref{aftd}, the amplitude is again given by (\ref{cb10}), but when
we compute $\mod$ we have to replace the projectors $\Lambda^{\pm}_0$ with the
renormalized ones, according to (\ref{cr28}). With the same notation of
chapter \ref{c1} we obtain the following expression for the correction to
the decay rate:
\begin{equation}\label{cr36}
 \Delta \Gamma_W ( \ne  +  n  \rt  e^-  +  p ) \, = \,
 \frac{G_F^2 \tre}{2 \pi^3} \, \int \, d p_3 \, p_3^2 \, Q^2 \, \theta(Q)
 \lambda \, F_\nu (Q) \left( 1 - F(p_3) \right)
\end{equation}
where the parameter $\lambda$ is reported in the appendix \ref{aftd}. Note
that, in the non-relativistic limit for the nucleons, which we here adopt,
the second term in (\ref{cr34}) contribute to the correction $\Delta
\Gamma_W$ with a term which vanishes after the angular integration is
performed. \\ From relations (\ref{cr4})-(\ref{cr6}) it is easy to show
that
\begin{equation}\label{cr37}
  \lambda \; = \; \frac{1}{\om} \left\{ - \, \left( \hat{T}^B_u \, + \,
  \hat{T}^F_u \right) \; + \; \left( \hat{T}^{B \prime}_p \, + \,
  \hat{T}^{F \prime}_p \right) \; + \; m_e \, \left( \cpc_B \, + \,
  \cpc_F \right) \right\}
\end{equation}
The quantities $\hat{T}^B_u$, $\hat{T}^F_u$ can be immediately read off by
Eqs. (\ref{cr10}), (\ref{cr13}) by substituting $p_0 = E_3$. Furthermore
$T^{B \prime}_p$, $T^{F \prime}_p$, $c^\prime_B$, $c^\prime_F$ may be
calculated by differentiating Eqs. (\ref{cr9}), (\ref{cr12}), (\ref{cr11}),
(\ref{cr14}) with respect to $p_0$ prior the integration in $d x$ is
performed. After some algebra, the final result is
\bea
  \lambda &=& \frac{2 \alpha}{\pi} \int d k \, \frac{B(k)}{k} \; - \;
  \frac{\alpha \, \pi \, T^2}{6 \, E_3 \, p_3} \, \ln \,
  \frac{E_3 \, + \, p_3}{E_3 \, - \, p_3} \; + \nonumber \\
  &-& \frac{\alpha}{2 \pi \, E_3 \, p_5} \int d k \, \frac{k}{E} \, F(E)
  \, \left( \left(E_3 \, + \, E \right) \, \ln \, A \; - \;
  \left(E_3 \, - \, E \right) \, \ln \, B \right) \; + \nonumber \\
  &-& \frac{2 \alpha}{\pi} \int d k \, \frac{k^2}{E} \,
  \frac{1}{p_3^2 - k^2} \, F(E)  \label{cr38}
\eea
Note that the first term in (\ref{cr38}) is infrared divergent. This term,
together with another similar divergent one coming from considering vertex
corrections (see the following section), is cancelled by an opposite term
provided by the photon and emission rate discussed in section \ref{c3d}.
The collinear divergence still present in (\ref{cr30}) for $p_3 = k$,
together with a similar contribution from the vertex correction is also
compensated by  bremsstr\"{a}hlung diagrams. \\ The extension of (\ref{cr36})
to the other processes in \reac can be straightforwardly obtained by using
Table \ref{tc1.1}. \\ The results for $\Delta \Gamma_W / (\alpha / \pi)
\Gamma_B$ are plotted in Figure \ref{fwavefun}. Also these corrections are
negative for all the six reactions in \reac, and are monotonically
decreasing with decreasing $T$.

\section{Vertex correction}
\label{c3c}

\begin{figure}
\epsfysize=8.4cm
\epsfxsize=7.0cm
\centerline{\epsffile{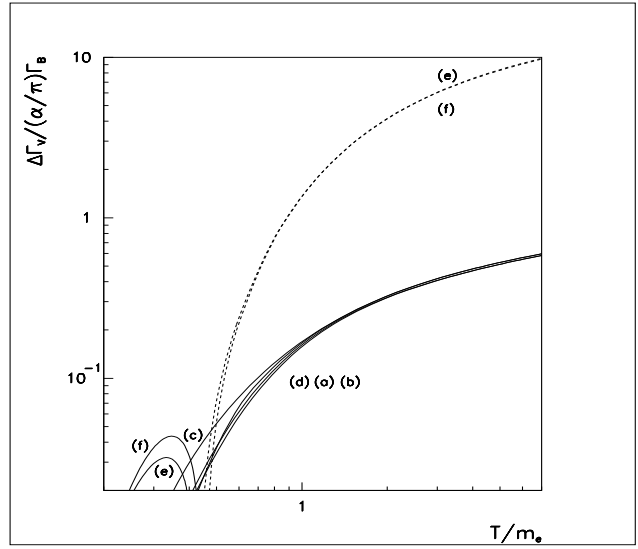}}
\caption{The ratio $\Delta \Gamma_V / (\alpha / \pi) \Gamma_B$. Dashed
lines correspond to negative values and labels refer to reactions in
\protect\reac .}
\label{fvert}
\end{figure}

The vertex correction to the Born rate for the process $\nu_e \, + \, n \,
\rt \, e^- \, + \, p$ is given by the diagram in Figure \ref{c3.2}, whose
amplitude is
\bea
M_V & = & i \, \frac{G_F}{\sqrt{2}} \, \int \, \frac{d^4 k}{(2 \pi)^4}
\, \ov{u}_p(p_4) \, (i e \gamma^\rho) \, S_p(p_4+k) \, \gamma^\mu
(\cv - \ca \gamma_5) \, u_n(p_2) \, {\times} \nonumber \\ & {\times} & i D_{\rho
\sigma}(k)
\, \ov{u}_e(p_3) \, (i e \gamma^\sigma) \, S_e(p_3-k) \, \gamma_\mu
(1 - \gamma_5) \, u_{\nu_e}(p_1) ~~~, \label{v1}
\eea
where $S_F$ ($F=e,p$) and $D_{\rho \sigma}$ are the real time formalism
fermion and photon propagators respectively, reported in the Appendix
\ref{aftd}. After some algebra, and using the Dirac equation for the
proton, $M_V$ can be cast in the form
\begin{equation}
M_V \; = \; \frac{i \, G_F \, e^2}{\sqrt{2} \, (2 \pi)^3} \, \left\{ I_1 \,
A_\mu \, B^\mu \; + \; \frac{1}{2} \, I_2^{\rho \lambda} \, A^\mu \,
B_{\rho \lambda \mu} \; + \;
\frac{1}{2} \, I_3^{\rho \lambda} \, A_{\rho \lambda \mu} \, B^\mu
\; + \; \frac{1}{4} \, I_{4 ~~ \sigma}^{~ \lambda} \,
A^{\rho \sigma \mu} \, B_{\rho \lambda \mu} \right\}
\label{v2}
\end{equation}
where
\bea
A^\mu & = & \ov{u}_p \, \gamma^\mu (\cv - \ca \gamma_5) \, u_n
\nonumber ~~~, \\
A^{\rho \sigma \mu} & = & \ov{u}_p \, \gamma^\rho \, \gamma^\sigma \,
\gamma^\mu (\cv - \ca \gamma_5) \, u_n
\nonumber ~~~, \\
B^\mu & = & \ov{u}_e \, \gamma^\mu (1 -  \gamma_5) \, u_{\nu_e}
\nonumber ~~~, \\
B^{\rho \sigma \mu} & = & \ov{u}_e \, \gamma^\rho \, \gamma^\sigma \,
\gamma^\mu (1 -  \gamma_5) \, u_{\nu_e}
\nonumber ~~~,
\eea
and
\bea
I_1 & = & - p_4{\cdot}p_3 \, \int d^4k \left( \frac{\delta(k^2)}{p_4{\cdot}k \, p_3{\cdot}k}
B(k_0)
\, + \, \frac{4 \delta(k^2 - m_e^2) \, F(k_0)}{(k + p_3)^2 \, (
(p_4+p_3+k)^2 - M^2)} \right) ~~~, \nonumber \\ I_2^{\rho \lambda} & = & 4
p_4^\rho
\, \int d^4k \,
\frac{\delta(k^2 - m_e^2) \, F(k_0) \, (p_3+k)^\lambda}{(k + p_3)^2 \,
((p_4+p_3+k)^2 - M^2)} ~~~, \nonumber \\ I_3^{\rho \lambda} & = & - 4 p_3
{\prime \rho} \, \int d^4k \,
\frac{\delta(k^2 - m_e^2) \, F(k_0) \, (p_3+k)^\lambda}{(k + p_3)^2 \,
((p_4+p_3+k)^2 - M^2)} ~~~, \nonumber \\ I_4^{\lambda \sigma} & = &  \int
d^4k \left( \frac{\delta(k^2) \, k^\lambda \, k^\sigma}{p_4{\cdot}k \, p_3{\cdot}k}
B(k_0) \, + \, \frac{4 \delta(k^2 - m_e^2) \, F(k_0) \, (p_3+k)^\lambda \,
(p_3+k)^\sigma }{(k + p_3)^2 \, ((p_4+p_3+k)^2 - M^2)} \right) \nonumber
~~~.
\eea
The order $\alpha$ vertex correction to the Born rate is provided by the
interference term of the diagram in Figure \ref{c3.2} with the tree level
diagram in Figure \ref{c1.1},
\begin{equation}
  M_V \, M_B^\ast \; + \; M_V^\ast \, M_B \nonumber ~~~,
\end{equation}
where $M_B$ is the Born expression (\ref{cb10})
\begin{equation}
  M_B \; =¤ \; i \, \frac{G_F}{\sqrt{2}} \, A_\mu \, B^\mu ~~~.
\end{equation}
After some lengthy calculations, the final result for the vertex correction
we obtain is
\bea
\Delta \Gamma_V (\nu_e + n \rightarrow e^- + p)= \frac{G_F^2 (\tre)}
{2 \pi^3} \frac{\alpha}{\pi} \int_0^\infty d p_3 ~ p_3
\int_0^\infty ~ d k ~ k ~Q^2~\theta(Q) \nonumber \\ {\times}
F_\nu(Q)~ \left[1-F(E_3) \right] ~\frac{F(E)}{E}
\left\{\frac{E}{E_3+E}\log A + \frac{E}{E_3-E}\log B -\frac{2
k p_3}{p_3^2 - k^2}\right\}.\label{v3} ~~~.
\eea
Note that we have only reported the non infrared divergent part (as for
$\Delta \Gamma_W$, the infrared one comes from the Bose terms only, and
will be cancelled by the  bremsstr\"{a}hlung contribution) \footnote{Note that
the result for the vertex correction quoted in \cite{Cambier} as Eq. (11)
has a missing factor $1/E$.}. \\ The expression in (\ref{v3}) is extended
to the other processes in \reac by using the Table \ref{tc1.1}, and the
plot for these corrections is reported in Figure \ref{fvert}. Note that for
the scattering reactions (a), (b), (c), (d) the vertex corrections are
positive in the relevant temperature range, while for (e), (f) the dominant
contribution is negative.

\section{Photon emission and absorption}
\label{c3d}

\begin{figure}
\epsfysize=8.4cm
\epsfxsize=7.0cm
\centerline{\epsffile{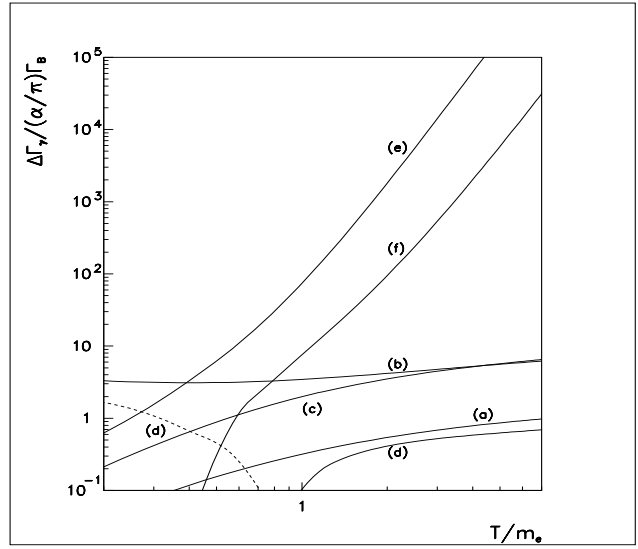}}
\caption{The ratio $\Delta \Gamma_\gamma / (\alpha / \pi) \Gamma_B$.
Dashed lines correspond to negative values and labels refer to reactions in
\protect\reac .}
\label{fbrem}
\end{figure}

In the primordial plasma, in which an electromagnetic component is present,
spontaneous and induced photon emission as well as photon absorption take
place, so that when considering a given process such as the one
corresponding to the diagram in Figure \ref{c1.1}, the  bremsstr\"{a}hlung
processes depicted by the diagrams in Figures \ref{c3.3} must be included
as well. These also cancel the infrared divergences due to the radiative
diagrams of Figure \ref{c3.1}, \ref{c3.2} corresponding to wavefunction and
vertex renormalizations. \\ The matrix element for the emission processes
in Figures \ref{c3.3} is the following:
\bea
M_\gamma^e & = & \frac{i \, G_F}{\sqrt{2}} \, \left\{ \ov{u}_p(p_4) \,
\gamma^\mu ( \cv - \ca \gamma_5) \, u_n(p_2) \, \ov{u}_e(p_3) \,
(i e \gamma^\alpha) \, S_e(p_3+k) \, \gamma_\mu (1 - \gamma_5) \,
u_{\nu_e}(p_1) \; + \right. \nonumber \\ & - & \left.
\ov{u}_p(p_4) \, (i e \gamma^\alpha)
\, S_p(p_4+k) \, \gamma^\mu (\cv - \ca \gamma_5) \, u_n(p_2) \,
\ov{u}_e(p_3) \, \gamma^\mu (1 - \gamma_5) \, u_{\nu_e}(p_1) \right\}
\epsilon_\lambda^\alpha(k) \nonumber
%\label{v4}
\eea
while a similar expression, $M_\gamma^a$, holds for the absorption
processes. In order to simplify the compuatation of the spin summed squared
matrix element, one can subtract from the beginning the infrared divergent
terms coming from the propagators. They are all proportional to the Born
$\mod$ and will cancel the divergences in $\Delta \Gamma_W$ and $\Delta
\Gamma_V$. Note also that, for kinematical reasons, only the $T=0$ part in
the propagators contribute to the amplitude. \\ After some extensive
calculations, we obtain the following finite contribution to the rate
$\Delta \Gamma_\gamma \, = \, \Delta \Gamma_\gamma^e \, + \, \Delta
\Gamma_\gamma^a$
\bea
\Delta  \Gamma_\gamma
(\nu_e + n \rightarrow e^- + p) = \frac{G_F^2 (\tre)} {2 \pi^3}
\frac{\alpha}{\pi} \int_0^\infty d p_3~\int_0^\infty
d k ~ \frac{p_3^2} {E_3}~B(k) \nonumber \\ {\times}~
\left[ 1-F(E_3) \right]~ \left\{ - \left[ \frac{2 E_3}{k}-
\frac{E_3^2}{k p_3}
\log\left(\frac{E_3 + p_3}{E_3 - p_3} \right)
\right]\left[ \widetilde{Q}^2_+ + \widetilde{Q}^2_- - 2
\widetilde{Q}^2\right] \right. \nonumber \\
- \left. \left[2\xi -\frac{E_3}{p_3} \log\left(\frac{E_3 +
p_3}{E_3 - p_3} \right) \right] \left[
\widetilde{Q}^2_+ - \widetilde{Q}^2_- \right] +\frac{k}{2
p_3} \log\left(\frac{E_3 + p_3}{E_3 - p_3}
\right) \left[\widetilde{Q}^2_+ + \widetilde{Q}^2_- \right]\right\},
\nonumber \\ \label{v5}
\eea
where $\widetilde{Q}^2_{\pm} \equiv (Q{\pm}k)^2 F_\nu(Q {\pm}k)~\theta(Q{\pm}k)$,
$\widetilde{Q}^2 \equiv Q^2 F_\nu(Q)~\theta(Q)$ and $S=1$ \footnote{Also
here we must note that our result corrects Eq. (13) in Ref. \cite{Cambier}
for the photon emission and absorption correction previously calculated.}.
\\ For the other processes in \reac we have to use the Table \ref{tc1.1}
and set $\xi=1$ for $\ne + n \leftrightarrow e^- + p$ and $n
\leftrightarrow e^- + \neb + p$, and $\xi=0$ for $ e^+ + n \leftrightarrow
\neb + p$. \\ The results for $\Delta \Gamma_\gamma / (\alpha / \pi)
\Gamma_B$ are reported in Figure \ref{fbrem}. It is easy to realize, from
this figure, that photon emission and absorption contribution to the
thermal radiative corrections for neutron decay and the inverse process is
the dominant one (and positive) in the temperature range relevant for
nucleosynthesis. This is simply understood since the inclusion of the
process $\gamma \, + \, n \;
\rt \; e^- \, + \, \neb \, + \,  p$ greatly increases the neutron decay
rates, otherwise strongly suppressed by phase space.

\section{Corrections to the equation of state}
\label{c3e}

The temperature dependent effective mass of electrons and positrons in the
primordial plasma affects the calculations of the thermodynamic quantities
like particle number density, energy density, pressure and entropy, which
directly enter in the evolution equations for the light element abundances
(see section \ref{pnc}), as well as in the weak decay rates as discussed in
section \ref{c3a}. In addition, also the induced effective photon mass has
to be taken into account. The modified dispersion relations of $e^{\pm}$,
$\gamma$ have been calculated in appendix \ref{aftd}, and they must be used
(instead of the vacuum dispersion relations) in the evaluations of the
quantities in (\ref{59})-(\ref{61}). \\ The relevant quantities to be
considered are the $e^{\pm}$, $\gamma$ energy densities, entering in Eqs.
(\ref{n22}) and (\ref{n36}), and pressure, which enter directly in the Eq.
(\ref{n36}) for entropy density. Corrections to electron and positron
number densities are also present in general, but, since in the electric
charge conservation equation (\ref{n24}) only their difference is relevant,
this results into a correction to the electron chemical potential. This is,
generally, a small quantity and the corrections to it can be usually
neglected.

\begin{figure}
\epsfysize=8.4cm
\epsfxsize=7.0cm
\centerline{\epsffile{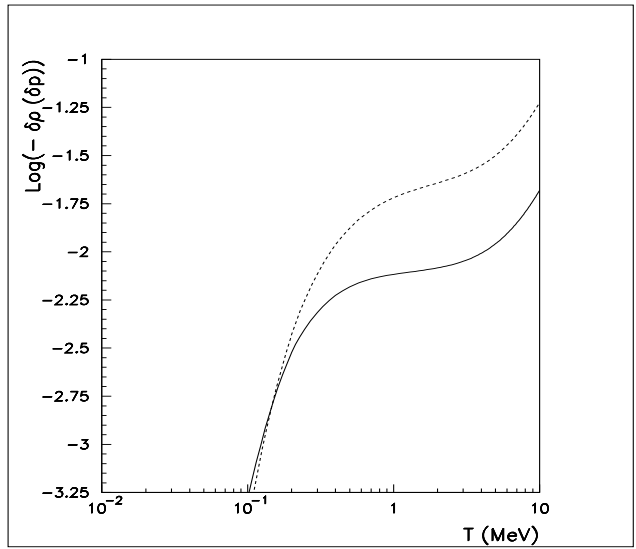}}
\caption{Radiative corrections to the energy density (solid line)
and pressure (dashed line) of the electromagnetic component of the
primordial plasma.}
\label{sta}
\end{figure}

The order $\alpha$ corrections to the photon energy density and pressure
are easily calculated by inserting Eq. (\ref{f21}) in the Eqs. (\ref{60}),
(\ref{61}) specialized to photons; with simple manipulations we obtain
\bea
\delta \rr_\gamma & \simeq & T^2 \frac{\alpha}{\pi} \, \int_0^\infty d p \,
\frac{p^2}{E} \, F_e(E) ~~~, \label{v6} \\
\delta \p_\gamma & \simeq & - \, \frac{1}{3} \, \delta \rr_\gamma
\label{v7} ~~~.
\eea
The expressions for order $\alpha$ corrections to $e^{\pm}$ energy density and
pressure are deduced by substituting Eq. (\ref{cr24}) into (\ref{60}),
(\ref{61}) specialized to $e^{\pm}$; we found ($\rr_e = \rr_{e^-} +
\rr_{e^+}$, $\p_e = \p_{e^-} + \p_{e^+}$)
\bea
\delta \rr_e & \simeq & \frac{2}{\pi^2} \, \int_0^\infty  d p \; p^2 \,
\mu \, \left( 1 \, - \, \frac{E}{T} \, F_e(-E)\right) \, F_e(E) ~~~,
\label{v8} \\
\delta \p_e & \simeq & \frac{2}{3 \pi^2} \, \int_0^\infty  d p \,
\frac{p^4}{E^2} \, \mu \, \left( 1 \, + \, \frac{E}{T} \, F_e(-E)\right)
\, F_e(E) \label{v9}
\eea
$(E = \sqrt{p^2 + m_e^2})$. \\ The corrections to the energy density and
pressure of the electromagnetic plasma ($\delta \rr = (30/\pi^2 T^4)(\delta
r_\gamma + \delta \rr_e)$, $(90/\pi^2 T^4) (\delta \p = \delta p_\gamma +
\delta \p_e)$) are plotted in Figure \ref{sta}. As we can see, its
contribution is of the same order of magnitude of the other thermal
radiative corrections.

As briefly mentioned above, corrections to the energy densities directly
influence the expansion rate of the Universe (see Eq. (\ref{n22})) and
eventually the primordial abundances through a change in the freeze out
temperature. Moreover corrections to $\rr$ and $\p$ also translate into
corrections to the entropy density according to relation (\ref{74}). On one
side, this traduces into a modification of the time-temperature relation
or, to be more specific, of the evolution equation for $T$ (\ref{n36}). On
the other side, the entropy of the $e^{\pm}$ plasma is transferred to photons
when $e^{\pm}$ pairs disappear, while neutrinos do not benefit of this entropy
releasing since at this epoch they are decoupled. Hence, this changes the
neutrino-to-photon temperature ratio, which directly affects the weak
rates. We have also calculated this effect, but we found that the
modification of $T_\nu/T$ is quite negligible in the range relevant for
BBN, much smaller than the precision of our goal.

\chapter{Calculations of Big Bang Nucleosynthesis. Results}
\label{sum}

In this thesis we have studied in details the fundamental processes
underlying primordial nucleosynthesis and performed a complete analysis of
the physical approximations that may be used as well as of corrections
which should be included to reach the 1\% reliability of the estimates. The
relevance of this study and, in particular, of the precise determination of
the primordial light element abundances, has been already pointed out in
the Introduction. In this final chapter we summarize the main results
obtained, give the accurate prediction for $^4He$ and outline the basic
ingredients of forthcoming numerical code computing light element
abundances, which is currently in progress.

\section{Primordial nucleosynthesis: cosmology and particle physics inputs}
\label{suma}

As discussed in chapter \ref{pn}, big bang nucleosynthesis is the result of
two competing factors: nuclear and subnuclear processes between the
particles in the primordial plasma, whose effect is to keep the species in
thermal equilibrium, and the expansion of the Universe, which tends to
hinder this equilibrium. It is then clear that the physics involved runs
over cosmology (and, in particular, thermodynamics in the expanding
Universe) and particle physics. Here we mainly summarize the basic inputs
for studying primordial nucleosynthesis as well as the physical assumptions
and approximations we used.

Standard cosmology is based on the observed homogeneity and isotropy of the
Universe. These directly lead to the cosmological equations (\ref{26}),
(\ref{27}) in the approximation of the Universe as a perfect fluid, to
which must be added the equation of state (\ref{28}) for the matter fields
present in the Universe must be added. Primordial nucleosynthesis takes
place when the curvature term $k/R^2$ in (\ref{26}) is completely
negligible, so that it is usually dropped out from the subsequent
considerations. \\ From the cosmological equations, the energy conservation
law (\ref{25}) follows, as well as entropy conservation (\ref{73}) for
species in kinetic equilibrium. The total energy density and pressure, for
species in equilibrium, are given by (\ref{79}), (\ref{80}). From these it
follows that non relativistic matter (primarily baryons) gives a small
contribution to the evolution in the RD era (in which BBN takes place),
since its energy density is exponentially suppressed. This approximation
will be used in several occasions. \\ The approach and the departure from
thermal equilibrium is fully described by the Boltzmann transport
equations, which involve the interaction rates of the particles present in
the plasma. It is found that when a ultrarelativistic or non relativistic
specie is completely decoupled from the heat bath, its distribution has the
same form as that of an equilibrium one, but with a temperature parameter
scaling, as the Universe expands, according to Eqs. (\ref{91}) or
(\ref{94}). For such decoupled species, entropy is then separately
conserved. \\ Further ingredients come from particle physics; let us focus
on the nucleosynthesis era, when the relevant particles present in the
plasma are photons, neutrinos, electrons and positrons, baryons. Photons
and $e^{\pm}$ pairs form the background electromagnetic plasma, and their
equilibrium is established by electromagnetic interactions. Baryons are
maintained in kinetic equilibrium by electromagnetic and strong nuclear
interactions, while neutrinos by weak interaction processes only. Chemical
equilibrium in the electromagnetic component of the plasma is established
by the annihilation reactions $e^+ e^- \rt \gamma \,
\gamma$, while the weak reactions in \reac are responsible for the nuclear
chemical equilibrium. Departure from chemical equilibrium happens as
follows. The electron neutrino decoupling temperature is calculated to be
$T_D^{\ne}
\simeq 2.3 \, MeV$, while for $\nm , \nt$ we have $T_D^{\nm , \nt} \simeq
3.5 \, MeV$. In nucleosynthesis calculations it is assumed that neutrinos
are completely decoupled from the background plasma before $e^+ e^-$
annihilations. Neutrino decoupling, however, occurs very close to $e^+ e^-$
annihilation, so that some residual interaction with the thermal plasma can
cause the neutrinos to be slightly heated by the resultant entropy release.
Careful studies have been conducted on this subject \cite{resid, Dicus82},
demonstrating that neutrino decoupling is {\it not} an instantaneous
process. Hence, the spectrum of the decoupled neutrinos deviates slightly
from the Fermi-Dirac form, causing the effective neutrino temperature to
increase with momentum. However this increase accounts for only 0.7 \%
(even at relatively high momenta), justifying the usual approximation of
instantaneous decoupling. Furthermore, the electron chemical potential,
which is constrained by electric charge conservation, as well as the baryon
chemical potential, is usually supposed to be a small quantity. This
assumption is practically justified a posteriori. Instead, the neutrino
chemical potentials are unconstrained, but they are set to zero in the
standard nucleosynthesis scenario. Note that a non zero chemical potential
for electron neutrinos can alter neutron-proton equilibrium, as well as
increase the expansion rate of the Universe, while chemical potentials for
other neutrino types can only speed up the expansion. In general, the
lowering of the $n/p$ ratio at freeze out may be compensated for by the net
speed up of the expansion rate, although exotic models exist in which large
lepton numbers can be generated (see \cite{Sarkrev} and references therein)
\footnote{These models are, however, constrained by BBN itself, since the
introduction of non vanishing chemical potentials for neutrinos can
substantially modify abundance predictions, destroying the agreement
between theory and observations. In practice, allowed neutrino degeneracy
cannot significantly altet the standard BBN model.}.

There are no further assumptions and approximations regarding
thermodynamics.

Electromagnetic and weak interaction reactions between elementary particles
are {\it calculated} in the framework of the electroweak Standard Model. In
particular, the crucial (for nucleosynthesis) weak processes in \reac have
been here evaluated as follows. \\ Born approximation is used as the
reference for the reaction rates: it is based on tree level calculations of
the probability amplitude in the infinite nucleon mass limit. We have then
relaxed these assumptions. Firstly we reported the relevant zero
temperature QED radiative and Coulomb corrections to the weak rates and
then considered finite nucleon mass effects, namely weak magnetism, phase
space modification and kinematical corrections due to the thermal motion of
the initial nucleon in the comoving reference frame. The results for these
corrections are plotted in Figures \ref{fzerot} , \ref{fkin}. Furthermore,
we have also studied in details thermal radiative corrections, explicitly
induced by the finite density of the plasma. In particular, we have
calculated electron effective mass and wavefunction renormalization
effects, vertex corrections and thermal photon emission and absorption
processes, and the obtained results are reported in Figures \ref{fmass},
\ref{fwavefun}, \ref{fvert}, \ref{fbrem}. With reference to the above
mentioned Figures, the results for each reaction channel can be summarized
as follows.

\begin{figure}
\epsfysize=8.4cm
\epsfxsize=7.0cm
\centerline{\epsffile{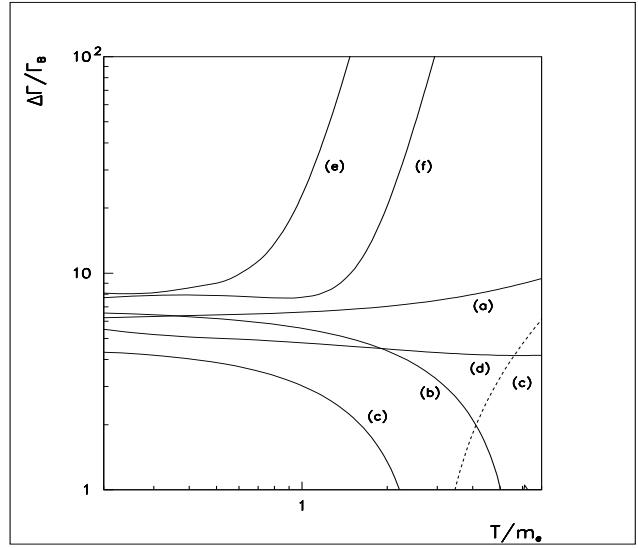}}
\caption{The total relative correction to the Born rates, in percent, are
shown for the six processes in \protect\reac .}
\label{ftchan}
\end{figure}

\begin{figure}
\epsfysize=8.4cm
\epsfxsize=7.0cm
\centerline{\epsffile{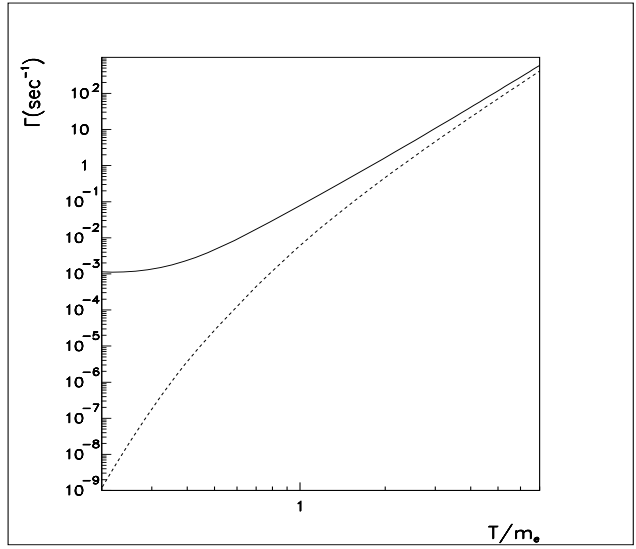}}
\caption{The total rates $\Gamma(n \rightarrow p)$  (solid line) and
$\Gamma(p \rightarrow n)$  (dashed line), including all radiative, finite
mass and thermal corrections.}
\label{ftotrat}
\end{figure}

\begin{figure}
\epsfysize=8.4cm
\epsfxsize=7.0cm
\centerline{\epsffile{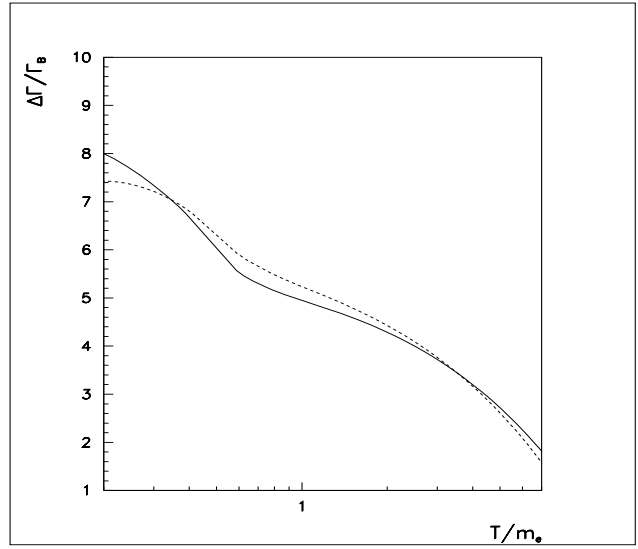}}
\caption{The total relative corrections $\Delta \Gamma /\Gamma_B$ for the
$n \rightarrow p$ (solid line) and $p \rightarrow n$ (dashed line)
processes, expressed in percent.}
\label{ftotcor}
\end{figure}

\begin{itemize}
\item[(a)] $\ne + n \rightarrow e^- + p$

For the crucial BBN temperature range, $0.1~MeV \leq T \leq 3.5 ~MeV$, the
two main corrections to the Born rate come from zero temperature radiative
and kinetic terms. The contribution $\Delta \Gamma_R$ is weakly depending
on $T$ and represents the dominant term, though for large temperature the
kinetic correction $\Delta \Gamma_K$ starts contributing significantly. The
two combined contributions correct the Born rate for a factor $6
\div 9 \%$, whilst thermal radiative ones are
of the order of $1 \%$.
\item[(b)] $e^- + p \rightarrow \ne + n $

For this channel radiative corrections are dominant at low temperature,
while the kinetic ones give a quite relevant effect in whole BBN
temperature range and correct the Born rate for a factor varying from $1
\%$ at low $T$ up to $3 \%$ for $T = 3 \div 4 ~MeV$. The radiative
contribution is quite rapidly decreasing with temperature, reaching large
negative values. This leads to a partial cancellation between $\Delta
\Gamma_R$ and $\Delta \Gamma_K$.
Thermal corrections are dominated by bremsstr\"{a}hlung contribution $\Delta
\Gamma_\gamma$ and can be as large as $2 \%$ of $\Gamma_B$ for large
temperature.
\item[(c)] $e^+ + n \rightarrow \neb + p $

For this process the radiative corrections have a behaviour quite similar
to channel $(b)$, though they are even more rapidly decreasing with
temperature. This again leads to a partial cancellation between $\Delta
\Gamma_R$ and the positive monotonically increasing $\Delta \Gamma_K$.
Thermal corrections are again mainly provided by photon emission/absorption
and monotonically increase with temperature up to a factor $2 \%$ of the
corresponding Born rate.
\item[(d)] $\neb + p \rightarrow e^+ + n $

The radiative and kinetic corrections sum up to a factor of about $5 \%$ of
the Born rate. Actually the opposite behaviour of $\Delta \Gamma_R$ and
$\Delta \Gamma_K$ conspires to give an almost constant total correction in
the whole interesting temperature range. Thermal effects are quite small,
contributing for less than $0.5 \%$.

\item[(e)] $n \rightarrow e^- + \neb + p$

For the neutron decay the radiative corrections are practically constant
and give the leading effect for small temperature, $T \leq m_e$. For larger
$T$ the thermal photon emission/absorption processes rapidly become
dominating over all other correction terms with a relative ratio to the
Born rate as large as $10^4$. However, this large correction is weakly
contributing to total $n \rightarrow p$ rate, since in the temperature
range $T > m_e$ the scattering processes $(a)$-$(d)$ are largely dominant
over decay and inverse decay.
\item[(f)] $e^- + \neb + p \rightarrow n$

Same considerations of the direct process $(e)$ hold for inverse neutron
decay. As for the direct channel, the kinetic contribution is negative down
to temperatures of the order $T \sim 0.2 ~MeV$. The contribution of
$\left|\Delta \Gamma_K\right|$ is however negligible  for both processes,
smaller than $1 \%$.
\end{itemize}
All results are summarized in Figure \ref{ftchan}, where we have shown the
total relative corrections in percent. For the processes $(a)$ and $(d)$
the correction is almost constant over the entire considered range for $T$,
and of the order of $6 \div 10 \%$ and $5 \div 6 \%$, respectively. The
positive kinetic contribution soften the deep decreasing of the radiative
terms for channels $(b)$ and $(c)$. Finally the large effect of thermal
bremsstr\"{a}hlung for neutron decay and inverse process $(e)$ and $(f)$ is
particularly evident.
\begin{figure}
\epsfysize=8.4cm
\epsfxsize=7.0cm
\centerline{\epsffile{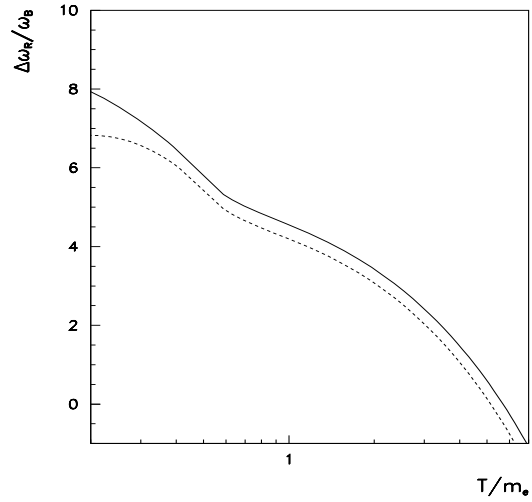}}
\caption{The zero temperature radiative corrections, for $n \rightarrow p$ 
(solid line) and $p \rightarrow n$
(dashed line) processes. Finite mass contributions coming from phase space
integration and weak magnetism are also included. The result is expressed
in percent.}
\label{ftzer}
\end{figure}
\begin{figure}
\epsfysize=8.4cm
\epsfxsize=7.0cm
\centerline{\epsffile{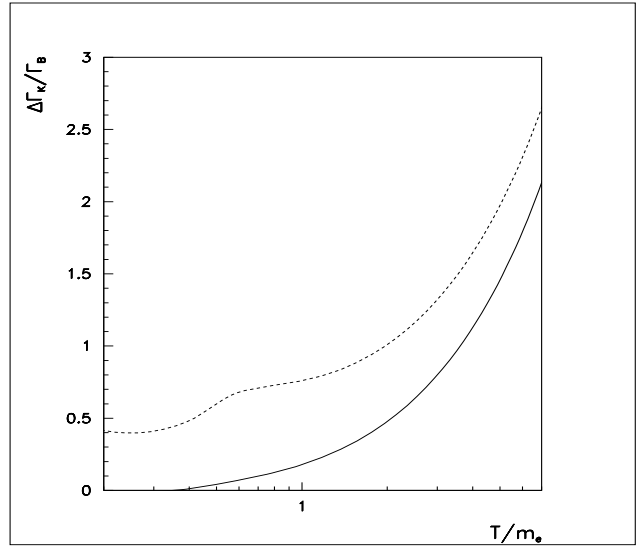}}
\caption{The kinematical contributions  to the $n \rightarrow p$
(solid line) and $p \rightarrow n$ (dashed line) processes, normalized to
the Born rates, in percent.}
\label{ftkin}
\end{figure}
\begin{figure}
\epsfysize=8.4cm
\epsfxsize=7.0cm
\centerline{
\epsffile{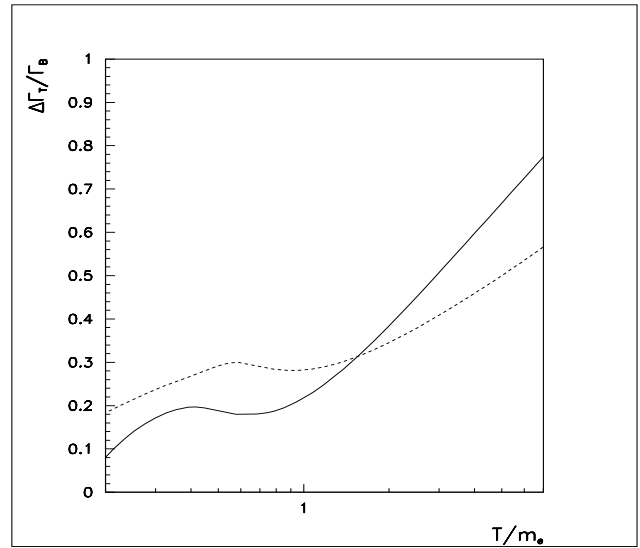}}
\caption{The whole thermal radiative corrections to
the $n \rightarrow p$ (solid line) and $p \rightarrow n$ (dashed line)
processes, expressed in percent.}
\label{fttherm}
\end{figure}
The cumulative ``zero temperature", kinematical and thermal radiative
corrections to the $n \rt p$ and $p \rt n$ rates are instead reported in
Figures \ref{ftzer}, \ref{ftkin}, \ref{fttherm}. We have shown the total
rates $\Gamma(n \rt p)$ and $\Gamma(p \rt n)$ in Figure \ref{ftotrat},
while the total relative correction $\Delta
\Gamma / \Gamma_B \equiv (\Gamma -
\Gamma_B)
/\Gamma_B$, in percent, are plotted in Figure \ref{ftotcor}.
The whole correction $\Delta \Gamma$ for $n \leftrightarrow p$ turn out to
be a positive decreasing function over the whole temperature range relevant
for BBN. The main contribution at low temperature for both total rates
comes from the radiative corrections, while for $T > 2 \div 3~MeV$ kinetic
contribution starts dominating. This is particularly evident by looking at
Figures \ref{ftzer} and \ref{ftkin}. While for radiative corrections the
effect on $n \rightarrow p$ total rate is larger than on the $p \rt n$ one,
$\Delta \Gamma_K$ shows an opposite behaviour. The competition of these two
corrections is then responsible for the presence of the inversion points in
Figure \ref{ftotcor} at $T \simeq 0.15 ~MeV$ and $T \simeq 2 ~MeV$.
Finally, the order of magnitude of the pure thermal radiative  corrections
\begin{equation}
\Delta \Gamma_T \equiv \Delta \Gamma_M + \Delta \Gamma_W +\Delta \Gamma_V
+\Delta \Gamma_\gamma~~~,
\label{tottherm}
\end{equation}
is sensibly smaller, but nevertheless they may contribute for a factor $0.2
\div 0.4 \%$ at the freeze-out temperature $T \sim 1 ~MeV$.
We have performed a fit of the numerical results for $\Gamma(n \rt p)$ and
$\Gamma(p \rt n)$. The fitting expressions are the following
\bea
\Gamma(n \rightarrow p) & = & \frac{1}{\tau_n^{exp}} \sum_{l=0}^{8} a_l
\left( \frac{T}{m_e} \right)^l~~~, \label{fitnp} \\
\Gamma(p \rightarrow n) & = & \frac{1}{\tau_n^{exp}} \exp
\left( -\frac{q~ m_e}{T} \right) ~\sum_{l=1}^{10} b_l
\left( \frac{T}{m_e} \right)^l~~~, \label{fitpn}
\eea
where for $n \rightarrow p$
\bea
&a_0 = 1~~;~~a_1=-0.0135609~~;~~a_2=1.227825~~;~~a_3=-19.344104~~;
 \nonumber \\
&a_4=85.9281797~~;~~a_5=-11.221606~~;~~a_6=14.4804529~~;~~a_7=-3.082314~~;
 \nonumber \\
&a_8=0.357466~~;~~a_9=-0.0200726~~;~~a_{10}=0.10049 {\cdot}10^{-3}~~;~~
 a_{11}=0.488642{\cdot}10^{-4}~~;
 \nonumber \\
&a_{12}=-0.237462{\cdot}10^{-5}~~,~~a_{13}=0.359274{\cdot}10^{-7}
\label{coeffnp}
\eea
\newpage
\noindent
while for $p \rightarrow n$
\bea
&b_1=20.884200~~;~~b_2=-71.802238~~;~~b_3=118.853956~~;~~
 b_4=-11.564594~~;
\nonumber \\
&b_5=45.057913~~;~~b_6=-3.316009~~;~~b_7=0.2744236~~;~~
 b_8=-0.75654{\cdot}10^{-2}~~;
 \nonumber \\
&b_9=-0.37916 {\cdot} 10^{-3}~~;~~b_{10}= 0.201124 {\cdot} 10^{-4}~~;~~ b_{11}=
0.791665{\cdot}10^{-6}~~;
 \nonumber \\
&b_{12}= -0.640452{\cdot}10^{-7}~~;~~b_{13}=0.104891{\cdot}10^{-8}~~;~~ q= 2.90377¤~~.
\label{coeffpn}
\eea
The fit has been obtained requiring that the fitting functions differ by
less than $0.1 \%$ from the numerical values. \\ Finally, note that mass
shift radiative corrections affect not only the weak interaction reaction
rates but also the energy density and pressure of $e^{\pm}$ and photons in the
Universe. We have considered the corrections to these quantities as well,
and the results are reported in Figure \ref{sta} \\ Last but not least, we
have to consider the strong nuclear processes intervening in primordial
nucleosynthesis. Unfortunately, there is no very reliable theoretical
computation for the interested nuclear reaction rates, so the strategy
adopted is quite different: experimental data on the processes involved are
used. However, a certain degree of uncertainty is introduced in this way,
mainly due to the fact that the typical energies of the reactions taking
place during BBN are quite different (too much low) from the available
experimental ones, so that it is necessary to take some extrapolations.
Moreover, a suitable thermal averaging of the measured cross sections
(usually made with a Boltzmann distribution) is required:
\begin{equation}\label{sigmv}
  < \sigma \, v > \; = \; \sqrt{\frac{8}{\pi M T^3}} \, \int d E
  \, \sigma(E) \, E \, e^{- E/T}
\end{equation}
($M$ is the reduced mass).All this conspires to produce uncertainties in
the obtained reaction rates as large as 10\%. Fortunately, the impact of
these on the predictions for light element abundances is extremely low
\cite{Villante}. This happens because, as we have already stressed, the
crucial reactions for nucleosynthesis are those fixing the $n/p$ ratio at
freeze out (especially for $^4He$ abundance), and such reactions are very
well theoretically evaluated, as discussed in this thesis. All the sources
of uncertainties in primordial nucleosynthesis predictions are, however,
currently well controlled by both MonteCarlo \cite{MC} simulations and
semi-analytical analysis \cite{Villante}.

\section{Prediction on the $^4He$ abundance}
\label{sumb}

The conditions for BBN to take place were reached when the temperature in
the Universe was in the range $10 \div 0.1 \, MeV$. Primordial
nucleosynthesis is, in fact, the final outcome of the decoupling of the
weak interactions which keep neutron and proton in thermal equilibrium and,
shortly after, of the onset of nuclear reactions which start producing
light nuclear species. At temperature $T >> 1~ MeV$ the neutron to proton
ratio is order unity and starts decreasing exponentially when the
temperature reaches the value of their mass difference. As the temperature
decreases, however, weak interactions are no longer fast enough to maintain
equilibrium and a substantial final neutron fraction survives down to the
phase of nucleosynthesis. All neutrons become practically bound in $^{4}He$
nuclei, due to the high binding energy per nucleon, which therefore
represent the dominant products of BBN. While the $^{4}He$ mass fraction is
weakly depending on the baryon to photon ratio $\eta$, it is strongly
dependent on the neutron fraction at the nucleon freeze-out. An accurate
theoretical prediction for the helium abundance, as well as for the other
light nuclei produced during BBN, will be the subject of the next section;
it can be obtained by using the standard BBN numerical code suitably
modified to take into account all the corrections considered in this
thesis. Here we report on the results for the expected variation of the
surviving neutron fraction $X_n$ induced by the whole effects $\Delta
\Gamma (n \leftrightarrow p)$. This allows for a simple estimate of the
corresponding variation of $X_{^4He}$; the baseline value for this quantity
obtained with the standard BBN code without the inclusion of any correction
is \cite{LT}
\begin{equation}\label{s1}
  X^0_{^4He} \; = \; 0.2411 ~~~.
\end{equation}
The neutron fraction is determined by the differential equation
\be
\frac{d X_n}{d T} = \frac{d t}{d T} \left[
\Gamma(p \rightarrow n) (1 - X_n) -
\Gamma(n \rightarrow p) X_n \right]~~~.
\label{s2}
\ee
Writing $X_n=X_n^0 + \delta X_n$, where $\delta X_n$ is the correction
induced by $\Delta \Gamma$, we have at first order
\begin{eqnarray}
\frac{d X_n^0}{d T} & = & \frac{d t}{d T} \left[ \Gamma_B(p \rightarrow n) (1 - X_n^0) -
\Gamma_B(n \rightarrow p) X_n^0 \right]~~~, \label{s3} \\
\frac{d}{d T} \delta X_n & = & - \frac{d t}{d T} \left[
\left( \Gamma_B(p \rightarrow n) + \Gamma_B(n \rightarrow p) \right)
\delta X_n  - \Delta
\Gamma(p \rightarrow n) \right. \nonumber \\
& + & \left. \left( \Delta \Gamma(p \rightarrow n)  + \Delta \Gamma(n
\rightarrow p) \right) X_n^0   \right]~~~.
\label{s4}
\end{eqnarray}
Notice that the zeroth order abundance $X_n^0$ has been defined as the one
obtained by the Born amplitudes rescaled by the {\it constant} factor
$961/886.7$, which provides at tree level the correct prediction for
neutron lifetime (see our discussion in section 3). The Born rates $
\Gamma_B(p \leftrightarrow n)$ in (\ref{s3}),(\ref{s4})
are therefore rescaled by the same factor. Equations (\ref{s3}),(\ref{s4})
have been numerically solved using our fitting function for $\Gamma( n
\leftrightarrow p)$ and a similar one for $\Gamma_B( n \leftrightarrow p)$,
which we do not report for brevity. We found for the asymptotic abundance
$\delta X_n\simeq 0.0024$, with a relative change, in percent, $\delta
X_n/X_n^0 = 1.6 \%$. \\ These results allows for a simple estimation of
corrections to $^4He$ mass fraction by means of the formula (\ref{n20})
\begin{equation}\label{s5}
  X_{^4He} \; \simeq \; 2 \, X_n(t_{ns}) ~~~.
\end{equation}
Note that $X_n(t_{ns})$ is evaluated at the time when nucleosynthesis
begins, $t_{ns} \simeq 180 \, s$; actually, by this time the neutron
abundance surviving at freeze out has been depleted by $\beta$-decay to
\begin{equation}\label{s6}
  X_n(t_{ns}) \; \simeq \; X_N^0 \, e^{- t_{ns}/\tau_n} ~~~,
\end{equation}
where $\tau_n$ is the neutron lifetime. Hence the corrections to $^4He$
mass fraction can be evaluated from
\begin{equation}\label{s7}
 \delta X_{^4He} \; \simeq \; 2 \; \delta X_n \, e^{- t_{ns}/\tau_n} ~~~.
\end{equation}
and using the results for $\delta X_n$ we find
\begin{equation}\label{s8}
  \delta X_{^4He} \; \simeq \; 0.004 ~~~~~~~~~~~~~~~~~~~~~~~~~~~
  \frac{\delta X_{^4He}}{X_{^4He}} \; \simeq \; 1.6 \, \% ~~~.
\end{equation}
The total correction, as mentioned in the previous section, is largely
dominated by ``zero temperature" radiative and finite nucleon mass
corrections, which give a {\it positive} contribution to $X_{^4He}$.

\section{Building the BBN code}
\label{sumc}

In this section we finally describe the way the primordial abundances of
light nuclei can be numerically evaluated and discuss how to build of a BBN
code \cite{Kawano} - \cite{wag73}. Such a code follows the evolution of
these abundances, and primarily tracks $p$, $n$, $D$, $T$, $^3He$, $^4He$,
$^6Li$, $^7Li$ and $^7Be$. The initial temperature can be chosen to be, for
example, $T = 10
\, MeV$, and hence for the initial abundances we can take the nuclear
statistical equilibrium values (\ref{n10}):
\begin{equation}\label{s9}
  X_A \; = \; \left( \frac{\zeta(3)}{\sqrt{8 \pi}} \right)^{A-1} \,
  \frac{g_A}{2} \, A^{\frac{5}{2}} \, \left( \frac{T}{m_N}
  \right)^{\frac{3}{2} (A-1)} \, \eta^{A-1} \, X_p^{Z} \, X_n^{A-Z} \,
  e^{\frac{B_A}{T}}
\end{equation}
(hereafter we use the same notation of chapter \ref{pn}). In fact, at
temperatures $\gapproxeq \, 1 \, MeV$, the nuclear rates are sufficiently
high to keep all abundances at their equilibrium values. It is usually made
the reasonable assumption that the elements are always maintained in
kinetic equilibrium by electromagnetic and strong interactions, while this
may be not true for chemical equilibrium; then the coefficients in the
nuclear rates depend only on $\eta$ and $T$. \\ The set of evolution
equations relevant for nucleosynthesis, which is the main body of the BBN
code, has been deduced in chapter \ref{pn}. Here, we report these equations
for completeness:
\bea
 \frac{1}{R} \, \frac{dR}{dt} & \simeq &
 \sqrt{\frac{8 \pi}{3 M_P^2}} \, \left[\rho_\gamma+\rho_e+\rho_\nu
 +\rho_B\right]^{1/2} \label{s10} \\
 \frac{1}{n_B} \, \frac{d n_B}{dt} & = & - 3 \, \frac{1}{R} \,
 \frac{dR}{dt}  \label{s11} \\
 \phi_e & \simeq & \frac{\pi^2}{2} \, \frac{n_B \, q_B}{T^3 \, f(z)}
 \label{s12} \\
   \frac{dT}{dt} & = & -\left[3\frac{1}{R} \frac{dR}{dt}
   \left(\rho_\gamma \, + \, \p_\gamma \, + \, \rho_e \, + \, \p_e \, +
   \, \Theta(T-T_D) \, (\rho_\nu \, + \, \p_\nu)   \, + \, \p_B \right)
   \; \right. \nonumber \\
   & + & \left.
 \frac{\partial \rho_e}{\partial \phi_e}\left(\sum_j
 \frac{\partial \phi_e}{\partial Y_j} \,\frac{d Y_j}{dt} - \, 3 \,
 \frac{1}{R} \frac{dR}{dt}\,
 n_B \,\frac{\partial \phi_e}{\partial n_B}\right)
 \; + \; n_B \,\sum_j
   \, \left(\Delta M_j \, + \, \frac{3}{2} T
   \right) \, \frac{d Y_j}{dt} \right]  \nonumber \\
   & {\times} & \left[ \frac{\partial \rho_e}{\partial T} \, + \,
 \frac{\partial \rho_e}{\partial \phi_e} \,
 \frac{\partial \phi_e}{\partial T}
 \; + \; \frac{d\rho_\gamma}{dT} \;
   + \; \Theta(T-T_D) \,\frac{d\rho_\nu}{dT} \; + \; \frac{3}{2}
   n_B \, \sum_j \, Y_j \right]^{-1} \label{s13} \\
 \frac{dY_i}{dt} & = & \sum_{j,k,l} \, N_i \left(
 \Gamma_{kl \rt ij} \, \frac{Y_l^{N_l} \, Y_k^{N_k}}{N_l! \, N_k !}
 \; - \; \Gamma_{ij \rt kl} \, \frac{Y_i^{N_i} \, Y_j^{N_j}}{N_i ! \, N_j
 !} \right) \equiv \Gamma_{i}(Y_j) \label{s14} ~~~.
\eea
As already mentioned, this set of equations does not include the evolution
equations for neutrinos. Here, however, we have explicited the effect of
not decoupled neutrinos on the involved quantities by means of the
theta-function $\Theta(T-T_D)$, $T_D$ being the decoupling temperature. The
relevant thermodynamic quantities can be read off by Eqs.
(\ref{n29})-(\ref{n34}), while $q_B$ is the baryon electric charge, as
defined by equation (\ref{n25}). Note that it has been assumed that the
electron chemical potential parameter $\phi_e$ is small, so that all terms
depending on $\phi_e$ \footnote{Note that, in general, both $\rho_e$,
$\p_e$ and $\rho_\gamma$, $\p_\gamma$ depend on the electron chemical
potential, the last two quantitiex through the effective finite photon mass
induced by the $e^+ e^-$ bath. Furthermore, observe that from the entropy
conservation also $\rho_\nu$, $\p_\nu$ would depend on $\phi_e$. However,
this dependence is effective only in Eqs. (\ref{s10}), (\ref{s11}), while
in Eq. (\ref{s13}) the presence of the theta-function makes it non
operative (neutrino entropy is related to the electron one only when they
are coupled to the plasma.} can be expanded at first order in this
parameter. \\ The other inputs are contained in the equations (\ref{s14}),
and are the weak interaction and strong nuclear reaction rates. The
corrected weak $n \leftrightarrow p$ rates are taken from our fits in Eqs.
(\ref{fitnp})-(\ref{coeffpn}). The complete nuclear reaction network is
composed of 88 reactions, which we report in appendix \ref{netw}; however,
in many cases, one can also consider a limited number of reactions,
introducing a negligible error in the computations. For brevity, we do not
report the expressions for the nuclear reaction rates which can be found in
\cite{Kawano}.

\subsection{Numerical tricks}

The basics equations to be followed for evaluating the primordial
abundances are (\ref{s11}) and (\ref{s14}). Eq. (\ref{s10}) can be directly
substituted into (\ref{s11}), while the expression (\ref{s12}) for the
electron chemical potential has to be used for calculating $\rho_e$, $\p_e$
but also $\rho_\gamma$, $\p_\gamma$ and $\rho_\nu$ (see the previous
footnote). Instead, by means of Eq. (\ref{s13}) we can translate the time
evolution into the temperature evolution of the interested quantities.

For numerical reasons, it is better to turn the variable $n_B$ into the
dimensionless quantity
\begin{equation}
  \tilde{h} \; = \; \frac{n_B}{T^3} ~~~,
\end{equation}
which evolves more slowly with $T$ than $n_B$. Furthermore, instead of $T$,
we use the variable
\begin{equation}
  z \; = \; \frac{m_e}{T}~~~.
\end{equation}
In terms of these news variables Eq. (\ref{s11}) and Eqs. (\ref{s14}) take
the form
\bea
  \frac{d \tilde{h}}{dz} & = &  \left[
  \frac{
  \left( 3 \hat{\p}_\gamma-\hat{\rho}_\gamma+ 3 \hat{\p}_e - \hat{\rho}_e
  + z \frac{\partial \hat{\rho}_\gamma}
  {\partial z}+ z \frac{\partial \hat{\rho}_e}
  {\partial z} + \frac{3}{2} \tilde{h} \sum_j Y_j  \right) \widetilde{H}
  + \tilde{h} \sum_j \left( z \Delta \widetilde{M}_j + \frac{3}{2} \right)
  \widetilde{\Gamma}_j}{3
  \left( \hat{\rho}_\gamma + \hat{\p}_\gamma + \hat{\rho}_e + \hat{\p}_e +
  \frac 43 \Theta(z_D-z)\hat{\rho}_\nu +\tilde{h} \sum_j Y_j \right) \widetilde{H}
  + \tilde{h} \sum_j \left( z \Delta \widetilde{M}_j + \frac{3}{2} \right)
  \widetilde{\Gamma}_j} \right] {\cdot} \nonumber \\
  & {\cdot} &  \frac{3 \tilde{h}}{z},
  \label{basic1}
\eea
\bea
\frac{dY_i}{dz} & = & \left[\frac
{4 \hat{\rho}_\gamma + 4 \hat{\rho}_e + 4 \Theta(z_D-z)\hat{\rho}_\nu
- z \frac{\partial \hat{\rho}_\gamma}{\partial z}
- z \frac{\partial \hat{\rho}_e}{\partial z}
+\frac 32 \tilde{h} \sum_j Y_j } {3 \left( \hat{\rho}_\gamma +
\hat{\p}_\gamma + \hat{\rho}_e + \hat{\p}_e + \frac 43
\Theta(z_D-z)\hat{\rho}_\nu +\tilde{h} \sum_j Y_j \right) \widetilde{H}
+ \tilde{h} \sum_j \left( z \Delta \widetilde{M}_j + \frac{3}{2} \right)
\widetilde{\Gamma}_j} \right] {\cdot} \nonumber \\
& {\cdot} & \frac{\widetilde{\Gamma}_i}{z}
\label{basic2}
\eea
with $\widetilde{H}$ denoting the dimensionless Hubble parameter
$\widetilde{H}=H/m_e$
\begin{equation}
  \widetilde{H} \, = \, \sqrt{\frac{8 \pi}{3}} \frac{m_e}{M_{P}} \frac{1}{z^2}
  \left[\hat{\rho}_\gamma + \hat{\rho}_e + \hat{\rho}_\nu + \tilde{h}
  \left( z \widetilde{M}_u +
  \sum_i \left( z \Delta \widetilde{M}_i + \frac{3}{2} \right) Y_i \right)
  \right]^{1/2} ~~~,
\end{equation}
and $z_D=me/T_D$. In writing these equations we have used the following
notations
\bea
\widetilde{M}_u \; = \; \frac{M_u}{m_e} ~~~~~~~~ &,& ~~~~~~~~
\Delta \widetilde{M}_i \; = \; \frac{\Delta M_i}{m_e}
~~~~~~~~,~~~~~~~~\widetilde{\Gamma}_i
\; = \; \frac{\Gamma_i}{m_e}~~~, \\
\hat{\rho}_\alpha\; = \; \frac{\rho_\alpha}{T^4 } ~~~~~~~~ &,& ~~~~~~~~
\hat{\p}_\alpha \; = \; \frac{\p_\alpha}{T^4 }  ~~~,
\eea
with $\alpha=e,\gamma,\nu$. \\ As mentioned above, $\rho_e$, $\p_e$ and
$\rho_\gamma$, $\p_\gamma$, depend on the electron chemical potential. In
the new numerical code we are writing, we use the following strategy. First
of all, we have checked that the influence of $\phi_e$ on the various
quantities of interest is very weak, justifying the expansion at first
order in $\phi_e$ leading to Eq. (\ref{s12}). By virtue of this fact, we
have then substituted Eq. (\ref{s12}) in the corrected expressions for
$\rho_e$, $\p_e$, $\rho_\gamma$, $\p_\gamma$ (given by Eqs. (\ref{n31})
- (\ref{n32}) plus Eqs. (\ref{v6}) - (\ref{v9})) and subsequently we have
performed a fit for these quantities as functions of $z$. To speed up the
running of the BBN code we have finally used these fits in the evolution
equations. The obtained values for the electron energy density and pressure
are fitted (in the relevant temperature range) by the expressions
\bea
\hat{\rho}_e & = & 1.145 ~+~ 0.033981~ z ~-~ 0.14543~ z^2 ~+~ 0.025507~ z^3
- (0.54168 {\times} 10^{-3})~ z^4 \nonumber \\
& - & (0.11263 {\times} 10^{-3})~ z^5 - (0.29742 {\times} 10^{-5})~ z^6 ~+~ (0.38331 {\times}
10^{-6})~ z^7 \nonumber \\ &+& (0.45263 {\times} 10^{-7})~ z^8 ~+~ (0.19241 {\times}
10^{-8})~ z^9 ~-~ (0.96597 {\times} 10^{-10})~ z^{10} \nonumber \\ &-& (0.19505 {\times}
10^{-10})~ z^{11} ~-~ (0.14079 {\times} 10^{-12})~ z^{12} ~~~,
\label{fit1}
\eea
\bea
\hat{\p}_e & = & \left( 0.3786 ~+~ 0.019126~z ~-~ 0.063895~ z^2 ~+~
0.032085~z^3 \right. \nonumber \\ &-& 0.0048501~z^4 ~-~ 0.00016611~z^5 ~+~
0.000082922~z^6 ~+~ (7.9884{\times} 10^{-6})~z^7 \nonumber \\ & - & (0.0619 {\times}
10^{-7})~z^8 ~-~ (1.9568 {\times} 10^{-7})~z^9 ~-~ (1.0921 {\times} 10^{-8})~z^{10}
\nonumber \\
&+& \left. (3.8564 {\times} 10^{-9})~z^{11} \right) ~ e^{- 0.13145~z^2} ~~~.
\label{fit2}
\eea
The expression for $z$ derivative of $\hat{\rho}_e$, entering in
(\ref{basic1}) and (\ref{basic2}), are obtained from (\ref{fit1}).
\\
Instead, in the considered temperature range, we have found that
$\hat{\rho}_\gamma$ takes values between 0.6580 and 0.6573 while
$\hat{\p}_\gamma$ varies from 0.2193 and 0.2187. For simplicity we have
then taken the average values of 0.6577 and 0.2190 as the ones appropriate
for $\hat{\rho}_\gamma$ and $\hat{\p}_\gamma$ respectively.

Finally, the most critical numerical part of the BBN code concerns the
solution method of the set of differential equations (\ref{s14}) for $Y_i$.
In fact, at high temperatures, nuclear reactions proceed in both forward
and reverse directions almost equally rapidly, so that the right-hand-side
of (\ref{s14}) is a small difference of large numbers, which causes severe
numerical problems. To avoid this, we have used the Gear method of backward
differentiation formulas for stiff problems, described for example in
\cite{gear}.

The implementation of the described code, using a Fortran Power Station, is
currently under study, and represents the next step towards a precise
determination of light element abundances in the Universe.

\begin{figure}
\epsfysize=8.4cm
\epsfxsize=7.0cm
\centerline{\epsffile{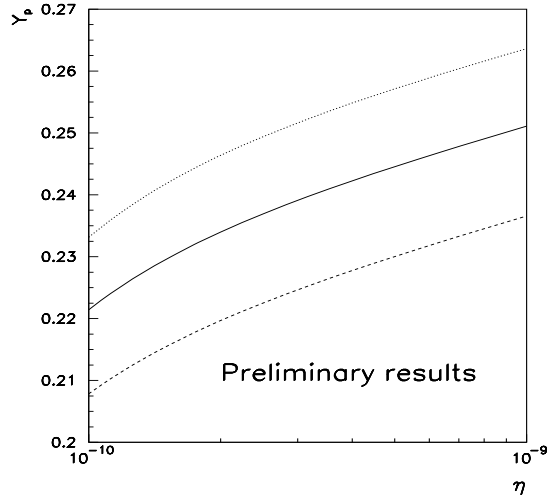}}
\caption{Our predictions for $^4He$ mass fraction $Y_p$ versus $\eta$.
The solid line refers to $N_\nu = 3$, while the dashed ones to $N_\nu = 2$
(below) and $N_\nu = 4$ (above).}
\label{figfinal1}
\end{figure}

\begin{figure}
\epsfysize=8.4cm
\epsfxsize=7.0cm
\centerline{\epsffile{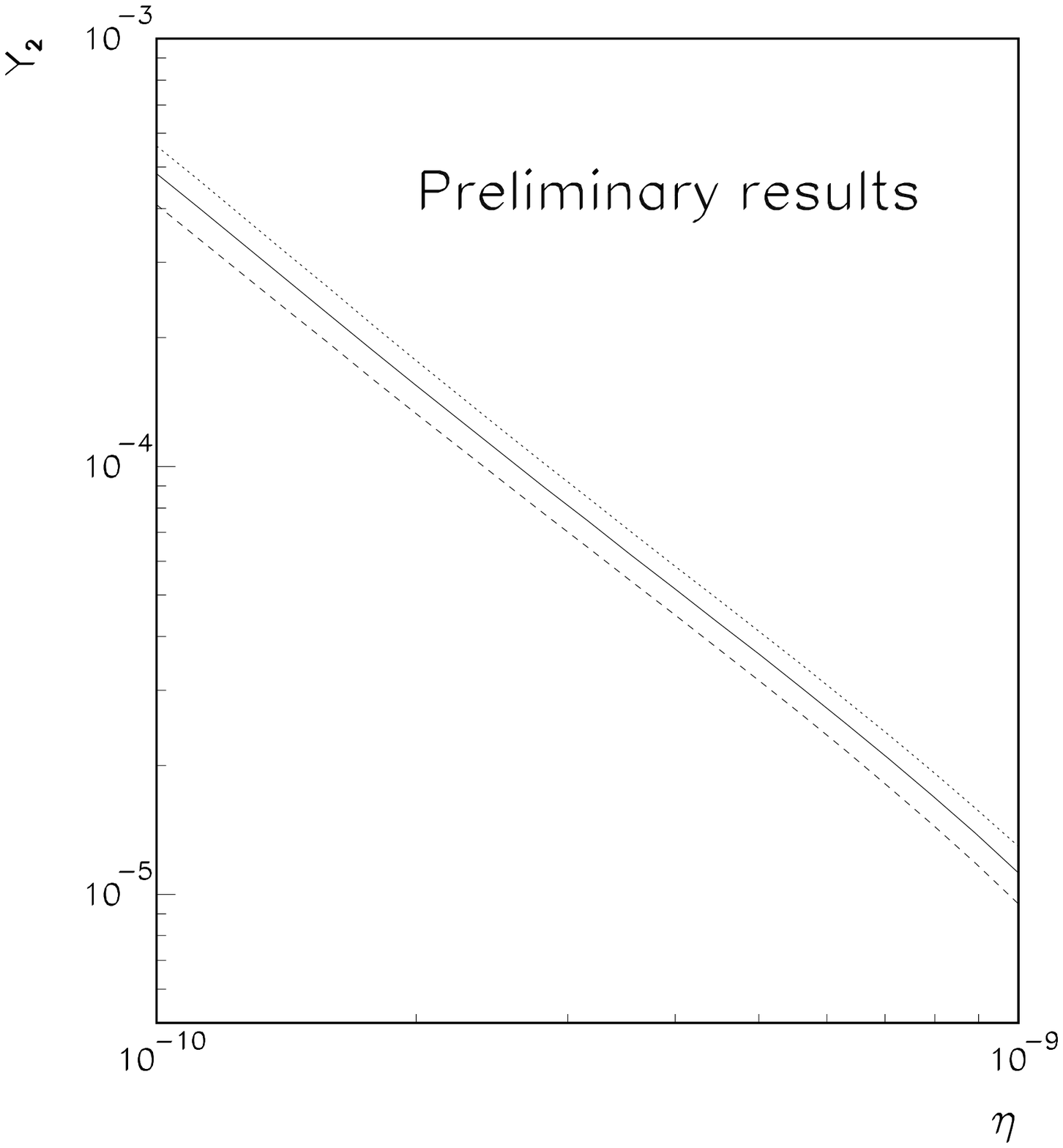}}
\caption{Our predictions for $D$ abundance  $Y_2$ versus $\eta$
(notations as in Figure \protect\ref{figfinal1} .}
\label{figfinal2}
\end{figure}

\begin{figure}
\epsfysize=8.4cm
\epsfxsize=7.0cm
\centerline{\epsffile{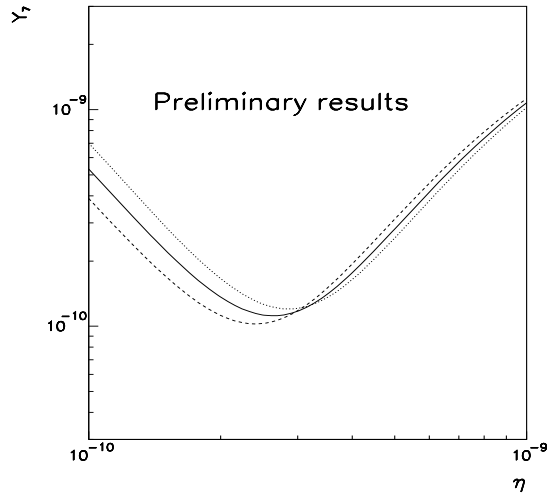}}
\caption{Our predictions for $^7Li$ abundance $Y_7$ versus $\eta$
(notations as in Figure \protect\ref{figfinal1} .}
\label{figfinal3}
\end{figure}

The preliminary results obtained by our code for $^4He$ mass fraction and
$D$, $^7Li$ abundances as functions of $\eta$ are shown in Figures
\ref{figfinal1}, \ref{figfinal2}, \ref{figfinal3}.

%------------------------------------------------------------------------
\newpage
$\left. \right. $
\newpage
\noindent {\huge {\bf Acknowledgments}}
\addcontentsline{toc}{chapter}{{Acknowledgments}}

\vspace{2cm}

\noindent I'm particularly indebted with Drs. Gianpiero Mangano, Gennaro
Miele and O. Pisanti for their continuous unvaluable help, patience and
encouragement during my Ph.D. thesis work. My sincere thanks for their
friendship are here expressed. A special thank goes also to Prof. Franco
Buccella and Prof. P. Strolin for their useful remarks and encouragement
and to my friends F. Acerra, S. De Simone, E. Piedipalumbo, P. Santorelli
and N. Tancredi.

%-----------------------------------------------------------------------

%-----------------------------------------------------------------------
\appendix

\chapter{Non equilibrium distribution of species}
\label{ane}

Decoupling of species from the background plasma is properly described by
the evolution of the particles's phase space distribution function $f(p^\mu
, x^\mu)$, given by the Boltzmann equation. In this appendix we follow the
approach of \cite{kolb} and restrict our analysis to an homogeneous and
isotropic plasma, described by the Robertson-Walker metric. In this case $f
= f(|{\bvec{p}}|,t)$ or equivalently $f = f(E,t)$ and the Boltzmann
equation is
\begin{equation}\label{ne1}
  E \, \frac{\partial f}{\partial t} \; - \; \frac{\dot{R}}{R} \,
  |{\bvec{p}}|^2 \, \frac{\partial f}{\partial E} \; = \; C[f] ~~~,
\end{equation}
where $C[f]$ is the collisional operator for the processes creating and
destroying the given particle specie. It is more convenient to use the
number density $n(t)$ rather than the distribution function, defined by
\begin{equation}\label{ne2}
  n(t) \; = \; g \, \int \frac{d^3 p}{(2 \pi)^3} \, f(E,t) ~~~,
\end{equation}
$g$ being the internal degrees of freedom. For this quantity the Boltzmann
equation becomes
\begin{equation}\label{ne3}
  \frac{d n}{d t} \; + \; 3 H \, n \; = \; g \, \int \frac{d^3 p}{(2 \pi)^3}
  \, \frac{1}{E} \, C[f] ~~~,
\end{equation}
with $H = \dot{R}/R$. Let us focus on a given particle $\psi$, interacting
with the other particles in the heat bath through the reaction $\psi + a +
b + ... \longleftrightarrow i + j + ...$; for this process the collisional
term can be written in the form
\newpage
\bea
g \, \int \frac{d^3 p}{(2 \pi)^3} \, \frac{1}{E} \, C[f] & = &  - \,
\int d \Pi_T \, (2 \pi)^4 \, \delta^4({\mathrm p}_\psi + {\mathrm p}_a +
{\mathrm p}_b ... - {\mathrm p}_i
- {\mathrm p}_j ...) \, {\times} \nonumber \\
& {\times} & \left( \ov{|M|^2}_{forward} \, \Phi_{forward} \, - \,
\ov{|M|^2}_{backward} \, \Phi_{backward} \right) \label{ne4} ~~~,
\eea
with
\bea
\Phi_{forward} & = & f_\psi \, f_a \, f_b \, ... \, (1 \, {\pm} \, f_i) \,
(1 \, {\pm} \, f_j) \, ... \\
\Phi_{backward} & = &  f_i \, f_j \, ... \, (1 \, {\pm} \, f_\psi) \,
(1 \, {\pm} \, f_a) \, (1 \, {\pm} \, f_b) \, ...
\eea
where +/- refers to bosons/fermions. The phase space density is given by
\bea
d \Pi_T & = & d \Pi_\psi \,  d \Pi_a \,  d \Pi_b \, ... \, d \Pi_i \,  d
\Pi_j ~~~, \\
d \Pi_\alpha & = & \frac{g_|\alpha}{(2 \pi)^3} \, \frac{d^3 p_\alpha}{2
E_\alpha} ~~~,
\eea
while the squared matrix elements $|M|^2$ are averaged over initial and
final spins and, assuming CP invariance, $\ov{|M|^2}_{forward} =
\ov{|M|^2}_{backward} = |M|^2$. \\
For illustrative purposes, we consider only situations in which Bose
condensation or Fermi degeneracy can be neglected, so that we may safely
disregard the stimulated emission or blocking factors in (\ref{ne4}): $(1
\, {\pm} \, f_\alpha) \simeq 1$. In these approximations, the evolution
equation (\ref{ne3}) simplifies to
\bea
\frac{d n_\psi}{d t} \; + \; 3 H \, n_\psi \; = \; & - & \int
d \Pi_T \, (2 \pi)^4 \, \delta^4({\mathrm p}_\psi + {\mathrm p}_a +
{\mathrm p}_b ... - {\mathrm p}_i
- {\mathrm p}_j ...) \, \ov{|M|^2} \, {\times} \nonumber \\
& {\times} & \left( f_\psi \, f_a \, f_b \, ... \, - \, f_i \, f_j \, ...\right)
~~~.  \label{ne7}
\eea
Note that the $3 H \, n_\psi$ term accounts for the dilution effect of the
expansion of the Universe, while the right hand side term accounts for the
number changing $\psi$ interactions. The dilution factor can be absorbed by
considering the particle number in the comoving volume $R^3$, proportional
to the quantity
\begin{equation}\label{ne8}
  Y \; =  \; \frac{n_\psi}{s} ~~~,
\end{equation}
where $s \propto R^{-3}$ is the entropy density. In fact, in this case the
left hand side of (\ref{ne7}) becomes simply
\begin{equation}\label{ne9}
  \frac{d n_\psi}{d t} \; + \; 3 H \, n_\psi \; = \; \frac{1}{s} \,
  \frac{d Y}{d t} ~~~.
\end{equation}
We are now in a position to describe the decoupling (i.e. the exit from
chemical equilibrium) of the given specie from the background plasma.
Restricting ourselves to the case in which this specie is stable, the only
number changing processes are of the type
\begin{equation}\label{ne10}
  \psi \; \ov{\psi} \; \longleftrightarrow \; X \;  \ov{X} ~~~.
\end{equation}
With $X$ we generically denote all the species into which $\psi$ can
annihilate. These particles are, of course, in thermal equilibrium and
hereafter we neglect, for simplicity, any chemical potential. Then we have
\begin{equation}\label{ne11}
  f_{X, \ov{X}} \; \simeq \; \exp\left\{- \, \frac{E_{X, \ov{X}}}{T}\right\}
\end{equation}
but, by means of energy conservation, the following condition
\begin{equation}\label{ne12}
  f_X \, f_{\ov{X}} \; \simeq \; \exp\left\{- \, \frac{E_\psi}{T}\right\} \,
  \exp\left\{- \, \frac{E_{\ov{\psi}}}{T}\right\} \; \simeq \; f_\psi^{EQ} \,
  f_{\ov{\psi}}^{EQ}
\end{equation}
holds, where with the index $EQ$ we have indicated the equilibrium
distributions. Hence the statistical factor in (\ref{ne7}) takes the form
\begin{equation}\label{ne13}
  f_\psi \, f_{\ov{\psi}} \; - \; f_X \, f_{\ov{X}} \; = \;
  f_\psi \, f_{\ov{\psi}} \; - \; f_\psi^{EQ} \, f_{\ov{\psi}}^{EQ}
\end{equation}
and the Boltzmann equation for the present case becomes
\begin{equation}\label{ne14}
  \frac{d n_\psi}{d t} \; + \; 3 H \, n_\psi \; = \; - \,
  < \sigma_{\psi\ov{\psi} \rightarrow X  \ov{X}} \, v > \, \left( n_\psi^2
  \, - ( n_\psi^{EQ} )^2 \right)
\end{equation}
where $<\sigma v>$ is the thermally averaged annihilation cross section
times velocity:
\bea
< \sigma_{\psi\ov{\psi} \rightarrow X  \ov{X}} v > & = & \frac{1}{(
n_\psi^{EQ} )^2} \, \int d \Pi_T \, (2 \pi)^4 \, \delta^4({\mathrm p}_\psi
+ {\mathrm p}_{\ov{\psi}} - {\mathrm p}_X - p_{\ov{X}}) \, {\times} \nonumber \\ &
{\times} &
\ov{|M|^2} \, \exp\left\{- \, \frac{E_\psi}{T}\right\} \,
  \exp\left\{- \, \frac{E_{\ov{\psi}}}{T}\right\}  \label{ne15} ~~~.
\eea
We stress that the rate of change of $\psi , \ov{\psi}$ is proportional to
the annihilation rate, while it tends to zero when the equilibrium is
approaching since creation processes balance destruction ones. \\ Another
impressive form of (\ref{ne14}) is obtained by considering the Boltzmann
equation for the quantity $Y$ and rewriting the $t$ dependence in terms of
the temperature or, better, in terms of the adimensional parameter $z =
m_\psi /T$:
\begin{equation}\label{ne16}
  \frac{z}{Y_{EQ}} \, \frac{d Y}{d z} \; = \; - \, \frac{\Gamma}{H} \,
  \left\{ \left( \frac{Y}{Y_{EQ}} \right)^2 \; - \; 1 \right\}
\end{equation}
\begin{equation}\label{ne17}
  \Gamma \; = \; n_{EQ} \, < \sigma \, v > ~~~.
\end{equation}
From these we immediately see that the rate of change of $\psi$ per
comoving volume is set by the factor $\Gamma / H$: when this is less than
order unity, the relative change in the number of $\psi$ in the comoving
volume becomes small, annihilations freeze out and the abundance of $\psi$
``freezes in".

\chapter{Neutrino-photon temperature relation}
\label{atnu}

\begin{figure}
\epsfysize=8.4cm
\epsfxsize=7.0cm
\centerline{\epsffile{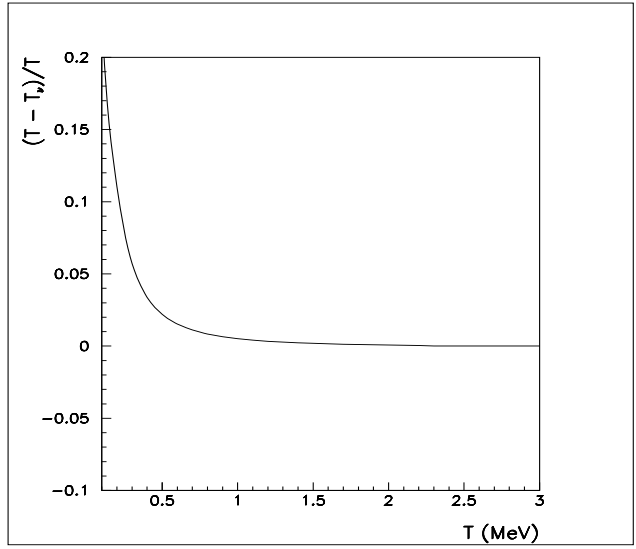}}
\caption{Neutrino - photon temperature difference.}
\label{nuplot}
\end{figure}

When neutrinos decouple (at $T_\nu^D$ around $1 \, MeV$) from the
background primordial plasma, their thermal distribution retains the
equilibrium Fermi-Dirac form with a temperature $T_\nu$ scaling
(approximately) as the inverse of the cosmic scale factor $R$. As discussed
in \ref{bbd1}, their entropy $S_\nu$ in the comoving volume is conserved
also after decoupling
\begin{equation}\label{tn1}
  S_\nu(T_\nu^D) \; = \; S_\nu(T_\nu) ~~~.
\end{equation}
On the other side, photons still remain in equilibrium with the $e^{\pm}$ pairs
until these disappear when become non relativistic (around $T\sim m_e$).
However, since the total entropy in the comoving volume is conserved as
well as neutrino entropy, it follows that the entropy $S_I$ of the species
still in equilibrium ($I = e^{\pm},\gamma$) is conserved too
\begin{equation}\label{tn2}
  S_I(T_D) \; = \; S_I(T) ~~~.
\end{equation}
Note that, until neutrinos decouple, $T_\nu = T$; consequently $T_\nu^D =
T_D$. According to Eqs. (\ref{72}) and (\ref{60}), (\ref{61}), the entropy
of a specie is explicitly given by
\begin{equation}\label{tn3}
  S_i \; = \; \frac{\gi}{6 \pi^2} \, I(x_i) \, R^3 \, T_i^3 ~~~,
\end{equation}
\begin{equation}\label{tn4}
  I(x_i) \; = \; \int_0^\infty d y \, \frac{(y^2 + 2 x_i y)^{1/2} \,
  (4 y^2 + 8 x_i y + 3 x_i^2)}{e^{x_i + y} \; {\pm} \; 1} ~~~,
\end{equation}
where $x_i = m_i/T_i$, $y = (E_i - m_i)/T_i$. Here, the effective ($T$
dependent) degrees of freedom are parametrized by $\gi(T) = (\gi / 6 \pi^2)
\, I(x_i)$. Substituting (\ref{tn3}) in (\ref{tn1}), (\ref{tn2}) and
dividing side by side these two equations, the following relation between
$T_\nu$ and $T$ is obtained \cite{tnt}
\begin{equation}\label{tn5}
  \left( \frac{T_\nu}{T} \right)^3 \; = \; \frac{g_\nu(T_D)}{g_I(T_D)} \,
  \frac{g_I(T)}{g_\nu(T)}
\end{equation}
or, more explicitly,
\begin{equation}\label{tn6}
  \frac{T_\nu}{T} \; = \; \left( \frac{2 \pi^4 \; + \; 15 \, I(x_e)}
  {2 \pi^4 \; + \; 15 \, I(x_e^D)} \right)^{\frac{1}{3}} ~~~.
\end{equation}
In the limit (considered in \ref{bbd1}) in which $g_I(T) \simeq 2$,
$g_I(T_D) \simeq 11/2$ ($g_\nu(T_D) = g_\nu(T)$), from (\ref{tn5}) the
relation (\ref{113}) follows. Instead (\ref{tn6}) gives the generalization
for $T_\nu$ as function of $T$ we looked for. In Figure \ref{nuplot} we
plot the relative difference in temperature between neutrinos and photons;
of course, before decoupling the two temperature coincide (the plotted
curve is evaluated from (\ref{tn6}) adopting the value $2.3 \, MeV$ for the
decoupling temperature; results practically unchange by shifting this
temperature in the range $2 \div 3 \, MeV$).

\chapter{Nuclear reaction network}
\label{netw}

In the following Tables \ref{net1} - \ref{net9} we show the complete
nuclear reaction network relevant for the calculations of the primordial
abundances of light elements \cite{wag67, wag69, wag73, Kawano}.

\begin{table}[H]
\caption{Nuclear $\beta$-decay reactions.}
\begin{center}
\begin{tabular}{|rcl|} \hline \hline

  $  ^3H $ & $ \rt $ & $  e^- + \neb + \,  ^3He

 $ \\ $  ^8Li  $ & $ \rt $ & $  e^- + \neb + \, 2^4He

 $ \\ $  ^{12}B  $ & $ \rt $ & $  e^- + \neb + \,  ^{12}C

 $ \\ $  ^{14}C  $ & $ \rt $ & $  e^- + \neb +  \, ^{14}N

 $ \\ $  ^8B  $ & $ \rt $ & $  e^+ + \ne +  \, 2^4He

 $ \\ $  ^{11}C  $ & $ \rt $ & $  e^+ + \ne +  \, ^{11}B

 $ \\ $  ^{12}N  $ & $ \rt $ & $  e^+ + \ne +  \, ^{12}C

 $ \\ $  ^{13}N  $ & $ \rt $ & $  e^+ + \ne +  \, ^{13}C

 $ \\ $  ^{14}O  $ & $ \rt $ & $  e^+ + \ne +  \, ^{14}N

 $ \\ $  ^{15}O  $ & $ \rt $ & $  e^+ + \ne +  \, ^{15}N

$ \\  \hline \hline
\end{tabular}
\end{center}
\label{net1}
 \end{table}

\begin{table}
\caption{Nuclear neutron-photon reactions.}
\begin{center}
\begin{tabular}{|rcl|} \hline \hline

       $  H  \;  $ & $ (n, \gamma) $ & $  \; D

 $ \\ $  D \;  $ & $ (n, \gamma) $ & $  \; ^3H

 $ \\ $  ^3He \;  $ & $ (n, \gamma) $ & $  \; ^4He

 $ \\ $  ^6Li \;  $ & $ (n, \gamma) $ & $  \; ^7Li

 $ \\ $  ^7Li \;  $ & $ (n, \gamma) $ & $  \; ^8Li

 $ \\ $  ^{10}B \;  $ & $ (n, \gamma) $ & $  \; ^{11}B

 $ \\ $  ^{11}B \;  $ & $ (n, \gamma) $ & $  \; ^{12}B

 $ \\ $  ^{12}C \;  $ & $ (n, \gamma) $ & $  \; ^{13}C

 $ \\ $  ^{13}C \;  $ & $ (n, \gamma) $ & $  \; ^{14}C

 $ \\ $  ^{14}N \;  $ & $ (n, \gamma) $ & $  \; ^{15}N

$ \\  \hline \hline
\end{tabular}
\end{center}
\label{net2}
 \end{table}

\begin{table}
\caption{Nuclear neutron-proton and proton-neutron reactions.}
\begin{center}
\begin{tabular}{|rcl|} \hline \hline

      $  ^3He \; $ & $ (n,p) $ & $  \; ^3H

 $ \\ $  ^7Be \; $ & $ (n,p) $ & $  \; ^7Li

 $ \\ $  ^{11}C \; $ & $ (n,p) $ & $  \; ^{11}B

 $ \\ $  ^{13}N \; $ & $ (n,p) $ & $  \; ^{13}C

 $ \\ $  ^{14}N \; $ & $ (n,p) $ & $  \; ^{14}C

 $ \\ $  ^{15}O \; $ & $ (n,p) $ & $  \; ^{15}N

 $ \\ $  ^{12}B \; $ & $ (p,n) $ & $  \; ^{12}C

$ \\  \hline \hline
\end{tabular}
\end{center}
\label{net3}
 \end{table}

\begin{table}
\caption{Nuclear neutron-$\alpha$ and $\alpha$-neutron reactions.}
\begin{center}
\begin{tabular}{|rcl|} \hline \hline

      $  ^6Li \; $ & $ (n, \alpha) $ & $  \; ^3H

 $ \\ $  ^7Be \; $ & $ (n, \alpha) $ & $  \; ^4He

 $ \\ $  ^{10}B \; $ & $ (n, \alpha) $ & $  \; ^7Li

 $ \\ $  ^{15}O \; $ & $ (n, \alpha) $ & $  \; ^{12}C

 $ \\ $  ^8Li \; $ & $ (\alpha,n) $ & $  \; ^{11}B

 $ \\ $  ^9Be \; $ & $ (\alpha,n) $ & $  \; ^{12}C

 $ \\ $  ^{10}B \; $ & $ (\alpha , n) $ & $  \; ^{13}N

 $ \\ $  ^{11}B \; $ & $ (\alpha , n) $ & $  \; ^{14}N

 $ \\ $  ^{12}B \; $ & $ (\alpha , n) $ & $  \; ^{15}N

 $ \\ $  ^{13}C \; $ & $ (\alpha , n) $ & $  \; ^{16}O

$ \\  \hline \hline
\end{tabular}
\end{center}
\label{net4}
 \end{table}

\begin{table}
\caption{Nuclear proton-photon reactions.}
\begin{center}
\begin{tabular}{|rcl|} \hline \hline

      $  D \; $ & $ (p, \gamma) $ & $  \; ^3He

 $ \\ $  ^3H \; $ & $ (p, \gamma) $ & $  \; ^4He

 $ \\ $  ^6Li \; $ & $ (p, \gamma) $ & $  \; ^7Be

 $ \\ $  ^7Be \; $ & $ (p, \gamma) $ & $  \; ^8B

 $ \\ $  ^9Be \; $ & $ (p, \gamma) $ & $  \; ^{10}B

 $ \\ $  ^{10}B \; $ & $ (p, \gamma) $ & $  \; ^{11}C

 $ \\ $  ^{11}B \; $ & $ (p, \gamma) $ & $  \; ^{12}C

 $ \\ $  ^{11}C \; $ & $ (p, \gamma) $ & $  \; ^{12}N

 $ \\ $  ^{12}C \; $ & $ (p, \gamma) $ & $  \; ^{13}N

 $ \\ $  ^{13}C \; $ & $ (p, \gamma) $ & $  \; ^{14}N

 $ \\ $  ^{14}C \; $ & $ (p, \gamma) $ & $  \; ^{15}N

 $ \\ $  ^{13}N \; $ & $ (p, \gamma) $ & $  \; ^{14}O

 $ \\ $  ^{14}N \; $ & $ (p, \gamma) $ & $  \; ^{15}O

 $ \\ $  ^{15}N \; $ & $ (p, \gamma) $ & $  \; ^{16}O

$ \\  \hline \hline
\end{tabular}
\end{center}
\label{net5}
 \end{table}

\begin{table}
\caption{Nuclear proton-$\alpha$ and $\alpha$-proton reactions.}
\begin{center}
\begin{tabular}{|rcl|} \hline \hline

      $  ^6Li \; $ & $ (p, \alpha) $ & $  \; ^3He

 $ \\ $  ^7Li \; $ & $ (p, \alpha) $ & $  \; ^4He

 $ \\ $  ^9Be \; $ & $ (p, \alpha) $ & $  \; ^6Li

 $ \\ $  ^{10}B \; $ & $ (p, \alpha) $ & $  \; ^7Be

 $ \\ $  ^{12}B \; $ & $ (p, \alpha) $ & $  \; ^9Be

 $ \\ $  ^{15}N \; $ & $ (p, \alpha) $ & $  \; ^{12}C

 $ \\ $  ^8B \; $ & $ (\alpha , p) $ & $  \; ^{11}C

 $ \\ $  ^{10}B \; $ & $ (\alpha , p) $ & $  \; ^{13}C

 $ \\ $  ^{11}B \; $ & $ (\alpha , p) $ & $  \; ^{14}C

 $ \\ $  ^{11}C \; $ & $ (\alpha , p) $ & $  \; ^{14}N

 $ \\ $  ^{12}N \; $ & $ (\alpha , p) $ & $  \; ^{15}O

 $ \\ $  ^{13}N \; $ & $ (\alpha , p) $ & $  \; ^{16}O

$ \\  \hline \hline
\end{tabular}
\end{center}
\label{net6}
 \end{table}

\begin{table}
\caption{Nuclear $\alpha$-photon reactions.}
\begin{center}
\begin{tabular}{|rcl|} \hline \hline

      $  D \; $ & $ (\alpha , \gamma ) $ & $  \; ^6Li

 $ \\ $  ^3H \; $ & $ (\alpha , \gamma ) $ & $  \; ^7Li

 $ \\ $  ^3He \; $ & $ (\alpha , \gamma ) $ & $  \; ^7Be

 $ \\ $  ^6Li \; $ & $ (\alpha , \gamma ) $ & $  \; ^{10}B

 $ \\ $  ^7Li \; $ & $ (\alpha , \gamma ) $ & $  \; ^{11}B

 $ \\ $  ^7Be \; $ & $ (\alpha , \gamma ) $ & $  \; ^{11}C)

 $ \\ $  ^{12}C \; $ & $ (\alpha , \gamma ) $ & $  \; ^{16}O

$ \\  \hline \hline
\end{tabular}
\end{center}
\label{net7}
 \end{table}

\begin{table}
\caption{Nuclear deuterium-nucleon reactions.}
\begin{center}
\begin{tabular}{|rcl|} \hline \hline

      $  D \; $ & $ (d,n) $ & $  \; ^3He

 $ \\ $  D \; $ & $ (d,p) $ & $  \; ^3H

 $ \\ $  ^3H \; $ & $ (d,n) $ & $  \; ^4He

 $ \\ $  ^3He \; $ & $ (d,p) $ & $  \; ^4He

 $ \\ $  ^9Be \; $ & $ (d,n) $ & $  \; ^{10}B

 $ \\ $  ^{10}B \; $ & $ (d,p) $ & $  \; ^{11}B

 $ \\ $  ^{11}B \; $ & $ (d,n) $ & $  \; ^{12}C

$ \\  \hline \hline
\end{tabular}
\end{center}
\label{net8}
 \end{table}

\begin{table}
\caption{Nuclear three-body reactions.}
\begin{center}
\begin{tabular}{|rcl|} \hline \hline

      $  ^3He \; $ & $ (^3He,2p) $ & $  \; ^4He

 $ \\ $  ^7Li \; $ & $ (d,n \, \alpha) $ & $  \; ^4He

 $ \\ $  ^7Be \; $ & $ (d,p \, \alpha) $ & $  \; ^4He

 $ \\ $  ^4He \; $ & $ (\alpha \, n,\gamma) $ & $  \; ^9Be

 $ \\ $  ^4He \; $ & $ (2\alpha,\gamma) $ & $  \; ^{12}C

 $ \\ $  ^8Li \; $ & $ (p,n \, \alpha) $ & $  \; ^4He

 $ \\ $  ^8B \; $ & $ (n,p \, \alpha) $ & $  \; ^4He

 $ \\ $  ^9Be \; $ & $ (p,d \, \alpha) $ & $  \; ^4He

 $ \\ $  ^{11}B \; $ & $ (p,2 \alpha) $ & $  \; ^4He

 $ \\ $  ^{11}C \; $ & $ (n,2 \alpha) $ & $  \; ^4He

 $ \\  \hline \hline
\end{tabular}
\end{center}
\label{net9}
 \end{table}

\chapter{Nuclide mass excess}
\label{mex}

In this appendix we report the measured mass excess $\Delta m_i$ for the 26
nuclides used in BBN calculations \cite{ndata}.

\begin{table}[b]
\caption{Nuclear mass excess $\Delta m_i$ in $KeV$ (referred to $^{12}C$).}
\begin{center}
\begin{tabular}{|rl|rl|} \hline \hline
$n$ & 8071.323 ${\pm}$ 0.002 & $^{11}B$ & 8667.984 ${\pm}$ 0.420   \\
\hline
$p$ & 7288.969 ${\pm}$ 0.001 & $^{11}C$ & 10650.531 ${\pm}$ 0.952  \\
\hline
$D$ & 13135.720 ${\pm}$ 0.001 & $^{12}B$ & 13368.901 ${\pm}$ 1.400  \\
\hline
$^3H$ & 14949.794 ${\pm}$ 0.001 & $^{12}C$ & 0.0  \\
\hline
$^3He$ & 14931.204 ${\pm}$ 0.001 & $^{12}N$ & 17338.083 ${\pm}$ 1.000  \\
\hline
$^4He$ & 2424.911 ${\pm}$ 0.001 & $^{13}C$ & 3125.011 ${\pm}$ 0.001 \\
\hline
$^6Li$ & 14086.312 ${\pm}$ 0.475 & $^{13}N$ & 5345.456 ${\pm}$ 0.270  \\
\hline
$^7Li$ & 14907.673 ${\pm}$ 0.473 & $^{14}C$ & 3019.892 ${\pm}$ 0.004 \\
\hline
$^7Be$ & 15769.489 ${\pm}$ 0.472 & $^{14}N$ & 2863.417 ${\pm}$ 0.001 \\
\hline
$^8Li$ & 20946.195 ${\pm}$ 0.488 & $^{14}O$ & 8006.456 ${\pm}$ 0.075 \\
\hline
$^8B$  & 22921.002 ${\pm}$ 1.107 & $^{15}N$ & 101.438 ${\pm}$ 0.001 \\
\hline
$^9Be$ & 40818.362 ${\pm}$ 62.471 & $^{15}O$ & 2855.388 ${\pm}$ 0.503 \\
\hline
$^{10}B$ & 12050.761 ${\pm}$ 0.370 & $^{16}O$ & -4736.998 ${\pm}$ 0.001
 \\  \hline \hline
\end{tabular}
\end{center}
\label{excess}
 \end{table}

\chapter{Finite nucleon mass corrected matrix element}
\label{matrix}

In this appendix we report the result for the spin summed squared modulus
of Eq. (\ref{cm1})
\begin{eqnarray}
\mod \; = & 16 \, ( &
\ca^2 \, (M_1^2 \, M_2^2  \, + \,  2 \, M_1^2 \, M_3^2  \, + \,
M_2^2 \, M_3^2  \, + \,  2 \, M_1^2 \, M_2 \, M_4  \, + \, 2 \, M_2 \,
M_3^2 \, M_4  \, +
\nonumber \\ & + &
M_1^2 \, M_4^2  \, + \,  2 \, M_2^2 \, M_4^2  \, +
\,  M_3^2 \, M_4^2  \, - \, 2 \, M_1^2 \, s  \, - \,  2 \, M_2^2 \, s  \, +
\nonumber \\ & - &
2 \, M_3^2 \, s  \, - \,  2 \, M_4^2 \, s  \, + \,  2 \, s^2  \, - \, M_1^2
\, t  \, - \, M_2^2 \, t  \, +
\nonumber \\ & - &
M_3^2 \, t  \, - \,  2 \, M_2 \, M_4 \, t  \, - \,  M_4^2 \, t  \, + \,  2
\, s \, t  \, + \, t^2)  \, +
\nonumber \\
& + & \cv^2 \, (M_1^2 \, M_2^2  \, + \,  2 \, M_1^2 \, M_3^2  \, + \, M_2^2
\, M_3^2  \, - \,  2 \, M_1^2 \, M_2 \, M_4  \, - \, 2 \, M_2 \, M_3^2
\, M_4 \, +
\nonumber \\ & + &
M_1^2 \, M_4^2  \, + \,  2 \, M_2^2 \, M_4^2  \, + \, M_3^2 \, M_4^2  \, -
\, 2 \, M_1^2 \, s  \, - \,  2 \, M_2^2 \, s  \, +
\nonumber \\ & - &
2 \, M_3^2 \, s
\, - \,  2 \, M_4^2 \, s  \, + \,  2 \, s^2  \, - \,  M_1^2 \, t  \, - \,
M_2^2 \, t  \, +
\nonumber \\ & - &
M_3^2 \, t  \, + \,  2 \, M_2 \, M_4 \, t  \, - \, M_4^2 \, t  \, + \,  2
\, s \, t  \, + \,  t^2)  \, +
\nonumber \\
& + & 2 \,  \ca \, f_{ps} \, ( \, - \, M_1^2 \, M_2 \, M_3^2  \, + \, M_2^3
\, M_3^2  \, + \, M_2 \, M_3^4  \, + \,  M_1^4 \, M_4
\, - \, M_1^2 \, M_3^2 \, M_4  \, +
\nonumber \\ & - &
M_2^2 \, M_3^2 \, M_4  \, - \, M_1^2 \, M_2 \, M_4^2  \, + \, M_1^2 \,
M_4^3  \, + \,  M_1^2 \, M_2 \, s
\, - \,  M_2 \, M_3^2 \, s  \, +
\nonumber \\ & - &
M_1^2 \, M_4 \, s  \, + \, M_3^2 \, M_4 \, s  \, - \,  M_2 \, M_3^2 \, t
\, - \,  M_1^2 \, M_4 \, t)
\, +
\nonumber \\
& - & 2 \, \ca \,  \frac{f_2}{M_N} \, ( M_1^2 \, M_2^3  \, - \,  M_2^3 \,
M_3^2 \, +
\,  M_1^2 \, M_2^2 \, M_4 \, - \,  M_2^2 \, M_3^2 \, M_4  \, - \,  M_1^2 \,
M_2 \, M_4^2  \, +
\nonumber \\ & + &
M_2 \, M_3^2 \, M_4^2  \, - \,  M_1^2 \, M_4^3  \, +
\, M_3^2 \, M_4^3  \, - \,  M_1^2 \, M_2 \, t  \, - \,  M_2^3 \, t  \, +
\nonumber \\ & - &
M_2 \, M_3^2 \, t \, - \,  M_1^2 \, M_4 \, t \, - \,  M_2^2 \, M_4 \, t  \,
- \,  M_3^2 \, M_4 \, t  \, - \,  M_2 \, M_4^2 \, t  \, +
\nonumber \\ & - &
M_4^3 \, t
\, + \,  2 \, M_2 \, s \, t  \, + \,  2 \, M_4 \, s \, t
\, + \,  M_2 \, t^2  \, + \,  M_4 \, t^2)   \, +
\nonumber \\
& + & 2 \, \cv \, f_3 \,( \,
- \, M_1^2 \, M_2 \, M_3^2  \, + \, M_2^3 \, M_3^2  \, + \,  M_2 \,
M_3^4  \, - \,  M_1^4 \, M_4  \, + \, M_1^2 \, M_3^2 \, M_4  \, +
\nonumber \\ & + &
M_2^2 \, M_3^2 \, M_4 \, - \,  M_1^2 \, M_2 \, M_4^2 \, - \,  M_1^2 \,
M_4^3 \, + \,  M_1^2 \, M_2 \, s  \, - \,  M_2 \, M_3^2
\, s  \, +
\nonumber \\ & + &
M_1^2 \, M_4 \, s  \, - \,  M_3^2 \, M_4 \, s  \, - \, M_2 \, M_3^2 \, t
\, + \,  M_1^2 \, M_4 \, t)   \, +
\nonumber \\
& + & 2 \,  \cv \, \ca \, ( \, - \, M_1^2 \, M_2^2  \, + \,  M_2^2 \, M_3^2
\, +
\, M_1^2 \, M_4^2 \, - \,  M_3^2 \, M_4^2  \, + \, M_1^2 \, t  \, +
\nonumber \\ & + &
M_2^2 \, t  \, + \, M_3^2 \, t  \, + \,  M_4^2 \, t
\, - \, 2 \, s \, t \, - \,  t^2)  \, +
\nonumber \\
& + & 2 \,  \cv \, \frac{f_2}{M_N} \, (M_1^2 \, M_2^3  \, + \,  M_1^2 \,
M_2 \, M_3^2 \, - \,  M_2 \, M_3^4  \, - \, M_1^4
\, M_4  \, - \,  M_1^2 \, M_2^2 \, M_4 \, +
\nonumber \\ & + &
M_1^2 \, M_3^2 \, M_4
\, - \, M_2 \, M_3^2 \, M_4^2  \, + \,  M_3^2 \, M_4^3  \, - \,  M_1^2
\, M_2 \, s \, + \, M_2 \, M_3^2 \, s  \, +
\nonumber \\ & + &
M_1^2 \, M_4 \, s  \, - \, M_3^2 \, M_4 \, s \, - \, M_1^2 \, M_2 \, t  \,
- \,  M_2^3 \, t  \, + \, M_2^2 \, M_4 \, t  \, +
\nonumber \\ & - &
M_3^2 \, M_4 \, t  \, + \,  M_2 \, M_4^2 \, t
\, - \,  M_4^3 \, t  \, + \, M_2 \, t^2  \, + \,  M_4 \, t^2) )
\label{cm3}
\end{eqnarray}
In the above formula, neutrino mass has not been set to zero to give a
general expression holding for all reactions. For the two-body scattering
processes in \reac it is
\begin{eqnarray*}
 s & = & \left( p_1 \, + \, p_2 \right)^2 \; = \;
 \left( p_3 \, + \, p_4 \right)^2 ~~~ , \\
 t & = & \left( p_1 \, - \, p_3 \right)^2 \; = \;
 \left( p_2 \, - \, p_4 \right)^2 ~~~ .
\end{eqnarray*}
As in section (\ref{c1a}), $p_1$ and $p_3$ are the initial and final lepton
momentum while $p_2$ and $p_4$ are the initial and final nucleon momentum,
respectively. Instead for $n \rt p \, e^- \, \neb$ it is
\begin{eqnarray*}
 s & = & \left( - \, p_1 \, + \, p_2 \right)^2 \; = \;
 \left( p_3 \, + \, p_4 \right)^2 ~~~ , \\
 t & = & \left( - \, p_1 \, - \, p_3 \right)^2 \; = \;
 \left( - \, p_2 \, + \, p_4 \right)^2 ~~~ .
\end{eqnarray*}
while for $p \, e^- \, \neb \rt n$ it is
\begin{eqnarray*}
 s & = & \left( p_1 \, + \, p_2 \right)^2 \; = \;
 \left( - \, p_3 \, + \, p_4 \right)^2 \\
 t & = & \left( p_1 \, + \, p_3 \right)^2 \; = \;
 \left( - \, p_2 \, + \, p_4 \right)^2
\end{eqnarray*}

\chapter{Finite temperature and density QED}
\label{aftd}

\section{The Real Time Formalism}
\label{aftd1}

At finite temperatures, physical processes take place in a heat bath
consisting of a background of a plasma of particles and antiparticles. All
interaction processes must be therefore considered as occurring in a
thermal ``vacuum" state, which does not coincide with the usual Poincar\'{e}
invariant vacuum of zero temperature quantum field theory. Manifest Lorentz
invariance therefore breaks down due to the choice of the preferred frame
of the bath, and the zero temperature renormalization prescription cannot
be applied straightforwardly. The covariant formalism that enables to
consider the interactions of a given particle with the surrounding medium
is the Finite Temperature and Density Quantum Field Theory \cite{dolan,
ftft1, ftft2}. Here we adopt the Real Time formulation of this theory, in
which the Feynman rules for the vertices are identical to the corresponding
ones in the vacuum. The effect of the temperature and of the density is
taken into account in the expressions of the free particle propagators. For
fermions and bosons we have respectively
\bea
S_F(p) & = & \left( {\not{p}} \, + \, m \right) \, \left(
\frac{1}{p^2 \, - \, m^2} \; + \; i \, \Gamma_F(p) \right)
\label{f1} ~~~,  \\
D_{\mu \nu} (p) & = & - \, g_{\mu \nu} \, \left(
\frac{1}{p^2 \, - \, m^2} \; - \; i \, \Gamma_B(p) \right)
\label{f2} ~~~,
\eea
where
\begin{equation}\label{f3}
  \Gamma(p) \; = \; 2 \pi \, \delta(p^2 \, - \, m^2) \,
  \left( \theta(p {\cdot} u) \, n(p) \; + \; \theta(- p {\cdot} u) \, \ov{n}(p)
  \right) ~~~,
\end{equation}
and
\bea
n_F(p) & = & \frac{1}{e^{\beta ( |p{\cdot}u| - \mu )} \, + \, 1}
\label{f4} ~~~, \\
n_B(p) & = & \frac{1}{e^{\beta ( |p{\cdot}u| - \mu )} \, - \, 1}
\label{f5} ~~~,
\eea
($\beta = 1/T$) are the Fermi-Dirac and Bose-Einstein distribution
functions ($\ov{n}_F$ and $\ov{n}_B$ are the distribution for the
antiparticles, obtained replacing $\mu$ with $- \mu$). Note that in a
plasma, another 4-vector must be considered, the 4-velocity $u^\mu$ of the
medium, which in its rest frame is given by $u^\mu = (1, \bvec{0})$. \\ The
additional contributions in the above formulas (denoted with
$\Gamma_{F,B}$) pick up {\it real} particles through the mass shell
$\delta$-functions and are proportional to the particle densities in the
thermal bath. They take into account the role played by the medium in the
single particle propagation. \\ The heat bath we are interested in is the
primordial plasma at the epoch of BBN, i.e. at temperature around 1 $MeV$.
At these temperatures the $W^{\pm}$ and $Z^0$  gauge bosons degrees of freedom
are not excited, so we can safely neglect the temperature dependent term
$\Gamma_B(p)$ in $D_{\mu \nu}$ for these bosons. \\ Our primordial plasma
consists of nucleon, $e^{\pm}$ pairs, photons and neutrinos. The temperature
dependent term in the propagator of nucleons would be suppressed by a
Boltzmann factor smaller than $\exp(-100)$, compared with the ones for the
other particles, in the temperature range relevant for BBN, so it will be
neglected in the following as well. \\ The most important finite
temperature effects induced by the remaining particles in the heat bath
come from the QED interactions of $e^{\pm}$ and photons, which are of order
$\alpha$. Neutrinos only interact weakly, thus their contribution is at
least of order $G_F$.
\\ In this appendix we calculate the $e^{\pm}$ self-energy and wavefunction
renormalization at finite temperature and photon self-energy, which are
relevant for the thermal radiative corrections considered in chapter
\ref{c3}.

\section{Electron self-energy}
\label{aftd2}

\begin{figure}
\epsfysize=3.5cm
\epsfxsize=10.0cm
\centerline{\epsffile{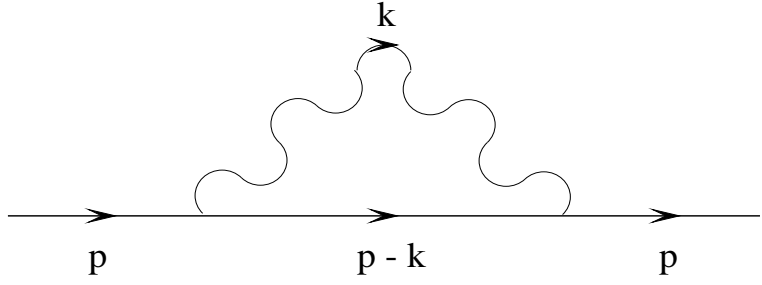}}
\vskip 0.5cm
\caption{Self-energy diagram for an electron in a plasma.}
\label{eself}
\end{figure}

The Feynman diagram for electron self-energy $\Sigma$ is reported in Figure
\ref{eself}; it gives
\begin{equation}\label{cr1}
  - i \, \Sigma \; = \; \int \frac{d^4 k}{(2 \pi)^4} \,
  \left( i \, e \, \gamma^\mu \right) \, i \, S_F(p-k) \,
  \left( i \, e \, \gamma^\nu \right) \, i \, D_{\mu \nu} (k)
\end{equation}
where $S_F(p-k)$ and $D_{\mu \nu} (k)$ are the (RTF) Fermi and Bose
propagators in a plasma, which are reported in (\ref{f1}), (\ref{f2}). Here
we are only interested in the temperature dependent part $\Sigma_T$ of the
self-energy; moreover, we neglect possible damping effects (which are
suppressed with respect to coherent ones) and then consider only its real
part:
\begin{eqnarray}
  \Re (\Sigma_T) \; = \; \Re (\Sigma_T^B) \; + \; \Re (\Sigma_T^F) & = &
  2 \, e^2 \, \int \frac{d^4 k}{(2 \pi)^4} \, \left(
  \left( \not{p} \, - \, \not{k} \, - \, 2 m_e \right) \,
  \frac{\Gamma_B(k)}{(p-k)^2 \, - \, m_e^2} \; + \right. \nonumber \\
  &-& \left. \left( \not{k} \,
  - \, 2 m_e \right) \, \frac{\Gamma_F(k)}{(p-k)^2} \right)
\label{cr2}
\end{eqnarray}
From Lorentz invariance, it follows that the self-energy has the general
form
\begin{equation}\label{cr3}
  \Re \Sigma_T \; = \; a \, \not{p} \; + \; b \, \not{u} \; + \; c
\end{equation}
where $a,b,c$ are functions of the only two invariants $p^2$,$p {\cdot} u$
(obviously $u^2 = 1$). In the following, when necessary, we shall adopt the
plasma rest frame, where $u^\mu = (1,\bvec{0})$. We then have
\begin{eqnarray}
  a &=& - \, \frac{1}{{\bvec{p}}^2} \left( T_p \, - \, p_0 \, T_u \right)
  \label{cr4} \\
  b &=& - \, \frac{1}{{\bvec{p}}^2} \left( (p_0^2 \, - \, {{\bvec{p}}^2} \,
  T_u \, - \, p_0 \, T_p \right)
  \label{cr5} \\
  c &=& \frac{1}{4} \, Tr \, \Re \Sigma_T  \label{cr6}
\end{eqnarray}
where
\begin{eqnarray}
  T_p &=& \frac{1}{4} \, Tr \left( \not{p} \, \Re \Sigma_T \right)
  \label{cr7} \\
  T_u &=& \frac{1}{4} \, Tr \left( \not{u} \, \Re \Sigma_T \right)
  \label{cr8}
\end{eqnarray}
For the Bose parts, after some calculations we obtain ($p = |\bvec{p}|$,
 $k = |\bvec{k}|$)
\begin{eqnarray}
  T_p^B &=& - \frac{\alpha}{2 \pi} \int d k \, k \, B(k) \,
  \left( - 4 \, + \, \frac{\eta^2 + 2 m_e^2}{2 k} \, \left(
  I_1(p_0 , p , k)  +  I_1(- p_0 , p , k) \right) \right)
  \label{cr9} \\
  & \left. \right. & \left. \right. \nonumber \\
  T_u^B &=& - \, \frac{\alpha}{2 \pi} \int d k \, B(k) \,
  \left( (p_0 \, - \, k) \, I_1(p_0 , p , k) \, +
  p_0 \, + \, k) \, I_1(- p_0 , p , k) \right)
  \label{cr10} \\
  & \left. \right. & \left. \right. \nonumber \\
  c_B &=& \frac{\alpha \, m_e}{\pi} \int d k \, B(k) \,
  \left( I_1(p_0 , p , k) \, + \, I_1(- p_0 , p , k) \right)
  \label{cr11}
\end{eqnarray}
where $\eta \, = \, p_0^2 \, - \, {\bvec{p}}^2 \, - m_e^2$. Instead for the
Fermi parts we get:
\begin{eqnarray}
  T_p^F &=& - \, \frac{\alpha}{2 \pi} \int d k \, \frac{k^2}{E} \,
  F(E) \, \left(
  I_2(p_0 , p , k) \, + \, I_2(- p_0 , p , k) \right)
  \label{cr12} \\
  & \left. \right. & \left. \right. \nonumber \\
  T_u^F &=& - \, \frac{\alpha}{2 \pi} \int d k \, k^2 \, F(E) \,
  \left( I_3(p_0 , p , k) \, - \, I_3(- p_0 , p , k) \right)
  \label{cr13} \\
  & \left. \right. & \left. \right. \nonumber \\
  c_F &=& \frac{\alpha \, m_e}{\pi} \int d k \, \frac{k^2}{E} \,
  F(E) \, \left( I_3(p_0 , p , k) \, + \, I_3(- p_0 , p , k) \right)
  \label{cr14}
\end{eqnarray}
with $E \, = \, \sqrt{{\bvec{p}}^2 \, + \, m_e^2}$. The quantities $I_i$
are defined by \footnote{We have chosen a reference frame in which the
electron momentum lies along the z axis ($x \, = \, \cos \, \theta$).}
\begin{eqnarray}
  I_1(p_0 , p , k) &=& \int^1_{-1} \frac{d x}{p_0 \, - \, \frac{\eta}{2 k}
  \, - \, p x} \; = \; \frac{1}{p} \, \ln \,
  \frac{\left| p_0 \, - \, \frac{\eta}{2k} \, + \, p \right|}
  {\left| p_0 \, - \, \frac{\eta}{2k} \, - \, p \right|} \\
  & \left. \right. & \left. \right. \nonumber \\
  I_2(p_0 , p , k) &=& \int^1_{-1} \, d x \,
  \frac{2 (p_0 E \, - \, p k x)}{\eta \, + \, 2 m_e^2 \, - \, 2 p_0 E \,
  + \, 2 p k x} \; = \nonumber \\
  & = & \frac{\eta \, + \, 2 m_e^2}{2 p k} \, \ln \,
  \frac{\left| \eta \, + \, 2 m_e^2 \, - \, 2 p_0 E \, + \, 2 p k \right|}
  {\left| \eta \, + \, 2 m_e^2 \, - \, 2 p_0 E \, - \, 2 p k \right|} \\
  & \left. \right. & \left. \right. \nonumber \\
  I_3(p_0 , p , k) &=& \int^1_{-1}
  \frac{2 \, d x }{\eta \, + \, 2 m_e^2 \, - \, 2 p_0 E \,
  + \, 2 p k x} \; = \nonumber \\
  & = &  \frac{1}{p k} \, \ln \,
  \frac{\left| \eta \, + \, 2 m_e^2 \, - \, 2 p_0 E \, + \, 2 p k \right|}
  {\left| \eta \, + \, 2 m_e^2 \, - \, 2 p_0 E \, - \, 2 p k \right|}
\end{eqnarray}
In the above formulas we have treated $p_0$ and $\bvec{p}$ as two
independent variables. Since the expressions in (\ref{cr9})-(\ref{cr14})
are already of order $\alpha$, in these we can substitute the vacuum
dispersion relation $p_0^2 \, = \, {\bvec{p}}^2 \, + \, m_e^2$, thus
obtaining the following results for the coefficients $a,b,c$
\begin{eqnarray}
  a & = & \frac{\alpha \, \pi \, T^2}{3 \, p^2} \, \left( - 1 \, + \,
  \frac{\omega}{2p} \, \ln \, \frac{\omega \, + \, p}{\omega \, - \, p}
  \right) \; + \nonumber \\
  & + & \frac{\alpha}{2 \pi p^2} \int \, d k \, \frac{k^2}{E} \, F(E) \,
  \left( - 4 \, + \, \frac{\omega E \, + \, m_e^2}{p k} \, \ln \, A \,
  + \, \frac{\omega E \, - \, m_e^2}{p k} \, \ln \, B \right)
\label{cr15} ~~~, \\
  & \left. \right. & \left. \right. \nonumber \\
  b & = & \frac{\alpha \, \pi \, T^2}{3 \, p^2} \, \omega \,
  \left( 1 \, - \,
  \frac{m_e^2}{2 \omega p} \, \ln \, \frac{\omega \, + \, p}{\omega \, - \, p}
  \right) \; + \nonumber \\
  & - & \frac{\alpha \, \om}{2 \pi p^2}
  \int  d k \, \frac{k^2}{E} \, F(E) \,
  \left( - 4 \, + \, \frac{m_e^2}{p k} \left( \frac{E \, + \omega}{\omega}
  \, \ln \, A \, + \, \frac{E \, - \omega}{\omega}
  \, \ln \, B \right) \! \right)
\label{cr16} ~~~, \\
  & \left. \right. & \left. \right. \nonumber \\
  c & = & \frac{\alpha \, m_e}{\pi \, p} \, \int \, d k \, \frac{k}{E} \,
  F(E) \, \left( \ln \, A \; - \; \ln \, B \right) \label{cr17} ~~~,
\end{eqnarray}
\begin{equation}
  A \; = \; \frac{\omega E \, + \, m_e^2 \, + p k}
  {\omega E \, + \, m_e^2 \, - p k} \;\;\;\;\;\;\;
  , \;\;\;\;\;\;\;\;
  B \; = \; \frac{\omega E \, - \, m_e^2 \, + p k}
  {\omega E \, - \, m_e^2 \, - p k} ~~~,
\label{cicci}
\end{equation}
with $\omega \, = \, \sqrt{{\bvec{p}}^2 \, + \, m_e^2}$.

From the calculated electron self-energy, we can now proceed to evaluate
the mass shift correction.

If we denote with $\psi$ the wavefunction of the electron in the plasma, it
satisfies the correct Dirac equation
\begin{equation}\label{cr18}
  \left( \not{p} \; - \; m_e \; - \; \Sigma \right) \, \psi \; = \; 0 ~~~,
\end{equation}
with $\Sigma$ in (\ref{cr3}). We can rewrite this in the form
\begin{equation}\label{cr19}
  \left( \not{\tilde p} \; - \; {\tilde m} \right) \, \psi \; = \; 0 ~~~,
\end{equation}
with
\begin{eqnarray}
  {\tilde p}_\mu &=& (1 \, - \, a) \, p_\mu \; - \; b \, u_\mu
  \label{cr20} \\
  {\tilde m} &=& m_e \; + \; c  \label{cr21}
\end{eqnarray}
By left-multiplying Eq. (\ref{cr19}) by $\not{\tilde p} \, + \, {\tilde m}$
we then deduce the correct dispersion relation for an electron in a medium
\begin{equation}\label{cr22}
  {\tilde p}^2 \; - \; {\tilde m}^2 \; = \; 0 ~~~,
\end{equation}
or
\begin{equation}\label{cr23}
  p_0^2 \; - \; {\bvec{p}}^2  \;  - \; \frac{2b}{1 - a} \, p_0 \; + \;
  \frac{b^2 \,  - \, (m_e \, + \, c)^2}{(1 - a)^2} \; = \; 0 ~~~.
\end{equation}
At first order in $\alpha$ we finally obtain the energy shift
\begin{equation}\label{cr24}
  p_0 \; \simeq \; \sqrt{p^2 + m_e^2} \; + \; \left( \ac \,
  \frac{m_e^2}{\omega} \; + \; \bc \; + \; \cc \, \frac{m_e}{\omega}
  \right) \; = \; \om \; + \; \mu ~~~,
\end{equation}
where $\ac$, $\bc$, $\cc$ are $a$, $b$, $c$ evaluated in
 $p_0 \, = \,\om$.

\section{Wavefunction renormalization}
\label{aftd3}

In general, the Dirac spinors corresponding to a particle propagating in a
plasma are different from the ones travelling in vacuum. This results in
effective temperature dependent energy projection operators $\Lambda^{\pm}_R$
which replace the usual vacuum ones $\Lambda^{\pm}_0$ in the computation of
spin summed squared amplitudes
\begin{equation}\label{cr28}
  \Lambda^+_0 \; = \; \frac{\not{p} \, + \, m_e}{2 \om} \; \longrightarrow
  \; \Lambda^+_R ~~~,
\end{equation}
and similarly for the negative energy projector. Thus, we first have to
calculate the effective energy projection operators and then evaluate the
corresponding corrections for the decay rates for the reaction in \reac.

An electron propagating through a plasma is described by the equation of
motion in (\ref{cr18}), so that the field propagator is given by
\begin{equation}\label{cr29}
  G \; = \; \frac{1}{\not{p} \, - \, m_e \, - \, \Sigma} \; = \;
  \frac{1}{\not{\tilde{p}} \, - \tilde{m}} \; = \;
  \frac{\not{\tilde{p}} \, + \tilde{m}}{\tilde{p}^2 \, - \, \tilde{m}^2}
  ~~~.
\end{equation}
We identify the particle states as corresponding to the energy poles in the
propagator; the wavefunction renormalization can then be read off by
evaluating the residue of $G$ at the pole. For definiteness, let us focus
on the positive energy projection operator; a similar procedure will hold
for the negative energy one. If we expand $G$ in (\ref{cr29}) around the
positive energy pole, say $\omr$, we obtain
\beas
  \left( \tilde{p}^2 \; - \; \tilde{m}^2 \right)^{-1} & = &
  \left( p^2 \, - \, m_e^2 \, - \, 2 \, a \, p^2 \, -  \, 2 \, b \, p {\cdot} u
  \, - \, 2 \, c \, m \right)^{-1} \; \simeq  \\
  & \simeq & \left\{ p_0^2 \, - \, \om^2 \, - \, 2 \, \left( \ac \, + \,
  \apc \, (p_0 - \omr) \right) (p_0^2 - {\bvec{p}}^2) \, +
  \right. \nonumber \\
  &+&  \left. 2 \, \left( \bc \, + \, \bpc \, (p_0 - \omr) \right) \,
  p {\cdot} u \, - \,
  2 \, m \, \left( \cc \, + \, \cpc \, (p_0 - \omr) \right) \right\}^{-1}
  \; \simeq \\
  & \simeq & \left\{ p_0^2 \, - \, \om^2 \, - \, 2 \, \ac \, p_0^2 \, -
  \, 2 \, \apc \, \om^2 \, (p_0 - \omr) \, +
  \right. \nonumber \\
  &+&  2 \, {\bvec{p}}^2 \,
  \left( \ac \, + \, \apc \, (p_0 - \omr) \right) \, - \, 2 \, \bc \,
  p {\cdot} u \, - \, 2 \, \bpc \, \hat{p} {\cdot} u \, (p_0 - \omr) \, +
  \nonumber \\
  &-& \left. 2 \, m
  \, \left( \cc \, + \, \cpc \, (p_0 - \omr) \right) \right\}^{-1} ~~~.
\eeas
Here we have used the ``hat" to denote that a given quantity is evaluated
at $p_0 = \omr$, while a ``prime" is used to indicate differentiation with
respect to $p_0$. Note that throughout the entire calculation we retain
only terms which are of the first order in $\alpha$. Substituting the
relation $\bc \, p {\cdot} u \, \simeq \, \bc \, \hat{p} {\cdot} u \, + \, \bc \, u_0
\, (p_0 - \omr)$
 in the above formula, we get
\beas
  \left( \tilde{p}^2 \; - \; \tilde{m}^2 \right)^{-1} & = &
  \left\{ (1 - 2 \ac) \, \left[ p_0^2 \, - \, \left( \om^2 \, + \,
  2 a \, m_e^2 \, + \, 2 \bc \, \hat{p} {\cdot} u \, + \, 2 \cc \, m \right)
  \right] \; + \right. \\
  & - & \left. (p_0 - \omr) \, \left[ 2 \bc \, u_0 \, + \, 2 \apc \,
  m_e^2 \, + \, 2 \bpc \, \hat{p} {\cdot} u  \, + \, 2 \cpc \, m \right]
  \right\}^{-1} ~~~.
\eeas
By requiring that $G$ (and then
 $\left( \tilde{p}^2 \; - \; \tilde{m}^2 \right)^{-1}$) has a pole for
$p_0 = \omr$ we deduce that
\begin{equation}
  \omr^2 \; = \; \om^2 \, + \,
  2 a \, m_e^2 \, + \, 2 \bc \, \hat{p} {\cdot} u \, + \, 2 \cc \, m
  \nonumber ~~~,
\end{equation}
which exactly corresponds to the energy shift in (\ref{cr24}). Around
 $p_0 = \omr$ we then obtain
\begin{equation}
  \left( \tilde{p}^2 \; - \; \tilde{m}^2 \right)^{-1} \; \simeq \;
  \frac{1}{p_0 - \omr} \, \frac{1}{2 \omr} \left( 1 \, + \, 2 \ac \, + \,
  \frac{\bc}{\om} \, u_0 \, + \, \apc \, \frac{m_e^2}{\om} \, + \,
  \bpc \, \frac{\hat{p} {\cdot} u}{\om} \, + \, \cpc \, \frac{m_e}{\om} \right)
  \nonumber ~~~,
\end{equation}
thus the expansion around the positive energy pole $p_0 = \omr$ of the
propagator $G$ in (\ref{cr29}) finally gives
\begin{eqnarray}\label{cr30}
  G & \simeq & \frac{1}{p_0 - \omr} \, \left\{ \left(
  1 \, + \, \ac \, + \,
  \frac{\bc}{\om} \, u_0 \, + \, \apc \, \frac{m_e^2}{\om} \, + \,
  \bpc \, \frac{\hat{p} {\cdot} u}{\om} \, + \, \cpc \, \frac{m_e}{\om} \right)
  \, \frac{\not{\hat{p}} + m_R}{2 \omr} \; + \right. \nonumber \\
  &-& \left. \frac{\bc}{2 \om} \,
  \left( \not{u} \, + \, \frac{\om}{m_e} \right) \right\} ~~~.
\end{eqnarray}
Here, we also report an alternative writing of the same expression
\begin{eqnarray}\label{cr31}
  G & \simeq & \frac{1}{p_0 - \omr} \, \left\{ \left(
  1 \, + \, \ac \, + \,
  \frac{\bc}{\om} \, (u_0 - 1) \, + \, \apc \, \frac{m_e^2}{\om} \, + \,
  \bpc \, \frac{\hat{p} {\cdot} u}{\om} \, + \, \cpc \, \frac{m_e}{\om} \right)
  \, \frac{\not{\hat{p}} + m_R}{2 \omr} \, + \right. \nonumber \\
  &-& \left. \frac{\bc}{2 \om^2} \,
  \left( \left( \om \, \not{u} \, - \, \not{p} \right) \, + \,
  \frac{{\bvec{p}}^2}{m_e} \right) \right\}
\end{eqnarray}
The quantity $m_R$ is the effective mass corresponding to the effective
energy $\omr$ in (\ref{cr24})
\begin{equation}\label{cr32}
  m_R \; \simeq \; m_e \, + \, \delta m \; \simeq \; m_e \, + \,
  \ac \, m_e \, + \, \bc \, \frac{\om}{m_e} \, + \, \cc ~~~.
\end{equation}
In the particularly relevant case in which $u_\mu = (1, \bvec{0})$ (plasma
rest frame) we obtain
\begin{equation}\label{cr33}
  G \; \simeq \; \frac{1}{p_0 - \omr} \, \left\{ \left(
  1 \, + \, \ac \, + \,
  \apc \, \frac{m_e^2}{\om} \, + \,
  \bpc \, + \, \cpc \, \frac{m_e}{\om} \right)
  \, \frac{\not{\hat{p}} + m_R}{2 \omr} \, - \, \frac{\bc}{2 \om^2} \,
  \left( {\bvec{p}} {\cdot} {\bvec{\gamma}}  \, + \,
  \frac{{\bvec{p}}^2}{m_e} \right) \right\}
\end{equation}
The positive energy projection operator  is then simply the residue of $G$
at the corresponding pole; for example, in the plasma rest frame we have
\begin{equation}\label{cr34}
  \Lambda^+_R \; = \; \left( 1 \; + \; \lambda \right) \,
  \frac{\not{\hat{p}} \, + \, m_R}{2 \omr} \; - \; \frac{\bc}{2 \om^2}
  \left( {\bvec{p}} {\cdot} {\bvec{\gamma}} \; + \; \frac{{\bvec{p}}^2}{m_e}
  \right) ~~~,
\end{equation}
with
\begin{equation}\label{cr35}
  \lambda \; = \; \ac \; + \; \apc \, \frac{m_e^2}{\om} \; + \; \bpc
  \; + \; \cpc \, \frac{m_e}{\om} ~~~.
\end{equation}
For the negative energy projector, one has simply to make the substitutions
$\om \rt - \om$, $ \omr \rt - \omr$, $\bvec{p} \rt - \bvec{p}$ in all the
quantities appearing in (\ref{cr34}), (\ref{cr35}) (in particular these
substitutions have to be made also in the expressions for $\ac$, $\apc$,
...).

Some considerations are now in order. We mainly note that differently from
the usual vacuum case, wavefunction renormalization at finite temperature
is not simply given by a multiplicative renormalization factor (in the
vacuum case $\Lambda_R \, = \, (1 + Z_2) \, \Lambda_0$), but also
introduces an additional term proportional to $\bc$ and with a different
spinorial structure. This is a pure matter effect, which is given by the
presence of the medium 4-velocity $u_\mu$. This dependence is, however,
also present in the multiplicative renormalization factor (the first term
between $\{ ... \}$ in (\ref{cr30}) or (\ref{cr31}) or in $\lambda$ in
(\ref{cr34})); we point out that, in the plasma rest frame, the peculiar
medium dependence of $\lambda$ only occurs through the energy derivative of
$b$, differently to what happens in the general case in which also a term
proportional to $b$ is present (see the first term in (\ref{cr31})). One
can easily check that in the limiting case of no medium, we recover the
usual expression for the renormalized projection operator. We note,
however, that this is not true \cite{EMMP} for the result quoted in
\cite{Sawyer}, although an approach similar to ours is used. Other
different approaches to the problem of wavefunction renormalization have
been proposed in literature \cite{Donoghue}-\cite{Kobes}, but the final
results striking differ. For example, in \cite{Donoghue} the authors start
introducing finite temperature spinors chosen as the solution of the
(nonlinear) Dirac equation (\ref{cr18}) whose corresponding creation and
annihilation operators are assumed to satisfy ordinary, zero temperature,
anticommutation relations. Expanding the propagator in terms of these
spinors they obtain a wavefunction renormalization factor which is only
multiplicative. In particular, this multiplicative factor is exactly the
same as our factor $(1 \, + \, \lambda)$ in (\ref{cr34}), but the
additional term proportional to ${\hat{b}}$ is absent. The essential
difference with our approach is the assumption made in \cite{Donoghue} on
the canonical spinor basis to be used; here, however, we do not make any
hypothesis on the renormalized field to be used, but simply recover the
particle content and the corresponding projector operators from the poles
of $G$ and their residues, respectively. Comparative analysis of different
approaches have been carried out in Refs. \cite{Chapman, EMMP}.

\section{Photon self-energy}
\label{aftd4}

\begin{figure}
\epsfysize=5.0cm
\epsfxsize=10.0cm
\centerline{\epsffile{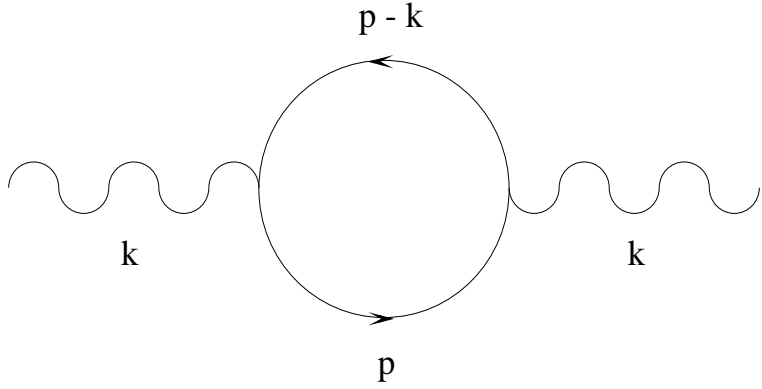}}
\caption{Self-energy diagram for a photon in a plasma.}
\label{pself}
\end{figure}

The Feynman diagrams for photon self-energy $\Pi_{\mu \nu}$ is shown in
Figure \ref{pself}; it gives
\begin{equation}\label{f10}
  i \, \Pi_{\mu \nu} \; = \; - \,  \int \frac{d^4 p}{(2 \pi)^4} \,
  Tr \left\{ \left( i \, e \, \gamma^\mu \right) \, i \, S_F(p) \,
  \left( i \, e \, \gamma^\nu \right) \, i \, S_F(p-k) \right\} ~~~,
\end{equation}
where again $S_F(p)$ is the electron propagator in the plasma given by
(\ref{f1}). As for the electron self-energy, we are interested only in the
real part of $\Pi_{\mu \nu}$ which, after some algebra, is given by
\bea
\Re (\Pi_{\mu \nu}) & = & - 2 e^2 \, \int \frac{d^3 p}{(2 \pi)^3} \,
\frac{1}{2E} \, \left( \frac{2 p_\mu p_\nu \, - \, \p_\mu k_\nu \, - \,
p_\nu k_\mu \, + \, p{\cdot}k g_{\mu \nu}}{k^2 \, - \, 2 p{\cdot}k} \, + \right.
\nonumber \\
& + & \left. \frac{2 p_\mu p_\nu \, + \, \p_\mu k_\nu \, + \, p_\nu k_\mu
\, - \, p{\cdot}k g_{\mu \nu}}{k^2 \, + \, 2 p{\cdot}k} \right) \, \left( n_e \, + \,
\ov{n}_e \right) \label{f11}
\eea
where $E = \sqrt{{\bvec{p}}^2 + m_e^2}$. It is now convenient to express
$\Pi_{\mu \nu}$ in terms of form factors by using Lorentz and gauge
invariance, which give the following constraints
\begin{equation}\label{f12}
  k^\mu \, \Pi_{\mu \nu} \; = \; k^\nu \, \Pi_{\mu \nu} \; = \; 0 ~~~.
\end{equation}
In presence of a medium, $\Pi_{\mu \nu}$ may be expressed in terms of the
tensor $g_{\mu \nu}$ and the 4-vectors $k_\mu$ and $u_\mu$. Then, noting
that
\begin{equation}
  \tilde{g}_{\mu \nu} \; \equiv \; g_{\mu \nu} \; - \; \frac{k_\mu k_\nu}
  {k^2} \nonumber ~~~,
\end{equation}
and
\begin{equation}
  \frac{\tilde{u}_\mu \tilde{u}_\nu}{\tilde{u}^2} ~~~, \nonumber
\end{equation}
with $\tilde{u}_\mu = \tilde{g}_{\mu \nu} u^\nu$, are the only two tensors
orthogonal to $k_\mu$ that can be built with $g_{\mu \nu}$, $k_\mu$ and
$u_\mu$, the most general form of $\Pi_{\mu \nu}$ (assuming parity
conservation) results
\begin{equation}\label{f13}
  \Pi_{\mu \nu} \; = \; \Pi_T \, R_{\mu \nu} \; + \; \Pi_L \, Q_{\mu \nu}
~~~,
\end{equation}
with
\begin{equation}\label{f14}
  R_{\mu \nu} \; = \; \tilde{g}_{\mu \nu} \; - \,
  \frac{\tilde{u}_\mu \tilde{u}_\nu}{\tilde{u}^2}
  ~~~~~~~~~~~ , ~~~~~~~~~~~~~~
  Q_{\mu \nu} \; = \; \frac{\tilde{u}_\mu \tilde{u}_\nu}{\tilde{u}^2}
  ~~~.
\end{equation}
The quantities $\Pi_T$, $\Pi_L$ are scalar functions of the only two
invariants $k^2$, $k{\cdot}u$ and in terms of $\Pi_{\mu \nu}$ are given by
\bea
\Pi_L & = & Q^{\mu \nu} \, \Pi_{\mu \nu} \label{f15} ~~~, \\
2 \Pi_t & = & R^{\mu \nu} \, \Pi_{\mu \nu} \label{f16} ~~~.
\eea
It is now simple to explicitly evaluate these form factors from
(\ref{f11}); at first order in $\alpha$, substituting the vacuum dispersion
relation $\omega^2 - {\bvec{k}}^2 = 0$ in the expression (\ref{f11}), after
some algebra we obtain
\bea
\Re (\Pi_L) & \simeq & O(\alpha^2) \label{f17} ~~~, \\
\Re (\Pi_T) & \simeq & \frac{2 \alpha}{\pi} \, \int_0^\infty \, d p \,
\frac{p^2}{\sqrt{p^2 \, + \,  m_e^2}} \, \left( n_e \, + \, \ov{n}_e
\right) \; + \; O(\alpha^2) \label{f18} ~~~.
\eea
From the equation of motion for the 4-vector potential field $A^\mu$
\begin{equation}\label{f19}
  \left( - k^2 \, g_{\mu \nu} \; + \; \Pi_{\mu \nu} \right) \, A^\nu
  ~~~,
\end{equation}
which more usefully can be written more usefully in the form
\begin{equation}\label{f20}
  \left( k^2 \, - \, \Pi_T \right) \, R_{\mu \nu} \; + \;
  \left( k^2 \, - \, \Pi_L \right) \, Q_{\mu \nu} \; = \; 0 ~~~,
\end{equation}
we can deduce the perturbed dispersion relation of the transverse modes,
which is
\begin{equation}\label{f21}
  \omega^2 \, - \, {\bvec{k}}^2 \; = \; \Pi_T ~~~,
\end{equation}
with $\Pi_T$ given, at first order in $\alpha$, in (\ref{f18}).

\end{document}